\documentclass[11pt,a4paper]{article}
\pdfoutput=1

\usepackage[margin=1in]{geometry} 
\usepackage{authblk} 
\usepackage{adjustbox}
\usepackage{mathrsfs}
\usepackage{amsmath,amssymb,amsbsy,amsfonts,latexsym,graphicx}
\usepackage[colorlinks=true, linkcolor=blue, citecolor=blue, urlcolor=blue]{hyperref}
\usepackage{color, array, subfigure}
\usepackage{cite}
\usepackage{xcolor}

\allowdisplaybreaks

\title{Black Hole Chemistry Knows Extra Dimensions}

\author[1]{Kyung Kiu Kim\thanks{kimkyungkiu@kookmin.ac.kr}}
\author[2]{Jeongwon Ho\thanks{freejwho@gmail.com}}
\author[3]{Seungjoon Hyun\thanks{sjhyun@yonsei.ac.kr}}
\author[3]{Taehyeon Song\thanks{sthing@yonsei.ac.kr}}

\affil[1]{College of General Education, Kookmin University, Seoul 02707, Korea}
\affil[2]{Department of Physics, Institute of Basic Science, Sungkyunkwan University, Suwon 16419, Korea}
\affil[3]{Department of Physics, Yonsei University, Seoul 03722, Korea}

\begin{document}

\maketitle

\begin{abstract}
We study black hole chemistry in the eight-dimensional Einstein-Yang-Mills-Maxwell theory, where the stabilized size of the compact internal space dynamically determines the four-dimensional thermodynamic pressure. Analyzing AdS black holes, we show that larger extra dimensions lower phase-transition temperatures and broaden the range of thermodynamically excluded black hole sizes. Additionally, we investigate a geometric representation of the thermodynamic volume by defining a volume-constrained minimal surface using a lapse-weighted volume functional. This codimension-two surface minimizes its area under the constraint that its enclosed volume exactly matches the thermodynamic volume. This surface prescription naturally reproduces the event horizon for dyonic and Schwarzschild-AdS black holes. We also apply this to the Kerr--AdS black hole and find an off-horizon surface numerically and analytically for finite and small angular momenta. In addition, we interpret the thermodynamic volume in terms of an effective number of disentanglers in a MERA system.

\vspace{1em}
\end{abstract}

\section{Introduction}

Black hole chemistry is the extended thermodynamics obtained by identifying the cosmological constant with bulk pressure \cite{Kastor:2009wy, Dolan:2010ha, Dolan:2011xt, Cvetic:2010jb, Kubiznak:2012wp, Kubiznak:2016qmn}. In this framework, the black hole mass is interpreted as enthalpy rather than internal energy \cite{Kastor:2009wy}, and the thermodynamic volume is defined as the quantity conjugate to pressure. This viewpoint provides a novel thermodynamic description of the Hawking--Page transition \cite{Hawking:1982dh}, which is interpreted as a liquid-solid phase transition in black hole chemistry \cite{Kubiznak:2014zwa}. In the context of AdS/CFT, the same transition is related to the confinement-deconfinement transition of the dual gauge theory \cite{Witten:1998zw}. It was also found that the Reissner-Nordstr\"om-AdS black hole exhibits a critical point and a Small/Large black hole phase transition analogous to the liquid-gas transition of the Van der Waals fluid \cite{Kubiznak:2012wp}. Since then, black hole chemistry has been applied to a wide range of systems, including reentrant phase transitions \cite{Altamirano:2013ane, Frassino:2014pha}, triple points \cite{Altamirano:2013uqa, Wei:2014hba, Dehghani:2020blz}, heat engines, non-standard critical points, superfluid transitions, and molecular microstate structures. See \cite{Altamirano:2014tva} for a review and, for example, \cite{Johnson:2014yja, Dolan:2014vba, Hennigar:2016xwd, Wei:2015iwa, Mancilla:2024spp} for related developments.  Extensions beyond the standard AdS setting have also been studied, including accelerating backgrounds \cite{Astorino:2016ybm, Anabalon:2018ydc, Anabalon:2018qfv}, flat spacetime \cite{Wu:2022xmp}, de Sitter spacetime \cite{Dolan:2013ft, Mbarek:2018bau, Simovic:2018tdy}, and soliton backgrounds \cite{Mbarek:2016mep, Quijada:2023fkc}.

Despite these successes, identifying pressure with the cosmological constant raises a conceptual question. The cosmological constant is a parameter of the theory, and therefore its variation is not, in general, an ordinary variation within a fixed theory.  It is therefore desirable to realize the quantity playing the role of the cosmological constant dynamically in explicit gravitational systems. One way to address this issue is to introduce a dynamically generated cosmological constant, as in \cite{Henneaux:1984ji, Teitelboim:1985dp, Creighton:1995au}\footnote{In these works, the cosmological constants appear as the energy-momentum of higher-form gauge fields without extra dimensions.}. The purpose of the present paper is to provide another simple realization in which the stabilized size of compact extra dimensions determines the effective four-dimensional cosmological constant.  In this way, the variation of the thermodynamic pressure can be related to the variation of an ordinary modulus of the higher-dimensional geometry.

If we don't consider any additional physical realization, as in the previous case, the holographic interpretation of this variation is subtle and profound. In AdS/CFT \cite{Maldacena:1997re}, the cosmological constant is related to the rank of the gauge group, the Newton constant, or equivalently, to the number of degrees of freedom in the dual field theory. Thus, changing the cosmological constant is not simply identified in the large-$N$ conformal field theory. Several approaches have been proposed to clarify this issue, including the inclusion of $1/N$ corrections and higher-derivative terms \cite{Karch:2015rpa, Sinamuli:2017rhp}, the introduction of a chemical potential conjugate to the central charge \cite{Cong:2021fnf}\footnote{An earlier proposal for a chemical potential conjugate to the number of M2-branes is given in \cite{Kim:2018nzy}.}, and the variation of the spatial volume of the dual CFT \cite{Ahmed:2023snm}. See \cite{Mann:2024sru} for a recent review. These developments indicate that the pressure-volume sector of black hole chemistry may encode more than a formal thermodynamic extension. It is therefore useful to construct explicit bulk models in which the effective cosmological constant has a clear dynamical origin

In this work, we consider an eight-dimensional Einstein--Yang--Mills--Maxwell theory with a cosmological constant. We decompose the spacetime into a four-dimensional Lorentzian spacetime and an Euclidean compact space. This compact space supports a Yang-Mills instanton configuration. The compact manifolds relevant for our consideration are $\mathbb{S}^4$, $\mathbb{CP}^2$, and $\mathbb{S}^2\times\mathbb{S}^2$.  Such instanton configurations on compact extra dimensions have been used to generate a stable Minkowski spacetime \cite{Randjbar-Daemi:1983xth} and to study cosmological evolution with dynamical compactification \cite{Kim:2018mfv, Kim:2023gbm, Ho:2023jqz, Ho:2023wcg}.  Here, we use the same basic compactification mechanism to study black holes in the black hole chemistry framework.

Our metric ansatz is
\begin{align}\label{metric ansatz 00}
ds^2 = g_{\mu\nu}(x)dx^\mu dx^\nu + L^2 e^{2f(x)}h_{ab}(y)dy^a dy^b \, ,
\end{align}
where $L$ is the length scale associated with the cosmological constant of the eight-dimensional theory.  The field $\mathcal{X}=e^{2f}$ measures the size of the compact internal space. After dimensional reduction to the four-dimensional Einstein frame, the effective potential for $\mathcal{X}$ is generated by the internal curvature, higher-dimensional cosmological constant, and the Yang-Mills instanton contribution.  Its stable extremum $\mathcal{X}_s$ determines the effective four-dimensional cosmological constant $\tilde{\Lambda}_{\rm eff}$. Depending on the higher-dimensional parameters, the resulting vacuum can be either anti-de Sitter or de Sitter. In the present work, the black-hole chemistry analysis is carried out mainly on the anti-de Sitter branch, where the thermodynamic pressure is identified as \begin{align}
\mathsf{P}_{\mathrm{th}}=-\frac{\tilde{\Lambda}_{\rm eff}}{8\pi G_4} \, .
\end{align}
Thus, the size of the compact internal space provides a dynamical origin for the thermodynamic pressure in the effective four-dimensional black hole system.

We analyze three classes of four-dimensional black holes in this framework: Schwarzschild--AdS, dyonic AdS, and Kerr--AdS. The Schwarzschild--AdS black hole exhibits the Hawking--Page transition. The dyonic and Kerr--AdS black holes exhibit the Small/Large black hole transition.  In each case, the usual phase diagram in the $(\mathsf{P}_{\mathrm{th}}, T)$ plane can be re-expressed as a phase diagram involving the stabilized extra-dimensional size.  We find that the transition temperature decreases as the size of the extra dimensions increases. Moreover, the range of excluded black hole sizes broadens as the size of the internal space increases.  Equivalently, the thermodynamic phase structure of the effective four-dimensional black hole is sensitive to the size of the compact extra dimensions.

This size restriction is reminiscent of the Gregory-Laflamme
instability, in which the relative scale between a black object and a
compact direction plays an important role \cite{Gregory:1993vy}.  The mechanism in the present work is different, however, since our restriction arises through the effective four-dimensional thermodynamic pressure rather than from a perturbative instability of a black string or brane.  Nevertheless, both phenomena illustrate that compact extra dimensions can impose nontrivial constraints on black objects. We return to this difference in Section \ref{secTpsc}.

The latter part of this work addresses the geometric meaning of the thermodynamic volume.  Since the compact modulus determines $\mathsf{P}_{\mathrm{th}}$, the conjugate quantity $\mathsf{V}_{\mathrm{th}}$ is the natural thermodynamic variable describing how the four-dimensional black hole responds to changes in the compact internal space. Beyond this geometric role, the thermodynamic volume invites a holographic reading. In the surface/state correspondence \cite{Miyaji:2015yva}, convex surfaces in asymptotically AdS geometries are mapped to quantum states defined by cutting the links of a tensor network, and in particular of the Multi-scale Entanglement Renormalization Ansatz (MERA) and its continuous version, cMERA. In that picture, the enclosed volume measures the number of disentanglers, while the area counts the cut links. Since our model supplies both a dynamical pressure and a well-defined representative surface for $\mathsf{V}_{\rm th}$, it provides a natural setting in which to test this identification. We find, however, that the lapse-weighted (LW) volume is observer-dependent, so that it should be interpreted as an effective number of disentanglers redshifted with respect to the boundary observer, in direct analogy to the relation between the Hawking and Tolman temperatures.

This paper is organized as follows.  In Section 2, we derive the four-dimensional reduced action in the Einstein frame and analyze the stable vacua for compact internal spaces $\mathbb{S}^4$. The other compact spaces($\mathbb{CP}^2$, and $\mathbb{S}^2\times\mathbb{S}^2$) are discussed in Appendix A. In Section 3, we study the phase diagrams of Schwarzschild-AdS, dyonic AdS, and Kerr-AdS black holes, and express the corresponding transition curves and excluded black hole sizes in terms of the extra-dimensional size.  We also discuss possible implications for black hole merging and fission in the presence of the extra dimensions.  In Section 4, we formulate the constrained minimal-area prescription for the thermodynamic-volume surface, show that it reproduces the event horizon in the static dyonic AdS case, and construct its nontrivial rotating continuation in Kerr-AdS.  We also compare the Kerr-AdS thermodynamic volume with the LW volumes enclosed by the horizon and the ergosphere, and interpret both the off-horizon surface and its LW volume in terms of the number of cut links and effective disentanglers in a (c)MERA network. Section 5 contains our conclusions and future directions. In addition, the Appendix contains some details needed for calculations in the main body of the paper.

\section{4-dimensional Reduced Action in the Einstein Frame}\label{effective models}

This section is devoted to deriving a $4$-dimensional reduced action from the 8-dimensional action. We will consider $\mathbb{S}^4$ as the compact manifold. In \cite{Ho:2023jqz, Ho:2023wcg}, the authors show that the only compact Einstein manifolds that are consistent with this instanton configuration and our metric ansatz are $\mathbb{S}^4$, $\mathbb{CP}^2$, and $\mathbb{S}^2\times \mathbb{S}^2$. See Appendix A for the $\mathbb{S}^2 \times \mathbb{S}^2$ and $\mathbb{CP}^2$ cases.

\subsection{Reduced action for $\mathbb{S}^4$}

Let us start with the Einstein-Yang-Mills-Maxwell system with a cosmological constant. The gauge group is chosen as $SU(2)\times U(1)$; the action is then given by
\begin{align}\label{original action}
S_{tot} = S_{grav} + S_{gauge}~,
\end{align}
where
\begin{align}\label{action}
&S_{grav}=\frac{1}{16 \pi  \mathbf{G}_{(8)}} \int d^8 x \sqrt{-G} \left(  R - 2 \Lambda \right)  \nonumber\\
&S_{gauge}=\int  d^8 x \sqrt{-G} \left(
\frac{1}{4 g_{\text{YM}}^2} \text{Tr} F_{M N} F^{M N} 
- \frac{1}{4 e^2} \mathcal{F}_{M N} \mathcal{F}^{M N} \right)\,.
\end{align}
Here, $e$ is the coupling of the $U(1)$ gauge field and $g_{\text{YM}}$ is that of the $SU(2)$ gauge field. All the fields associated with $SU(2)$ are matrix-valued. For instance, $F_{MN}\equiv F_{MN}^i \tau^i$, where $\tau^i$ is the generator of $SU(2)$ group. The normalization of the generators is taken as $\text{Tr}\, \tau^i\tau^j = -\delta^{ij}$. The equations of motion are given by
\begin{align}\label{eom00}
&D_M F^{M N} = 0,~  \nabla_M \mathcal{F}^{M N} =  0,\nonumber\\
&R_{MN} -\frac{1}{2}\left( R - 2 \Lambda \right)G_{MN} = 8\pi \mathbf{G}_{(8)}  \left(T_{MN}^{U(1)}+T_{MN}^{SU(2)} \right)\,,
\end{align}
where $D_M$ denotes the covariant derivative of the $SU(2)$ gauge field. The energy-momentum tensors are given as follows: 
\begin{align}
&T_{MN}^{U(1)} = \frac{1}{e^2} \left( \mathcal{F}_{M}{}^{P} \mathcal{F}_{NP}  - \frac{1}{4} G_{MN}  \mathcal{F}_{PQ} \mathcal{F}^{PQ} \right)\nonumber\\
&T_{MN}^{SU(2)} = \frac{1}{g_{\text{YM}}^2 }
\left(  \frac{1}{4} G_{M N} \text{Tr} F_{PQ} F^{PQ}  -  \text{Tr} F_{M}{}^{P} F_{NP} \right)  
\end{align}

We continue our study with the following ansatz to find the stable size of extra dimensions:  
\begin{align}
&{ds}^2=g_{\mu \nu }(x){dx}^{\mu }{dx}^{\nu } + L^2 e^{2f(x)}  h_{ab}(y) {dy}^a {dy}^b\label{metricansatz}\\
& A=A_{a}(y){dy}^{a} ,\;\mathcal{A}=\mathcal{A}_{\mu}(x){dx}^{\mu} \label{gaugeansatz}
\end{align} 
where $h_{ab}$ is the metric of $\mathbb{S}^4$ and $L$ is the length scale of $\Lambda$. Thus, we represent the cosmological constant in terms of this scale as 
\begin{align}
\Lambda=\frac{\mathbf{s}}{L^2}\,,
\end{align}
where $\mathbf{s}$ is the sign of $\Lambda$, which has a value $1$, $-1$ or $0$. In the case of $\mathbf{s}=0$, $L$ denotes the length scale related to the physical size of extra dimensions. The nonabelian gauge field $A$ is taken as an (anti) instanton solution on $\mathbb{S}^4$, so $F$ is (anti) self-dual.

Using the above metric and gauge field configurations, the equations of motion are decomposed as follows:
\begin{align}\label{eom-decompo}
R_{\mu\nu}^{(g)}-\frac{1}{2}R^{(g)}g_{\mu\nu} =& - \left( 4 \nabla^{(g)2}f+ 10\nabla^{(g)}f\cdot \nabla^{(g)}f -\frac{1}{2L^2}e^{-2 f}R^{(h)} \right)g_{\mu\nu}\nonumber\\
&+4\left(\nabla_{\mu}^{(g)}\nabla_{\nu}^{(g)}f + \partial_\mu f \partial_\nu f \right)-\frac{8\pi \mathbf{G}_{(8)}}{g_{\text{YM}}^2L^4}e^{-4f}\rho_n g_{\mu\nu}\nonumber\\
&-\frac{\mathbf{s}}{L^2}g_{\mu\nu}+\frac{8\pi \mathbf{G}_{(8)}}{e^2}\left(\mathcal{F}_\mu{}^\lambda\mathcal{F}_{\nu\lambda}-\frac{1}{4}\mathcal{F}_{\lambda\sigma}\mathcal{F}^{\lambda\sigma}g_{\mu\nu} \right)\nonumber\\
R_{ab}^{(h)}-\frac{1}{2}R^{(h)}h_{ab}=& 3 L^2 e^{2f}\left(\frac{1}{6}R^{(g)}-\nabla^{(g)2}f - 2 \nabla^{(g)}f\cdot\nabla^{(g)}f \right) h_{ab} \nonumber\\
&-\mathbf{s}\, e^{2f}h_{ab}-\frac{8\pi \mathbf{G}_{(8)}L^2}{4e^2}\mathcal{F}_{\mu\nu}\mathcal{F}^{\mu\nu}e^{2f}h_{ab}\nonumber\\
\nabla^{(g)}_\mu\left( e^{4f} \mathcal{F}^{\mu\nu}\right)=&0\,,
\end{align}
where all the indices are raised by $g^{\mu\nu}$ and $h^{ab}$, and $\rho_n = -\frac{1}{4}h^{ac}h^{bd}\text{Tr} F_{ab}F_{cd}$. In the $\mathbb{S}^4$ case, the field strength-square is given by $\rho_n=\frac{H^2}{48}\left(R^{(h)}\right)^2=\frac{|n|}{48}\left(R^{(h)}\right)^2$ for both instanton and anti-instanton, where $n$ is a winding number of the instanton configuration and $H$ is a constant parameter. This parameter is based on the $n$-winding extension of the one-instanton studied in \cite{Dolan:1980jb}. In addition, we use the following parameters and a curvature choice:
\begin{align}
\mathcal{R}\equiv R^{(g)},~~R^{(h)}=1,~~\kappa_4^2=\frac{8\pi \mathbf{G}_{(8)}}{\mathcal{V}L^4}=8\pi \mathbf{G}_{4},~~\ell_e^2\equiv \frac{4\pi \mathbf{G}_{(8)}}{e^2},~~\ell_H^2 =\frac{\pi \mathbf{G}_{(8)}|n|}{3g_{\text{YM}}^2}\,.
\end{align}
Here, $\ell_H^2$ is quantized. However, we assume that the ratio of the Newton constant to the Yang-Mills coupling is small, which is reasonable due to the weak gravity conjecture \cite{Arkani-Hamed:2006emk}. So, we will regard $\ell_H$ as a continuous parameter.

Together with the above ansatz (\ref{metricansatz}-\ref{gaugeansatz}) and the action (\ref{action}), we can integrate out the extra dimension part of the action and obtain the reduced action as follows:
\begin{align}\label{reduce action}
S_{tot}
=&\frac{\mathcal{V}L^4}{16 \pi \mathbf{G}_{(8)}} \int d^{4} x \sqrt{-g} e^{4f}\left(\mathcal{R}+\frac{1}{L^2}e^{-2f}R^{(h)}+12\nabla_{\mu}f\nabla^{\mu}f-\frac{2\mathbf{s}}{L^2} \right) \nonumber \\
&+\mathcal{V}L^4\int d^{4}x\sqrt{-g} e^{4f}\left(\frac{1}{4 g_{\text{YM}}^2 L^4} e^{-4f}\text{Tr} F_{ab} F^{ab}- \frac{1}{4 e^2} \mathcal{F}_{\mu \nu} \mathcal{F}^{\mu \nu} \right) \nonumber \\
=&\frac{1}{2\kappa_4^2} \int d^{4} x \sqrt{-g} \left\{e^{4f}\left( \mathcal{R} + 12\nabla_{\mu}f\nabla^{\mu}f-\ell_e^2\mathcal{F}_{\mu \nu} \mathcal{F}^{\mu \nu} \right)\right.\nonumber\\
&~~~~~~~~~~~~~~~~~~~~~~~~~~~~~~~~\left.-\frac{2\mathbf{s}}{L^2}e^{4f}+\frac{1}{L^2}e^{2f}  -\frac{\ell_H^2}{L^4}\right\}\,,
\end{align} 
where $\mathcal{V}=3\cdot 2^7\pi^2$ based on $R^{(h)}=1$\footnote{The physical volume of the extra dimension is written by $\mathcal{V}_{\text{phy}}\equiv\mathcal{V}L^4\mathcal{X}_s^2$. This is valid even with $\mathbf{s}=0$.}. The Einstein equation, the Maxwell equation, and the equation of motion of the scalar field $f$ from this reduced action coincide with the 8-dimensional equations (\ref{eom-decompo}). Furthermore, one may write down the action (\ref{reduce action}) in the Einstein frame by performing a Weyl transformation $g_{\mu\nu}=e^{-4f}\tilde{g}_{\mu\nu}$. The transformed action is
\begin{align}
\mathcal{I}=\frac{1}{2\kappa_4^2}\int d^{4} x \sqrt{-\tilde{g}}\left(\tilde{\mathcal{R}}-\ell_e^2 e^{4f} \tilde{\mathcal{F}}_{\mu \nu} \tilde{\mathcal{F}}^{\mu \nu} -12\tilde{\nabla}_{\mu}f\tilde{\nabla}^{\mu}f - V(\mathcal{X})\right)\,, \label{effactions4}
\end{align}
where the potential is given by
\begin{align}\label{PotentialEin}
V(\mathcal{X})= \frac{2\mathbf{s}}{L^2} e^{-4f} -\frac{1}{L^2}e^{-6f}
+\frac{\ell_H^2}{L^4}e^{-8f}=\frac{2\mathbf{s}}{L^2} \frac{1}{\mathcal{X}^2} -\frac{1}{L^2}\frac{1}{\mathcal{X}^3}
+\frac{\ell_H^2}{L^4}\frac{1}{\mathcal{X}^4}\,.
\end{align}
The tilde $``\sim"$ denotes the fields associated with the Einstein frame metric $\tilde{g}_{\mu\nu}$. The field $\mathcal{X}$ is defined as $\mathcal{X}=e^{2f}$, and it describes the size of the extra dimensions.

\begin{figure}
        \centering
        \begin{subfigure}{}
            \centering
            \includegraphics[width=45mm]{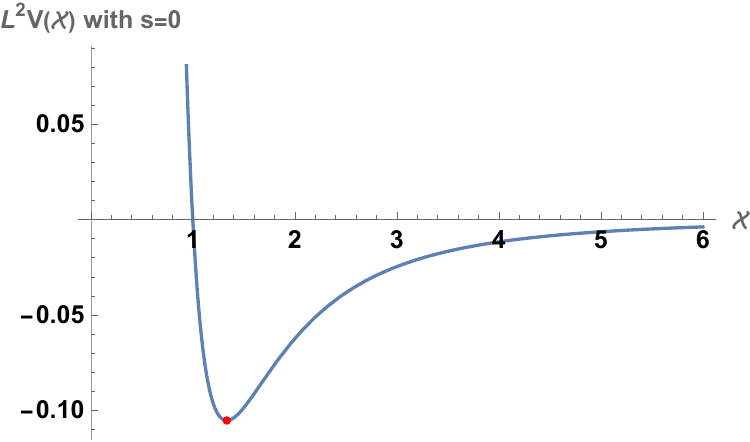}
        \end{subfigure}
        \begin{subfigure}{}
            \centering
            \includegraphics[width=45mm]{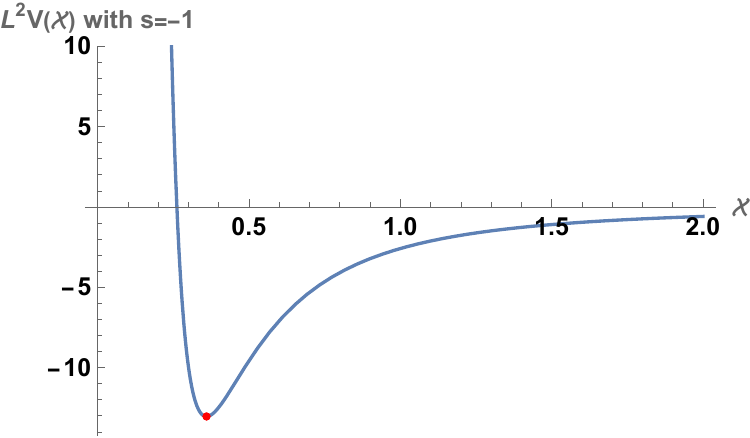}
         \end{subfigure}
         \begin{subfigure}{}
            \centering
            \includegraphics[width=45mm]{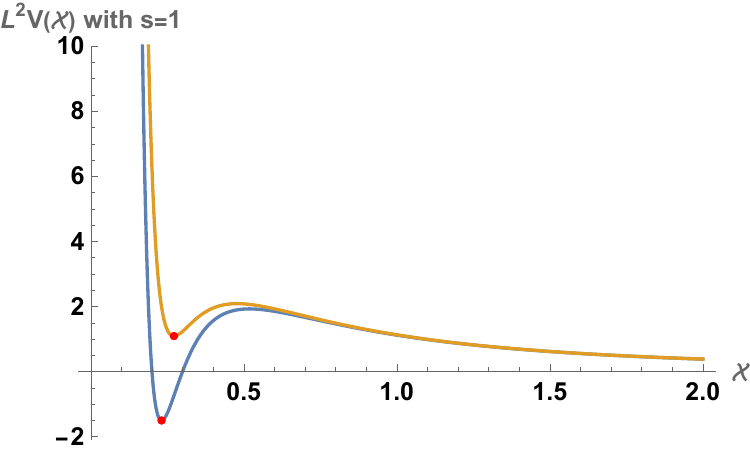}
         \end{subfigure}
         \caption{Typical shapes of the potential: The red dots denote the stable vacua for the constant warping factor $e^{2f}$.}\label{fig: potentials}
    \end{figure}

The potential (\ref{PotentialEin}) determines the stable vacua of the system. Let us examine the shape of the potential in various cases. For $\mathbf{s}=0$ and $\mathbf{s}=-1$, there exists only one extremum, corresponding to an AdS vacuum. Representative potentials for these cases are shown in Figure \ref{fig: potentials}. On the other hand, the $\mathbf{s}=1$ case exhibits two extrema. However, one of them is unstable. As shown in Figure \ref{fig: potentials}, the larger extremum corresponds to an unstable vacuum in terms of $\mathcal{X}$. In comparison, the smaller extremum represents a relevant vacuum as a classical solution. This can be either a de Sitter or anti-de Sitter vacuum, depending on the value of the parameter $\ell_H^2$. This situation is reminiscent of the anti-D3-brane uplift mechanism discussed in \cite{Kachru:2003aw}, although the origin of the uplift to the de Sitter vacuum is different. The stable vacua are summarized in Table \ref{table:2k}.

{}

\begin{table}[h!]
\caption{ {\bf Stable vacua for the reduced action (\ref{effactions4})}}
\vspace{3mm}
\begin{adjustbox}{width=\textwidth}
\begin{tabular}{c|c|c|c}
\hline\hline  
$\mathbf{s}$& $\mathcal{X}_s=e^{2f}$ &  $V(\mathcal{X}_s)=2\tilde{\Lambda}_{\text{eff}}$ &Sign of $V$\\ 
\hline 
& & & \\
$+1$& $\frac{3}{8}\left(1-\sqrt{1-\frac{64}{9}\frac{\ell^2_H}{L^2}}\right)$   & $\frac{2^8}{3^3 L^2\left(1-\sqrt{1-\frac{64}{9} \left(\frac{\ell _H}{L}\right)^2}\right)^3} {\left(1-\frac{\frac{16}{3}\left(\frac{\ell _H}{L}\right)^2}{ \left(1-\sqrt{1-\frac{64}{9} \left(\frac{\ell _H}{L}\right)^2}\right)}\right)}$ & $V>0$$\left( \frac{1}{2\sqrt{2}}<\frac{\ell_H}{L}<\frac{3}{8}\right)$\\
&for $\left(\frac{\ell_H}{L}\right)<\frac{3}{8}$ & & $V<0$  $\left(\frac{\ell_H}{L}<\frac{1}{2\sqrt{2}}\right)$  \\
\hline 
& & & \\
$0$& $\frac{4}{3}\frac{\ell^2_H}{L^2}$  & $-\frac{3^3 }{2^8 L^2}\left(\frac{L}{\ell _H}\right)^6 $ & $V<0$\\
& & & \\
\hline
& & &\\
$-1$& $\frac{3}{8} \left(\sqrt{\frac{64 \ell _H^2}{9 L^2}+1}-1\right)$   &$-\frac{2^8}{3^3 L^2} \frac{\frac{16}{3} \left(\frac{\ell _H}{L}\right)^2+\left(1-\sqrt{\frac{64}{9} \left(\frac{\ell _H}{L}\right)^2+1}\right)}{\left(1-\sqrt{\frac{64}{9} \left(\frac{\ell _H}{L}\right)^2+1}\right){}^4} $&  $V<0$\\
& & & \\
\hline\hline
\end{tabular}
\end{adjustbox}
\label{table:2k}
\end{table}

Here, one can observe a transition between de Sitter and anti-de Sitter spacetimes in the case with a positive $\Lambda(\mathbf{s}=1)$. Larger extra-dimensional sizes allow for dS spacetimes, while smaller extra-dimensional sizes correspond to AdS spacetimes. The critical size of the extra dimensions at which this transition occurs is $\mathcal{X} = 1/4$. This transition is governed by the parameter $\ell_H^2$ or the instanton number $n$.

Our paper focuses on the constant $\mathcal{X}$ case. In this case, the reduced action can be written as
\begin{align}\label{Action4-const-f}
\mathcal{I}=\frac{1}{2\kappa_4^2}\int d^{4} x \sqrt{-\tilde{g}}\left(\tilde{\mathcal{R}}-\ell_e^2 \mathcal{X}_s^2 \tilde{\mathcal{F}}_{\mu \nu} \tilde{\mathcal{F}}^{\mu \nu}  - V(\mathcal{X}_s)\right)\,,
\end{align}
where $\mathcal{X}_s$ denotes the stable size of the extra dimensions. In this constant $\mathcal{X}$ configuration, the expected geometries include dS and AdS vacua, as well as their charged and rotating black holes. These constitute the main ingredients of the present work.

However, this action is not consistently truncated for the case with constant $\mathcal{X}$. The equations of motion derived from the original action (\ref{effactions4}) for a constant field $\mathcal{X}$ do not permit a nontrivial function of $\tilde{\mathcal{F}}_{\mu\nu}\tilde{\mathcal{F}}^{\mu\nu}$; the only allowed configuration is the trivial case. Therefore, the constant $\mathcal{X}$ configuration imposes an additional constraint
\begin{align}\label{F2=0}
\tilde{\mathcal{F}}^{\mu\nu}\tilde{\mathcal{F}}_{\mu\nu}=0\,.
\end{align}
This constraint is difficult to satisfy for charged black holes. In four dimensions, the only way to fulfill this condition is by requiring the electric and magnetic charges to be equal. We will examine this case explicitly later, along with the corresponding thermodynamic properties.  In the following subsections, we describe how to treat $\mathbb{CP}^2$ or $\mathbb{S}^2\times\mathbb{S}^2$ as the extra compact space.

\section{Black Hole Phase Transitions and Excluded Sizes from Extra Dimensions}

In this section, we explore phase transitions of the system described by the reduced action (\ref{effactions4}), employing the thermodynamic volume within the framework of black hole chemistry. We subsequently analyze the influence of extra dimensions on these phase transitions.

\subsection{Schwarzschild black hole}\label{Schwar-section}

We begin by examining the phase transition associated with the Schwarzschild-AdS black hole. It is well established that the Hawking-Page phase transition occurs between thermal AdS and the Schwarzschild-AdS black hole. The metric of the black hole is given by
\begin{align}
&ds^2 = -U(r) dt^2 + r^2 \left(d\theta^2 + \sin^2\theta d\varphi^2 \right) + \frac{dr^2}{U(r)}\,,
\end{align}
where
\begin{align}
U(r) = 1 - \frac{2 \mathbf{G}_4 M}{r} - \frac{\tilde{\Lambda}_{\text{eff}}}{3}r^2\,.
\end{align}
The effective cosmological constant is obtained from the potential $V(\mathcal{X}_s)/2= \frac{\ell_H^2 +L^2\left( 2\mathbf{s}\mathcal{X}_s^2-\mathcal{X}_s\right)}{6 L^4 \mathcal{X}_s^4}$, where $\mathcal{X}_s$ represents the stable size of the extra dimensions, $e^{2f}$. Thus, the effective cosmological constant can be derived from the equation of motion $\partial_\mathcal{X}V|_{\mathcal{X}=\mathcal{X}_s}=0$, yielding:
\begin{align}\label{Lambda-eff}
\tilde{\Lambda}_{\text{eff}}= \frac{4 \mathbf{s} \mathcal{X}_s-1}{8 L^2 \mathcal{X}_s^3}\,.
\end{align}

We now outline the extended thermodynamics of the Schwarzschild-AdS black hole, treating the effective cosmological constant as a thermodynamic pressure. For a comprehensive review, see \cite{Altamirano:2014tva}. Key thermodynamic quantities—entropy, temperature, and mass parameter—are given by:
\begin{align}
S = \frac{\pi r_h^2}{\mathbf{G}_4}~,~T=\frac{1}{4 \pi }\left(\frac{1}{r_h}-3 \frac{\tilde{\Lambda}_{\text{eff}}}{3} r_h\right)~,~M= \frac{1}{\mathbf{G}_4}\left( \frac{r_h}{2}-\frac{1}{2} \frac{\tilde{\Lambda}_{\text{eff}}}{3} r_h^3\right)\,,
\end{align}
where $r_h$ denotes the horizon radius. In the context of black hole chemistry, the thermodynamic pressure is defined as:
\begin{align}\label{Pth-Lambda}
\mathsf{P}_{\mathrm{th}}= -\frac{\tilde{\Lambda}_{\text{eff}}}{8\pi \mathbf{G}_4}\,.
\end{align}
The first law of black hole thermodynamics is then derived as:
\begin{align}\label{first law M}
\delta M = T \delta S + \mathsf{V}_{\mathrm{th}} \delta \mathsf{P}_{\mathrm{th}}\,,
\end{align}
where the thermodynamic volume $\mathsf{V}_{\mathrm{th}}$ can be read as 
\begin{align}
\mathsf{V}_{\mathrm{th}}=\frac{4}{3}\pi r_h^3\,.
\end{align}
Using this definition, the Smarr relation can be written as
\begin{align}
M = 2 S T - 2 \mathsf{P}_{\mathrm{th}} \mathsf{V}_{\mathrm{th}}.
\end{align}

\begin{figure}
        \centering
        \begin{subfigure}{}
            \centering
            \includegraphics[width=60mm]{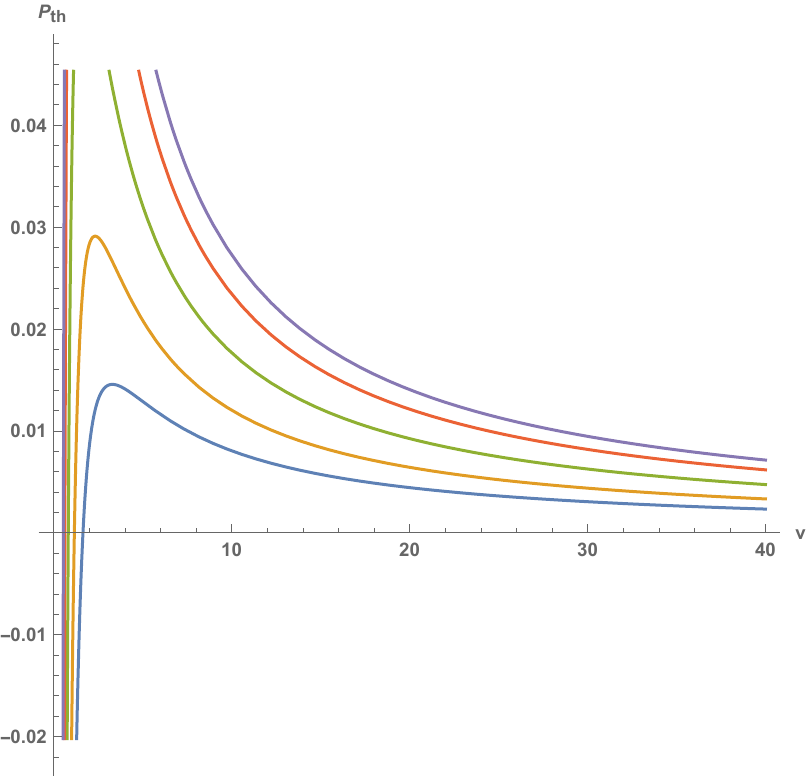}
        \end{subfigure}
        \begin{subfigure}{}
            \centering
            \includegraphics[width=60mm]{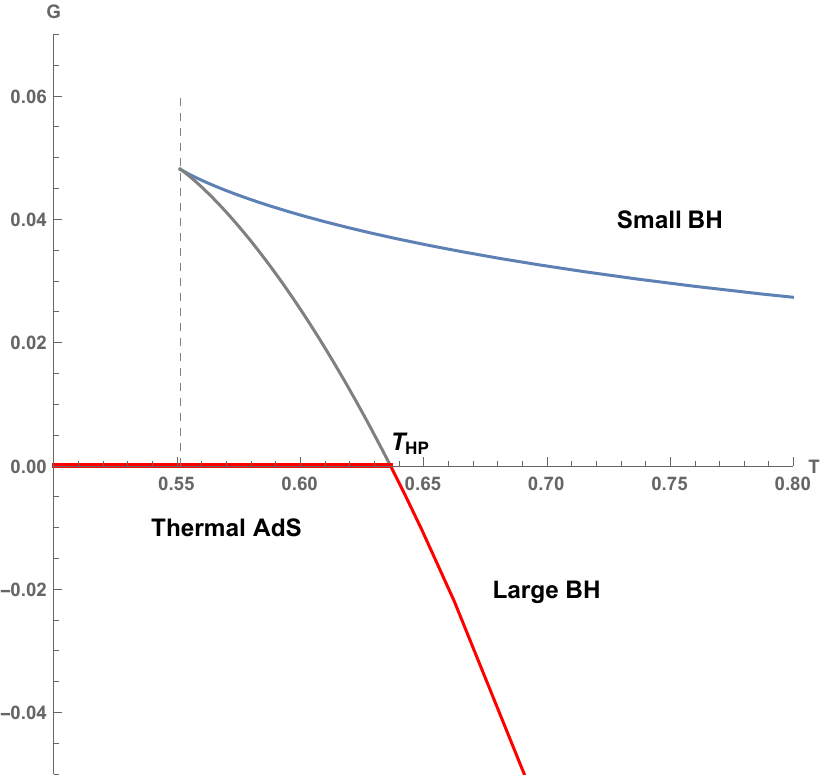}
         \end{subfigure}
         \caption{{\bf  $\mathsf{P}_{\mathrm{th}}-v$ curve (Left):} The thermodynamic pressure curves indicate that small black holes with $v<\frac{1}{\pi T}$ are unstable. For simplicity, we set $\mathbf{G}_4=1$. The upper curves correspond to higher temperatures.\\
{\bf Gibbs free energy of black holes and thermal AdS(Right):} This gravitational system exhibits a phase transition between thermal AdS and large black holes, occurring when the Gibbs free energy of the black holes approaches zero. The red curve demonstrates that this transition is of first order.}\label{sPth00}
    \end{figure}

To identify the internal energy, we rearrange equation (\ref{first law M}) as:
\begin{align}
\delta\left(M-\mathsf{V}_{\mathrm{th}}\mathsf{P}_{\mathrm{th}}\right)=T\delta S-P_{\mathrm{th}}\delta \mathsf{V}_{\mathrm{th}}\,.
\end{align}
This implies that the internal energy is $\mathcal{E}=M-\mathsf{V}_{\mathrm{th}}\mathsf{P}_{\mathrm{th}}$, indicating that the black hole mass corresponds to enthalpy. The Gibbs free energy $\mathcal{G}$, a key thermodynamic potential in the ($\mathsf{P}_{\mathrm{th}}, T$) phase space, is given by
\begin{align}
\mathcal{G} = \mathcal{E} + \mathsf{P}_{\mathrm{th}}\mathsf{V}_{\mathrm{th}} - S T  = M- ST\,.
\end{align}
To elucidate the phase structure, we express $\mathsf{P}_{\mathrm{th}}$ in terms of the thermodynamic volume and temperature. Defining the specific volume as $v\equiv \left(\frac{6 \mathsf{V}_{\mathrm{th}}}{\pi} \right)^{1/3}= 2 r_h$,  the pressure becomes:
\begin{align}
\mathsf{P}_{\mathrm{th}} =\frac{1}{\mathbf{G}_4} \left( \frac{T}{v}-\frac{1}{2 \pi  v^2}  \right)\,.
\end{align}
The system exhibits thermodynamic instability where $\frac{\partial \mathsf{P}_{\mathrm{th}}}{\partial v} > 0$, corresponding to solutions with $v<\frac{1}{\pi T}$. This instability underlies the well-known Hawking-Page transition. The ($\mathsf{P}_{\mathrm{th}}-v$) diagram is plotted to the left of Figure \ref{sPth00}, consistent with findings in prior studies such as \cite{Altamirano:2014tva}.

\begin{figure}
        \centering
        \begin{subfigure}{}
            \centering
            \includegraphics[width=60mm]{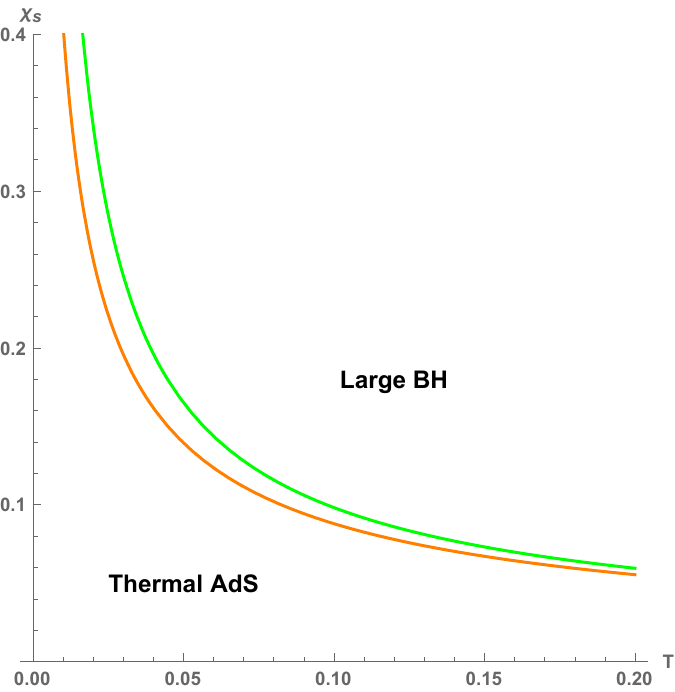}
        \end{subfigure}
        \begin{subfigure}{}
            \centering
            \includegraphics[width=60mm]{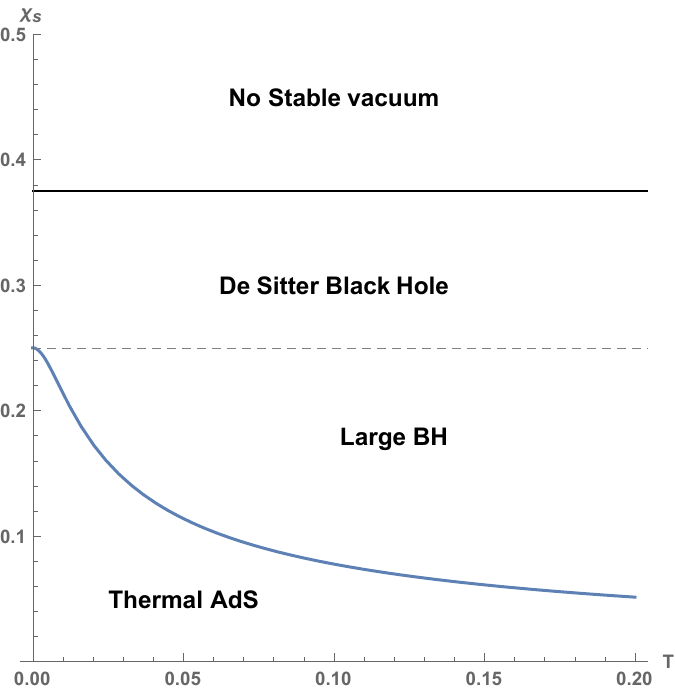}
         \end{subfigure}
         \caption{{\bf Phase diagram as a function of extra dimension size:} The left panel corresponds to $\mathbf{s}=-1$(Green) and $0$(Red), while the right panel corresponds to $\mathbf{s}=+1$. We set ${1}/{(64\pi \mathbf{G}_4 L^2)}=0.02$. The top region (``No stable vacuum'') drives an inflating spacetime since the effective potential is a slow-rolling potential without a stable extremum.}\label{sPD01}
    \end{figure}

\begin{figure}
        \centering
        \begin{subfigure}{}
            \centering
            \includegraphics[width=45mm]{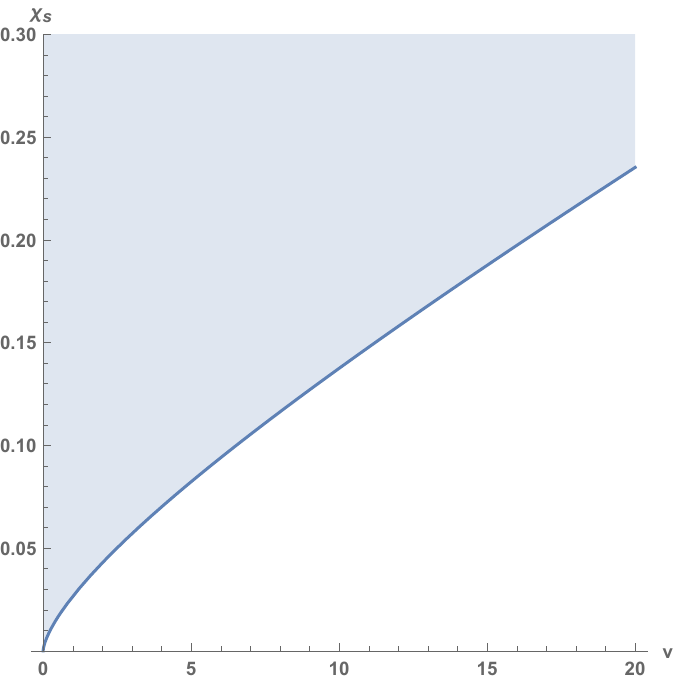}           
        \end{subfigure}
               \begin{subfigure}{}
            \centering
            \includegraphics[width=45mm]{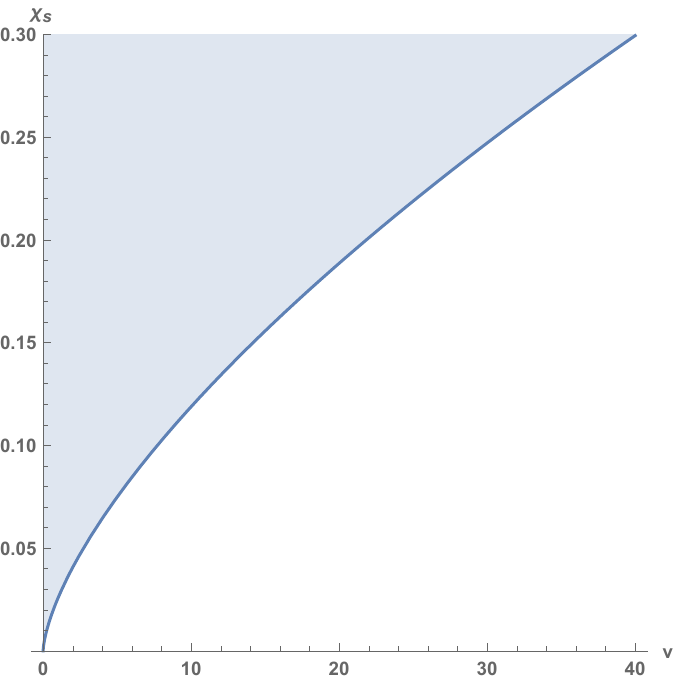}           
        \end{subfigure}
        \begin{subfigure}{}
            \centering
            \includegraphics[width=45mm]{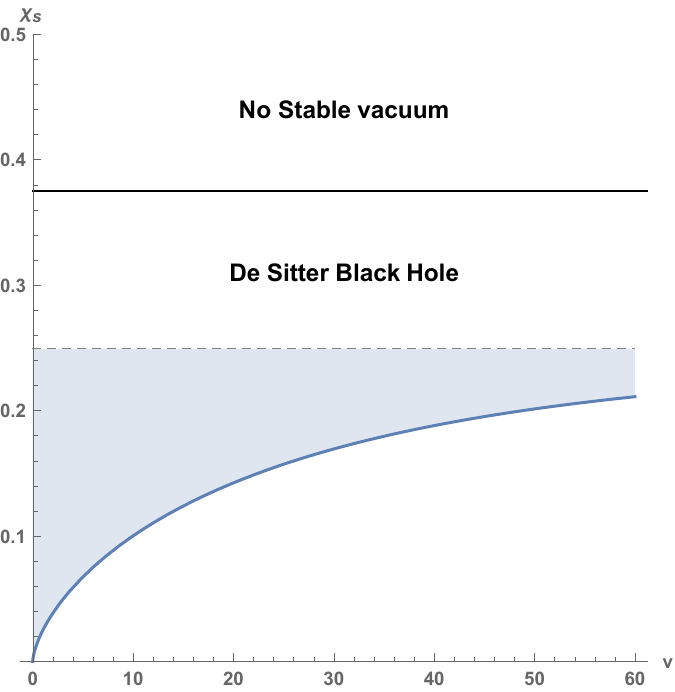}
         \end{subfigure}
         \caption{{\bf Thermodynamically disfavored black hole sizes:} Colored regions denote thermodynamically disfavored black hole sizes, with panels corresponding to $\mathbf{s}=-1$, $0$, and $+1$ from left to right. We set ${1}/{(64\pi \mathbf{G}_4 L^2)}=0.02$. The region not favored in the specified ensemble broadens as the extra dimension size increases. The de Sitter region is depicted as a blank white area.}\label{svx}
    \end{figure}

The Gibbs free energy of the Schwarzschild black hole is depicted on the right side of Figure \ref{sPth00}. This Schwarzschild black hole solution is a solution of the action (\ref{effactions4}) with vanishing charge and angular momentum. Additionally, another saddle of this reduced action exists. It is the thermal AdS solution, which exhibits zero free energy at the classical level. Consequently, the Hawking-Page transition occurs when the Gibbs free energy approaches zero, which corresponds to the condition \(\mathsf{P}_{\mathrm{th}} = \frac{3\pi T^2}{8\mathbf{G}_4} \). In Section 2, we established a relationship between the size of the extra dimensions (or the instanton number) and the effective cosmological constant, interpreted within the extended phase space as the thermodynamic pressure \(\mathsf{P}_{\mathrm{th}}\). This enables us to derive phase transition curves for \(\mathbf{s} = -1, 0\)(Green and red curves in the left panel of Figure \ref{sPD01}) and \(\mathbf{s} = +1\) (right panel of Figure \ref{sPD01}) as functions of the extra dimension size. Furthermore, we determine the restricted black hole sizes resulting from the Hawking-Page transition, with results presented in Figure \ref{svx}. The right panels of these figures include the de Sitter black hole region; however, thermodynamic stability analysis does not apply to this de Sitter case, which lies beyond the scope of the present study.

{}

{}

{}

\subsection{Dyonic black hole with $q=B$}\label{subsec:dyonic}

In this section, we extend the analysis of the neutral black hole case to the charged black hole case with a constant $\mathcal{X}$. Therefore, we consider the action (\ref{Action4-const-f}) with a constraint (\ref{F2=0}), as previously discussed. To derive black hole solutions, we propose the following configuration:
\begin{align}\label{Dyonic Ansatz}
&ds^2 = -U(r) dt^2 + r^2 \left(d\theta^2 + \sin^2\theta d\varphi^2 \right) + \frac{dr^2}{U(r)}\nonumber\\
&\tilde{A} = \tilde{A}_t(r) dt + B \cos\theta d\varphi\,.
\end{align}
Solving the equations of motion under the constraint (\ref{F2=0}), we obtain a dyonic black hole solution given by
\begin{align}
U(r) = 1 - \frac{2 \mathbf{G}_4 M}{r} - r^2\frac{\tilde{\Lambda}_{\text{eff}}}{3} + \frac{2 q^2 \ell_e^2}{r^2}~~,~~\tilde{A}_t(r)=\left(\frac{\mathbf{G}_4}{\ell_e^2}\mu-\frac{q}{r}\right)~~,~~B=q\,,\label{dyonicU}
\end{align}
where $\tilde{\Lambda}_{\text{eff}}$ is defined by equation (\ref{Lambda-eff}).

Owing to the magnetic-field contribution, this dyonic black hole exhibits a non-normalizable gauge-field configuration; nevertheless, its thermodynamics remains well-defined. Thus, it is legitimate to consider the black hole chemistry for this system. From a holographic perspective, the magnetic field corresponds to an external field on the boundary of the (2+1)-dimensional system, making the non-normalizable gauge field a natural consequence of the external field in the dual system. In addition, one can find the physical meaning of the constraint, $q=B$, through the holographic interpretation. The holographic interpretation of $q$ and $B$ is the charge density and the external magnetic flux in the (2+1)-dimensional system, respectively. An easy way to introduce this constraint is to consider a Chern-Simons theory. The Gauss constraint of such a Chern-Simons theory is given by $B = k q$, where $k$ is the Chern-Simons level. Therefore, this dyonic black hole can be identified with a dual bulk system to a Chern-Simons system with level $k=1$.

To construct the thermodynamics of this system, we find essential thermodynamic quantities for this black hole. These are entropy, temperature, and black hole mass, given as follows:
\begin{align}
S = \frac{\pi r_h^2}{\mathbf{G}_4}\,,~T=\frac{1}{4 \pi }\left(\frac{1}{r_h}-3r_h \frac{\tilde{\Lambda}_{\text{eff}}}{3}  -\frac{2 q^2 \ell _e^2}{r_h^3}\right),~M= \frac{1}{\mathbf{G}_4}\left( \frac{r_h}{2}-\frac{1}{2} \frac{\tilde{\Lambda}_{\text{eff}}}{3} r_h^3+ \frac{q^2 \ell _e^2}{r_h}\right)\,.
\end{align}
In addition, we adopt the thermodynamic pressure (\ref{Pth-Lambda}) again. Then, the variation of the mass parameter is derived as:
\begin{align}\label{variation-M}
\delta M = T \delta S + \mu\delta q - \mathcal{M}\delta B + \mathsf{V}_{\mathrm{th}} \delta \mathsf{P}_{\mathrm{th}}\,,
\end{align}
where the chemical potential $\mu$, magnetization $\mathcal{M}$ and thermodynamic volume $\mathsf{V}_{\mathrm{th}}$ are defined as
\begin{align}\label{Vth dyonic}
\mu = \frac{\ell^2_e}{\mathbf{G}_4}\frac{q}{r_h}~,~\mathcal{M}= -\frac{\ell_e^2}{\mathbf{G}_4}\frac{B}{r_h}= -\frac{\ell_e^2}{\mathbf{G}_4}\frac{q}{r_h}~,~\mathsf{V}_{\mathrm{th}}=\frac{4}{3}\pi r_h^3\,.
\end{align}
Here, the constraint $B=q$ was consistently imposed. Note that the sign of $\mathcal{M}$ is opposite to that of $B$. Also, the Smarr relation can be written as follows\footnote{See \cite{Hyun:2017nkb} for a different variation of the Smarr relation.}:
\begin{align}\label{Smarr-dyonic}
M = 2 S T - 2 \mathsf{P}_{\mathrm{th}} \mathsf{V}_{\mathrm{th}} + 2 q\,\mu = 2 S T - 2 \mathsf{P}_{\mathrm{th}} \mathsf{V}_{\mathrm{th}} + q\,\mu - \mathcal{M} B\,.
\end{align}

\begin{figure}
        \centering
        \begin{subfigure}{}
            \centering
            \includegraphics[width=60mm]{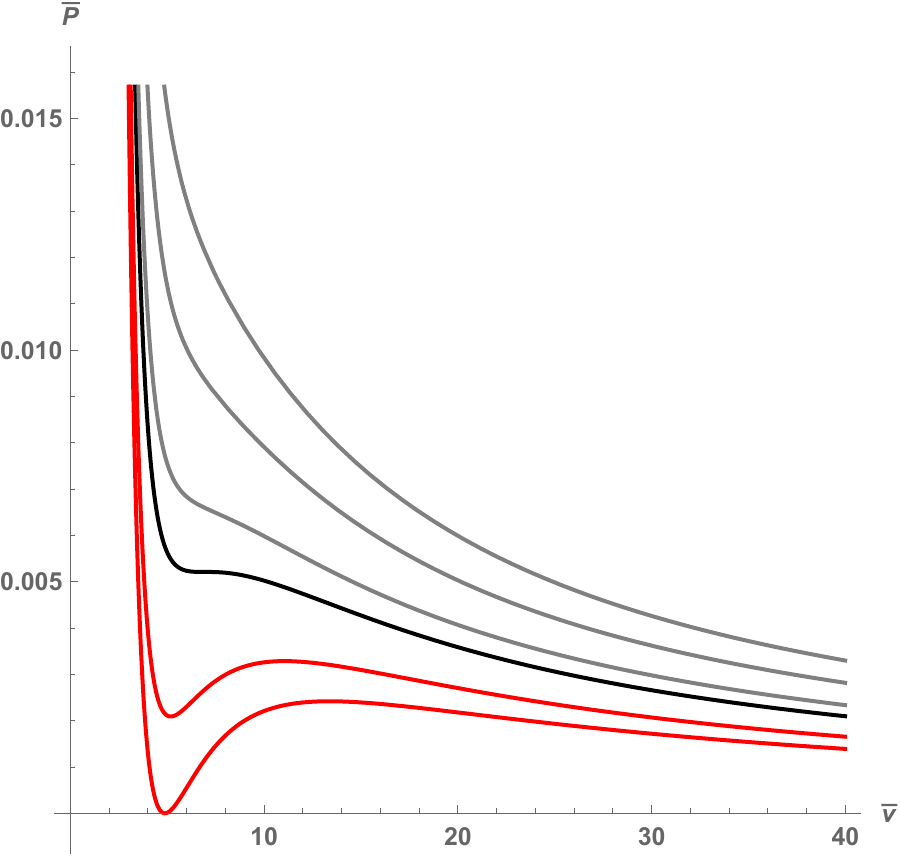}
        \end{subfigure}
        \begin{subfigure}{}
            \centering
            \includegraphics[width=60mm]{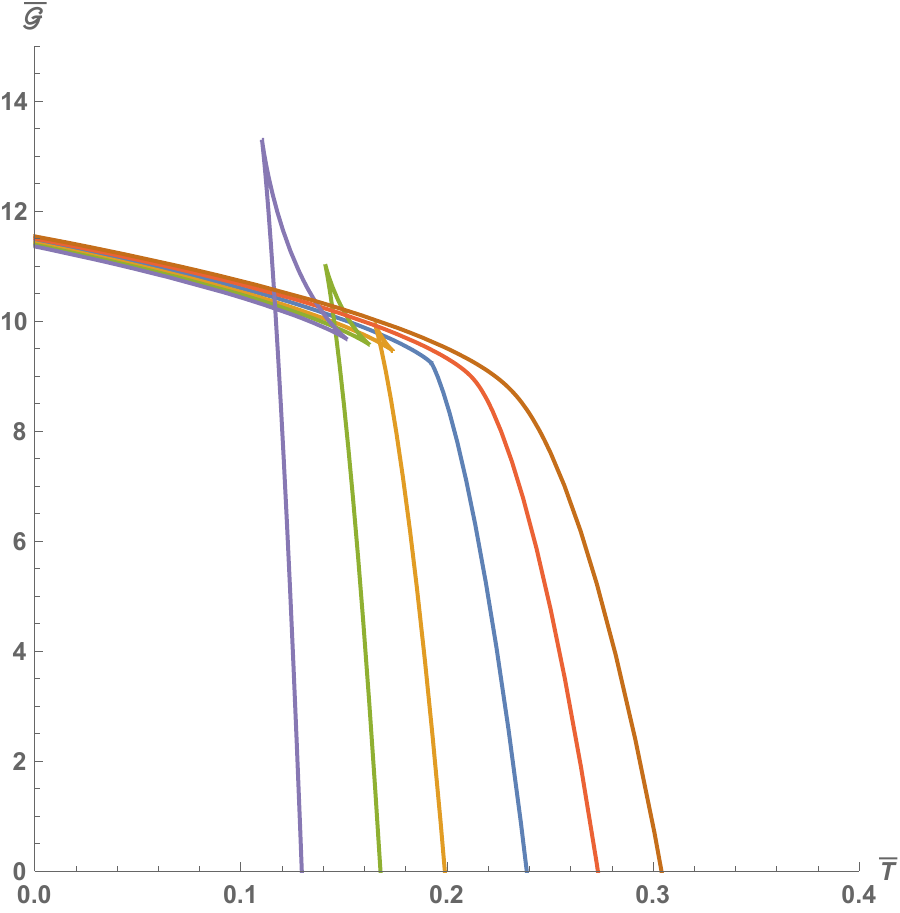}
         \end{subfigure}
         \caption{{\bf Thermodynamic pressure $\bar{P}(\bar{T},\bar{v})$ (Left):} Dimensionless quantities incorporating the charge parameter are defined in equations (\ref{dimensinoless-Exp-RN}) and (\ref{dimensinoless-Exp-RN-PTG}). Black hole solutions become unstable when $d\bar{P}/d\bar{v}>0$, enabling the Large/Small black hole phase transition. We set $\bar{T}=(\frac{0.4}{3\sqrt{3}},\frac{1}{3\sqrt{6}},\frac{1}{3\sqrt{3}},\frac{1.3}{3\sqrt{3}},\frac{1.5}{3\sqrt{3}})$ from the highest curve to the lowest curve.\\
{\bf Gibbs free energy (Right):} This figure illustrates the emergence of a swallowtail structure as the thermodynamic pressure ($\bar{P}$) decreases. We set $\bar{P}= (\frac{0.3}{192},\frac{0.5}{192},\frac{0.7}{192},\frac{1}{192},\frac{1.3}{192},\frac{1.6}{192})$ from the highest to the lowest curve.}
\label{Pth00}
    \end{figure}{}

The variation in equation (\ref{variation-M}) can be rearranged to clarify the internal energy and the first law as follows:
\begin{align}
\delta ( M - \mathsf{V}_{\mathrm{th}}\mathsf{P}_{\mathrm{th}} + \mathcal{M} B ) = T\delta S +\mu \delta q - \mathsf{P}_{\mathrm{th}}\delta \mathsf{V}_{\mathrm{th}} + B\delta\mathcal{M} 
\end{align}
yielding the internal energy:
\begin{align}
\mathcal{E} = M - \mathsf{V}_{\mathrm{th}}\mathsf{P}_{\mathrm{th}} + \mathcal{M} B\,.
\end{align}
Thus, the black hole mass $M$ is again identified with the enthalpy (adjusted for the magnetic contribution) rather than with the internal energy. We adopt the canonical ensemble with fixed charge and magnetic field, in which the Gibbs free energy serves as the appropriate thermodynamic potential. The Gibbs free energy is given as:
\begin{align}
\mathcal{G} = \mathcal{E} + \mathsf{P}_{\mathrm{th}}\mathsf{V}_{\mathrm{th}} - S T - \mathcal{M} B = M- ST.
\end{align}

Now, we rewrite the thermodynamic pressure using the Smarr relation (\ref{Smarr-dyonic}) in terms of the charge, temperature, and the specific volume\footnote{We employ the specific volume defined as $v\equiv \left(\frac{6 \mathsf{V}_{\mathrm{th}}}{\pi} \right)^{1/3}= 2 r_h$ as we mentioned earlier.} as follows:
\begin{align}
\mathsf{P}_{\mathrm{th}} =\frac{1}{\mathbf{G}_4} \left( \frac{4 q^2 \ell _e^2}{\pi  v^4} -\frac{1}{2 \pi  v^2} + \frac{T}{v} \right)\,.
\end{align}
This expression reveals the well-known Large/Small black hole phase transition \cite{Kubiznak:2012wp}. The only difference from the original work is that the charge is doubled due to the magnetic contribution. To analyze this thermodynamic system with non-zero charge, we introduce the thermodynamic variables per unit charge:
\begin{align}\label{dimensinoless-Exp-RN}
v= q \ell_e \bar{v}~,~ T = \frac{\bar{T}}{2\pi q \ell_e}~,~\mathsf{P}_{\mathrm{th}}=\frac{\bar{P}}{\pi \mathbf{G}_4 q^2 \ell_e^2}~,~\mathcal{G}=\frac{q \ell_e \bar{\mathcal{G}}}{8\mathbf{G}_4}\,.
\end{align}
Then, we have expressions with a fixed charge as follows:
\begin{align}\label{dimensinoless-Exp-RN-PTG}
\bar{P}=\frac{4}{\bar{v}^4}-\frac{1}{2\bar{v}^2} + \frac{\bar{T}}{2 \bar{v}}~~,~~\bar{T}= 2\bar{P} \bar{v} +\frac{1}{\bar{v}}-\frac{8}{\bar{v}^3}~~,~~\bar{\mathcal{G}} = \frac{24}{\bar{v}} + \bar{v} -\frac{2\bar{P}\bar{v}^3}{3}\,.
\end{align}
This pressure $\bar{P}$ is plotted in Figure \ref{Pth00} using the above expressions. This ($\bar{P}$-$\bar{v}$) diagram shows the typical behavior of the Van der Waals fluid.

\begin{figure}
        \centering
        \begin{subfigure}{}
            \centering
            \includegraphics[width=65mm]{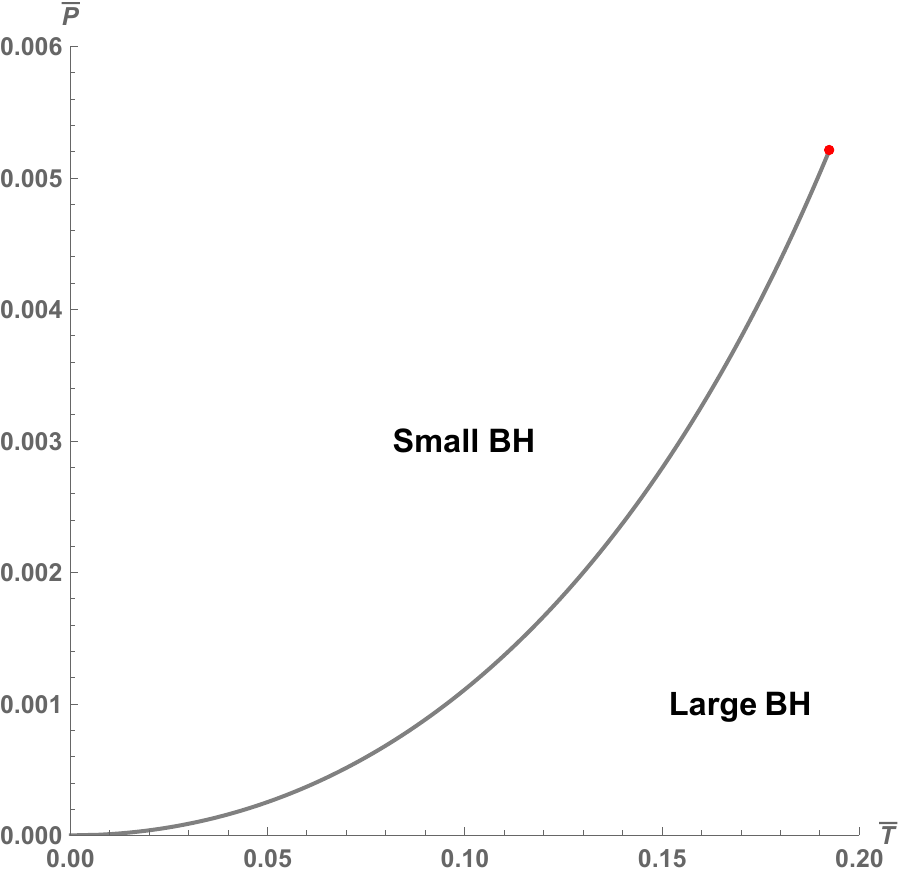}
        \end{subfigure}
       
         \caption{{\bf Phase diagram of dyonic black holes:} The figure shows phase diagrams for the dyonic black hole with $q=B$, illustrating Large/Small black hole transitions. Red dots mark the critical points. The dyonic case exhibits the typical first-order phase transition characteristic of AdS black holes. {\bf Kerr-AdS black holes also have qualitatively the same phase diagram, with $\bar{P}$ and $\bar{T}$ replaced by $\tilde{P}$ and $\tilde{T}$.}}\label{fig: PBQJ}
    \end{figure}

For the Large/Small black hole phase transition to occur, the $(\bar{P}$-$\bar{v})$ diagram in Figure \ref{Pth00} requires that the pressure curve exhibits two extrema. Or, equivalently, the following equation should have two roots in $\bar{v}$:
\begin{align}
\bar{v}^5\partial_{\bar{v}}\bar{P} = - 16 + \bar{v}^2\left(1-\frac{\bar{T}}{2}\bar{v}\right)=0\,.
\end{align}
It turns out that the condition for two roots is 
\begin{align}
\bar{T} < \frac{1}{3\sqrt{3}} \,.
\end{align}
In addition, for the Gibbs free energy $\bar{\mathcal{G}}$ to display a swallowtail structure, it must also have two extrema, which occurs when
\begin{align}
\bar{P} < \frac{1}{192} \,.
\end{align}
Thus, the critical point is $(\bar{T}_c, \bar{P}_c)=(1/3\sqrt{3},1/192)$. The Gibbs free energy $\bar{\mathcal{G}}$ is depicted in Figure \ref{Pth00}. Additionally, the phase boundary is determined numerically and presented in Figure \ref{fig: PBQJ}. This boundary curve resembles that of the RN black hole \cite{Kubiznak:2012wp} except for the magnetic charge contribution.

{}

\begin{figure}
       \centering
        \begin{subfigure}{}
            \centering
            \includegraphics[width=58mm]{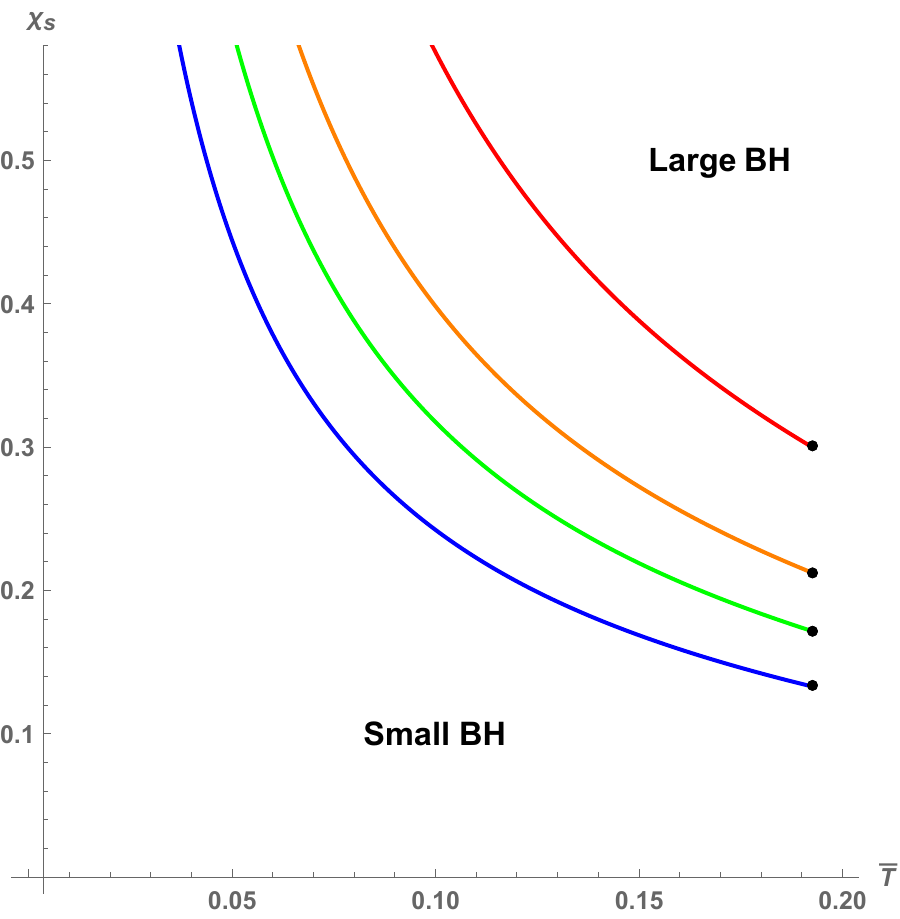}
        \end{subfigure}
        \begin{subfigure}{}
            \centering
            \includegraphics[width=58mm]{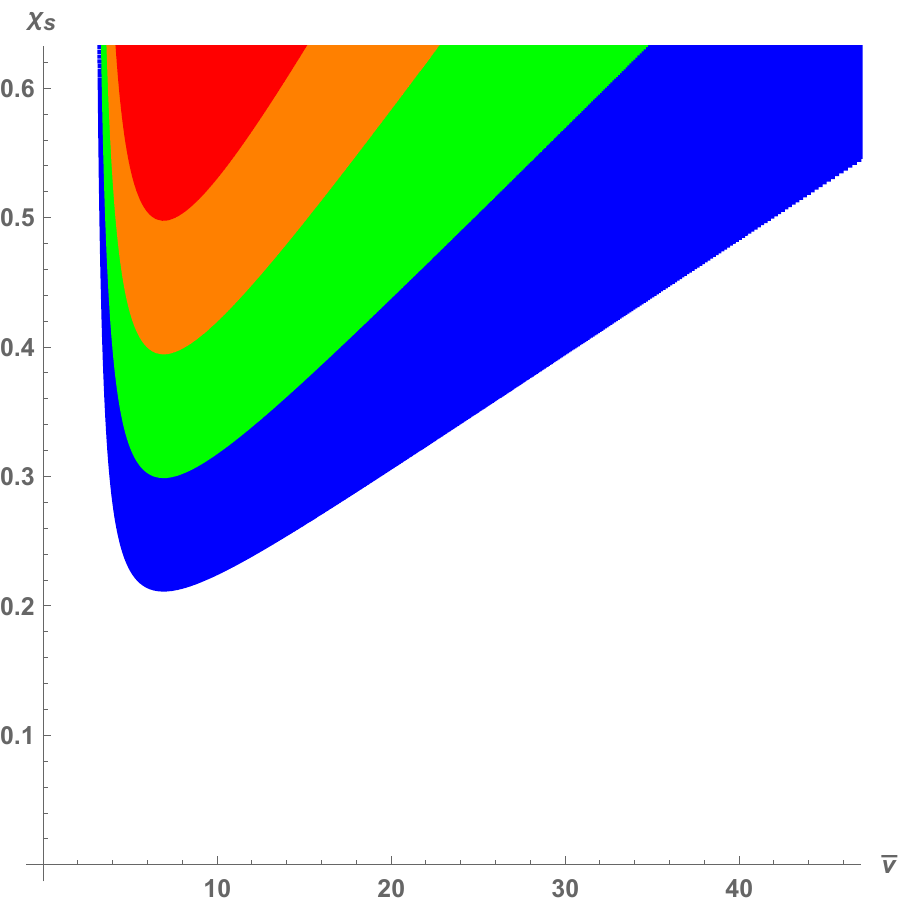}
         \end{subfigure}\\
          \centering
        \begin{subfigure}{}
            \centering
            \includegraphics[width=58mm]{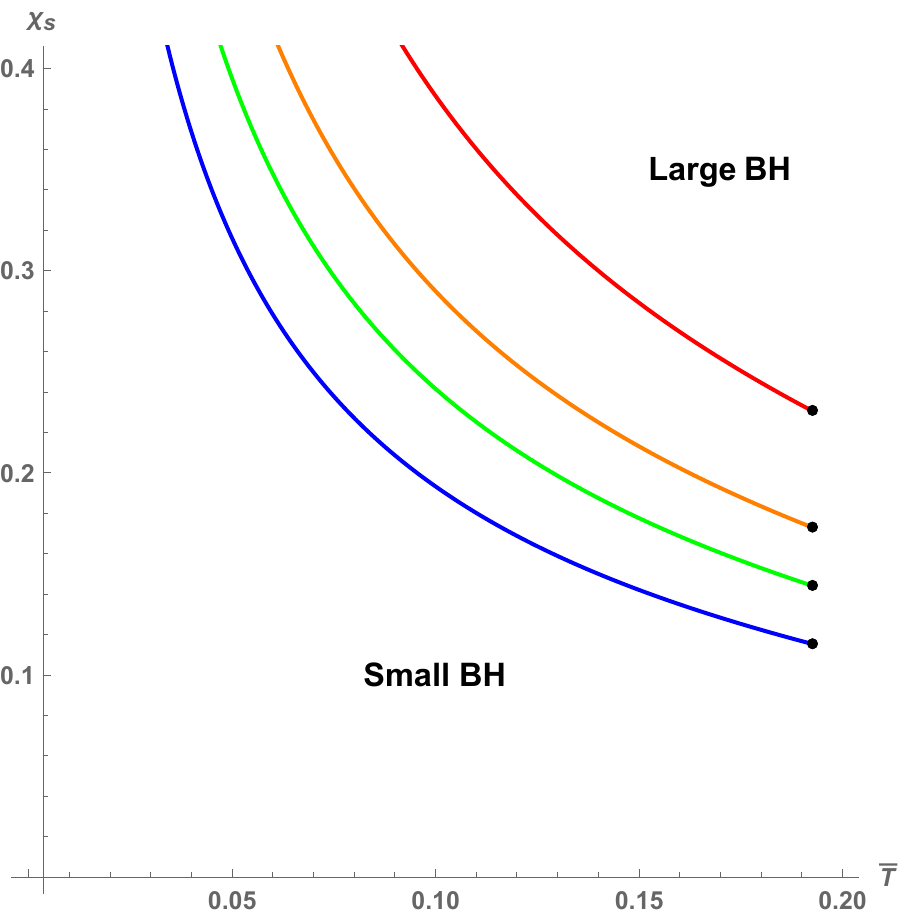}
        \end{subfigure}
        \begin{subfigure}{}
            \centering
            \includegraphics[width=58mm]{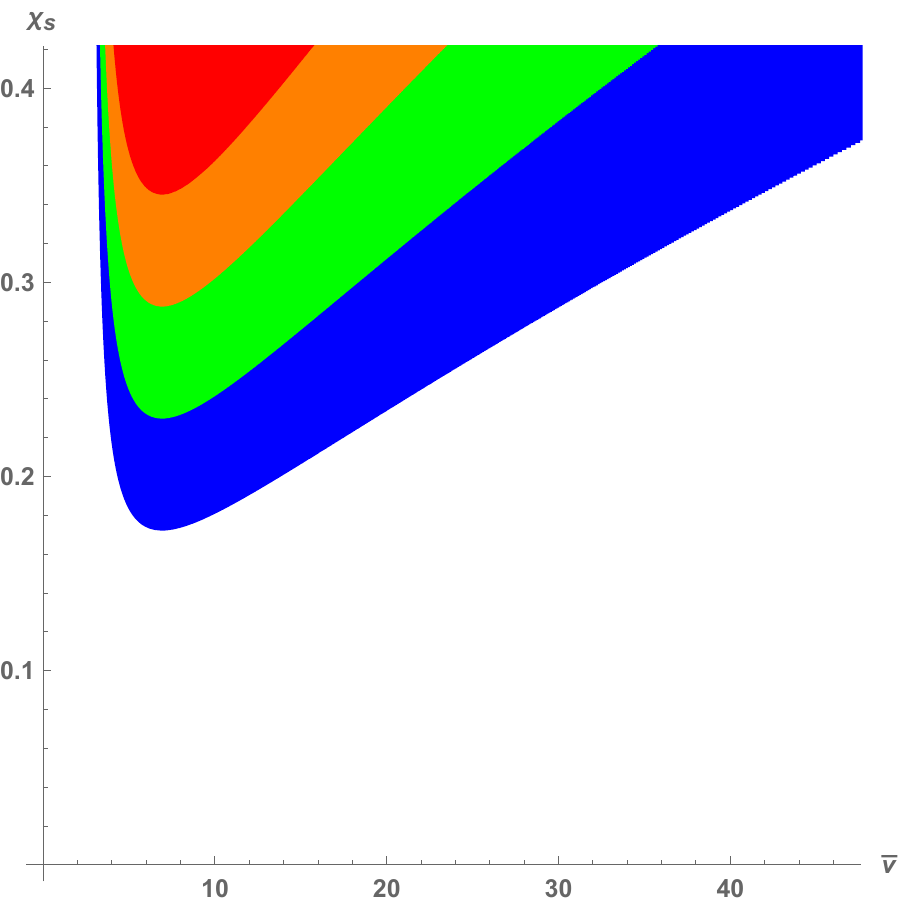}
         \end{subfigure}\\
          \centering
        \begin{subfigure}{}
            \centering
            \includegraphics[width=58mm]{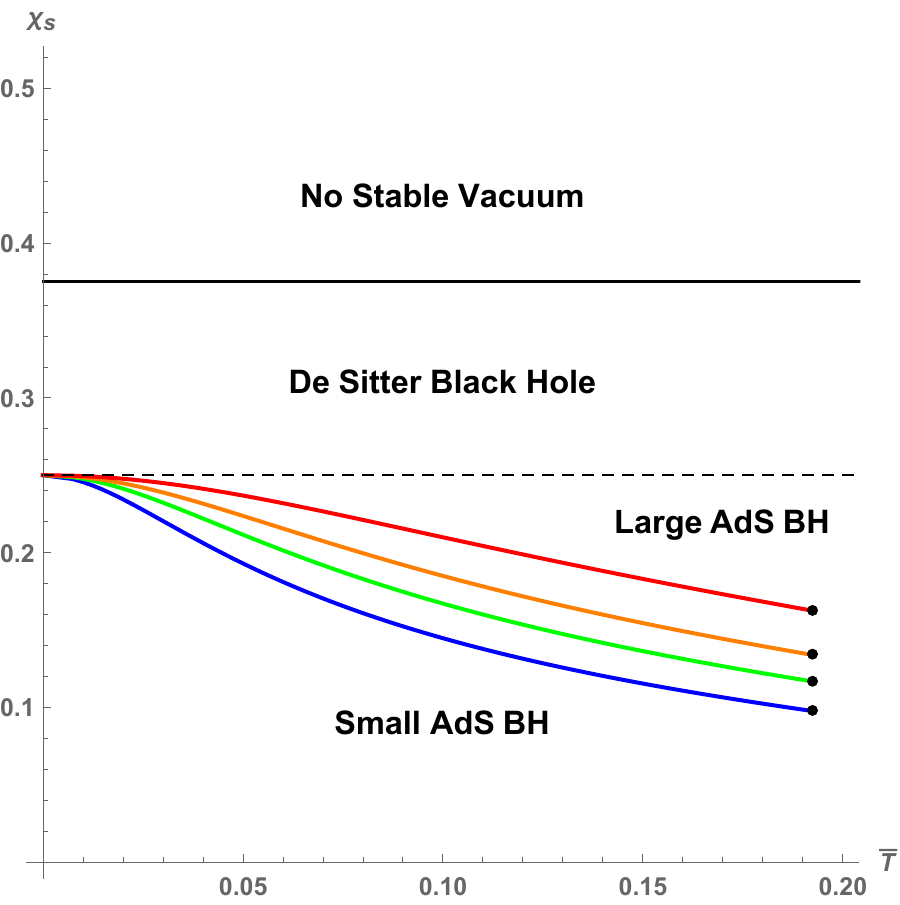}
        \end{subfigure}
        \begin{subfigure}{}
            \centering
            \includegraphics[width=58mm]{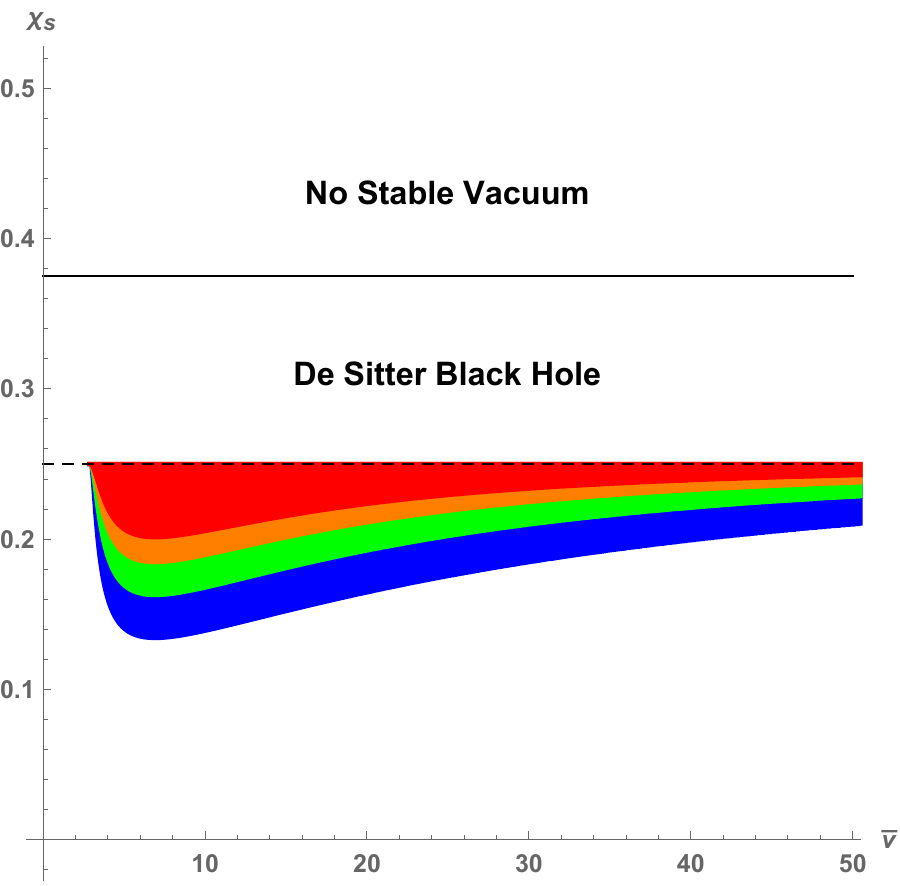}
         \end{subfigure}
         \caption{From top to bottom, the panels correspond to $\mathbf{s}=-1, 0, +1$, respectively. The left panels depict phase diagrams as functions of extra dimension size, while the right panels illustrate the thermodynamically disfavored range of the horizon size $\bar{v}=2 r_h/(q \ell_e)$. From the red to the blue curves or regions, we employ $\gamma=(0.06, 0.05, 0.04, 0.03)$ as defined in equation (\ref{P of chi}). {\bf Kerr-AdS black holes also have the same qualitative phase diagram, with $\bar{T}$ and $\bar{v}$ replaced by $\tilde{T}$ and $\tilde{v}$.}}\label{PD01}
    \end{figure}{}

We now establish the relationship between the thermodynamic pressure and the size of the extra dimensions. The extra dimension size is represented by the warping factor $ L^2 e^{2f} = L^2 \mathcal{X}$. For $ \mathbf{s} = 0 $, $ L$ serves merely as a length unit, allowing $ L = 1$ as a simplification. In contrast, for $ \mathbf{s} = \pm 1 $, $L^2 $ defines the length scale of the cosmological constant. The effective cosmological constant $\tilde{\Lambda}_{\rm eff}$ is the same as the Schwarzschild case (\ref{Lambda-eff}). 
Therefore, the pressure in terms of the extra dimension size is given by 
\begin{align}\label{P of chi}
\bar{P} = \gamma^3 \frac{1 - 4 \mathbf{s} \mathcal{X}_s}{\mathcal{X}_s^3} = \frac{q^2 \ell_e^2}{64 L^2} \frac{1 - 4 \mathbf{s} \mathcal{X}_s}{\mathcal{X}_s^3}\,.
\end{align}
Inverting this relation yields the dimensionless size:
\begin{align}\label{Xs-Dyonic}
\mathcal{X}_s = \frac{\gamma \left[ 2^{1/3} \left( \sqrt{81 \bar{P} + 768 \gamma^3 \mathbf{s}} + 9 \sqrt{\bar{P}} \right)^{2/3} - 8 \cdot 3^{1/3} \gamma \mathbf{s} \right]}{6^{2/3} \sqrt{\bar{P}} \left( \sqrt{81 \bar{P} + 768 \gamma^3 \mathbf{s}} + 9 \sqrt{\bar{P}} \right)^{1/3}}.
\end{align}
This expression enables the construction of phase diagrams for \(\mathbf{s} = 0, \pm 1\), presented in Figure \ref{PD01}. Additionally, we identify thermodynamically unstable regions for each phase diagram curve, illustrated in the right panels of Figure \ref{PD01}.

Significantly, each phase boundary curve delineates the excluded size range for black holes, providing insight into the size of extra dimensions. These excluded ranges are depicted in Figure \ref{PD01}. Larger extra dimensions correspond to broader ranges of excluded black hole sizes. For black hole formation to occur readily, the extra dimension size must be sufficiently small. The bottom panels of Figure \ref{PD01} illustrate the $\mathbf{s}=+1$ case (i.e., a positive $\Lambda$), revealing a transition between de Sitter and AdS dyonic black hole geometries. Near $\mathcal{X}_s=1/4$ within the AdS region, nearly all black hole sizes are thermodynamically disfavored. Exploring the behavior near this critical extra-dimension size presents an intriguing avenue for further investigation.

\subsection{Kerr-AdS Black Hole}

In this section, we investigate how the extra dimensions influence the black hole chemistry of the Kerr-AdS geometry. The conventional metric is:
\begin{align}\label{Kerr-AdS-metric}
ds^{2}=-\frac{\Delta_{r}}{\rho^{2}}\left({\rm d} t
-\frac{a \sin ^{2}\theta}{\Xi}  {\rm d}\phi\right)^{2}
+\frac{\rho^{2}}{\Delta_{r}} {\rm d} r^{2}+\frac{\rho^{2}}{\Delta_{\theta}} {\rm d} \theta^{2} 
+\frac{\Delta_{\theta}\sin^{2}\theta }{\rho^{2}}\left(a {\rm d} t
-\frac{r^{2}+a^{2}}{\Xi} {\rm d} \phi\right)^{2},
\end{align}
where
\begin{align}
&\rho^{2} =r^{2}+a^{2} \cos ^{2} \theta,  \qquad
\Delta_{r} =\left(r^{2}+a^{2}\right)\left(1- \frac{\tilde{\Lambda}_{\text{eff}}}{3} r^{2} \right)-2 \mathbf{G}_{4} m r, \\
&\Xi =1+ \frac{\tilde{\Lambda}_{\text{eff}}}{3} a^{2},  \qquad\qquad
\Delta_{\theta} =1 + \frac{\tilde{\Lambda}_{\text{eff}}}{3} a^{2} \cos ^{2} \theta.
\end{align}
The effective cosmological constant $\tilde{\Lambda}_{\text{eff}}$ is again given in (\ref{Lambda-eff}). As usual, we write $\tilde{\Lambda}_{\mathrm{eff}}= - 3 /{L}_{\mathrm{eff}}^2$ for the AdS spacetime.

The mass $M$ and angular momentum $J$ can be written in terms of the parameters $m$ and $a$ as\footnote{The inverse expression in terms of physical quantities is $\left\{m, a\right\}=\left\{ \left(1-{J^2}/(M^2 {L}_{\rm eff}^2)\right)^2M, {J}/{M} \right\}$.}
\begin{align}
 M=\frac{m}{\Xi^{2}},\quad J=\frac{am}{\Xi^{2}}\,.	
\end{align}
Given that the metric satisfies $\Delta_r(r_+)=0$ with the event horizon $r_+$, $m$ can be expressed in terms of other parameters. Thus, the mass and angular momentum are rewritten as:
\begin{align}
M=\frac{1}{2 \mathbf{G}_4 \Xi ^2 r_+}\left(a^2+r_+^2-\frac{1}{3} r_+^2 \left(a^2+r_+^2\right) \tilde{\Lambda }_{\text{eff}} \right),\label{M Kerr} \\
J=\frac{a}{2 \mathbf{G}_4 \Xi ^2 r_+}\left( a^2+r_+^2 -\frac{1}{3} r_+^2 \left(a^2+r_+^2\right) \tilde{\Lambda }_{\text{eff}}\right).
\label{MJ}
\end{align}
Other quantities, such as the Hawking temperature, entropy, and angular velocity of the Kerr-AdS black hole, are given as follows: 
\begin{align}
T= \frac{r_+}{4\pi\left(a^2+r_+^2\right)}\left(1-\frac{a^2}{r_+^2} -\frac{1}{3} \left(a^2+3 r_+^2\right) \tilde{\Lambda }_{\text{eff}}\right),~
S=\frac{\pi (r_+^{2}+a^{2})}{\mathbf{G}_{4} \Xi},~ 
\Omega ' =\frac{a\Xi}{r_+^{2}+a^{2}}\,.
\end{align}
Also, the angular speed in the appropriate frame is 
\begin{align}
\Omega =\Omega ' -\frac{\tilde{\Lambda}_{\text{eff}}}{3}a .
\end{align}
Using the pressure defined in \eqref{Pth-Lambda}, we vary the mass
in \eqref{M Kerr} at fixed $\mathbf{G}_4$ and express the result
in terms of variations of the entropy, angular momentum, and
pressure. This yields the extended first law,
\begin{align}\label{first law J}
\delta M
= T\delta S + \Omega\delta J
+ \mathsf{V}_{\mathrm{th}}\delta\mathsf{P}_{\mathrm{th}}\,.
\end{align}
The thermodynamic volume is
\begin{align}\label{Vth-J}
\mathsf{V}_{\mathrm{th}}=\frac{4 \pi  r_+^3}{3} \left( \frac{1+\frac{a^2}{r_+^2}}{\Xi^2}\right)\left(1-\frac{1}{2}\frac{a^2}{{L}_{\rm eff}^2} +\frac{a^2}{2 r_+^2}\right)\,.
\end{align}
This expression, derived in \cite{Kubiznak:2016qmn}, reduces to the thermodynamic volume of the dyonic case in equation (\ref{Vth dyonic}) as $a\to 0$. A conjectured inequality between this volume and the horizon area has been proposed \cite{Cvetic:2010jb}, and investigating its role in black hole chemistry presents an intriguing prospect. These thermodynamic quantities also satisfy the four-dimensional
Smarr relation,
\begin{align}\label{Smarr-Kerr}
M = 2TS + 2\Omega J
- 2\mathsf{P}_{\mathrm{th}}\mathsf{V}_{\mathrm{th}}\,.
\end{align}

Let us continue to discuss the black hole chemistry of the Kerr-AdS black hole. As in the previous case, the internal energy is given by $\mathcal{E}=M-\mathsf{P}_{\mathrm{th}} \mathsf{V}_{\mathrm{th}}$, consistent with the mass variation in equation (\ref{first law J}). The Gibbs free energy and its variation are
\begin{align}
\mathcal{G}= M - ST~~,~~\delta\mathcal{G}= - S\delta T + \mathsf{V}_{\mathrm{th}}\delta \mathsf{P}_{\mathrm{th}} + \Omega \delta J\,.
\end{align}
Thus, the Gibbs free energy serves as the key thermodynamic potential for describing phase transitions in the ($\mathsf{P}_{\mathrm{th}}$-$T$) space with fixed $J$.

Similar to the dyonic black hole case, we introduce dimensionless parameters adapted to the fixed angular momentum $J$, defined as
\begin{align}
&\tilde{r}_+=\frac{r_+}{\sqrt{\mathbf{G}_4 J}},\qquad
\tilde{a}=\frac{a}{\sqrt{\mathbf{G}_4 J}},\qquad
\tilde{T}= 4\pi \sqrt{\mathbf{G}_4 J}\,T,\qquad
\tilde{P}=8 \pi \mathbf{G}_4^2 J\, \mathsf{P}_{\mathrm{th}},
\nonumber\\
&\tilde{V}=\frac{\mathsf{V}_{\mathrm{th}}}{2\pi(\mathbf{G}_4 J)^{3/2}},\qquad
\tilde{\mathcal{G}}=\sqrt{\frac{\mathbf{G}_4}{J}}\mathcal{G},\qquad
\tilde{v}=\left(\frac{6\tilde{V}}{\pi} \right)^{1/3},\qquad\Xi=1-\frac{\tilde{a}^2\tilde{P}}{3}\,.
\end{align}
Here $\tilde{v}$ is the dimensionless counterpart of the specific volume $v=\left(\frac{6\mathsf{V}_{\mathrm{th}}}{\pi}\right)^{1/3}$. The temperature, thermodynamic volume, specific volume, and Gibbs free energy are then written as
\begin{align}\label{vPG-J}
\tilde{T}=& \frac{\tilde{r}_+}{\tilde{a}^{2}+\tilde{r}_+^{2}}\left[ 1-\frac{\tilde{a}^{2}}{\tilde{r}_+^{2}}+\frac{\tilde{P}}{3}\left(\tilde{a}^{2}+3\tilde{r}_+^{2}\right)\right]\,, \nonumber\\
\tilde{V}=& \frac{2\tilde{r}_+^{3}}{3} \left( \frac{1+\frac{\tilde{a}^{2}}{\tilde{r}_+^{2}}}{\Xi^{2}} \right) \left( 1-\frac{\tilde{P}\tilde{a}^{2}}{6}+\frac{\tilde{a}^{2}}{2\tilde{r}_+^{2}} \right),\,\tilde{v}= \left(\frac{6\tilde{V}}{\pi}\right)^{1/3},\nonumber\\
\tilde{\mathcal{G}}=& \frac{1}{2\Xi^{2}\tilde{r}_+} \left[\tilde{a}^{2}+\tilde{r}_+^{2}+\frac{\tilde{P}}{3}\tilde{r}_+^{2}\left(\tilde{a}^{2}+\tilde{r}_+^{2}\right)
\right] - \frac{\tilde{T}\left(\tilde{r}_+^{2}+\tilde{a}^{2}\right)}{4\Xi}\,.
\end{align}
These formulas are sufficient to generate the $(\tilde{P}-\tilde{v})$ curves and the Gibbs free energy $\tilde{\mathcal{G}}$ diagram for the Kerr-AdS black hole in the fixed-$J$ ensemble. The resulting behavior is the standard fixed-$J$ Gibbs free energy behavior of the Kerr-AdS black hole, exhibiting the characteristic swallowtail structure associated with the Large/Small black hole phase transition. Since these thermodynamic curves are qualitatively the same as the corresponding diagrams obtained in the dyonic black hole case (Figure \ref{Pth00} and \ref{fig: PBQJ}), we do not reproduce them separately.

We now find the relationship among the extra-dimension size and the excluded sizes of the black holes under phase transitions. Using equation (\ref{Lambda-eff}), one can transform the phase diagram into diagrams based on the extra-dimension size. The scaled pressure is given by\footnote{This is the same form as the dyonic case (\ref{P of chi}) with $\gamma$ replaced by $\tilde{\gamma}$.} 
\begin{align}
\tilde{P} =\tilde{\gamma}^{3} \left( \frac{1-4 \mathbf{s}\mathcal{X}_s}{\mathcal{X}_s^3}
\right),\qquad\tilde{\gamma}^{3}=\frac{ \mathbf{G}_4 J}{8L^2}.
\end{align}
The resulting phase boundaries and excluded specific-volume regions are also qualitatively the same as those obtained in the dyonic black hole case (Figure \ref{PD01}). Therefore, we do not reproduce the corresponding extra-dimensional phase diagrams for the Kerr-AdS case.

\subsection{Thermodynamic phase space constraints from compactification scale}\label{secTpsc}

\begin{table}[h]
\centering
\renewcommand{\arraystretch}{2.0} 
\caption{ {\bf Conjugate Volume and Pressure with small J and $\mathcal{X}_s$}}
\vspace{3mm}
\begin{tabular}{|l|l|}
\hline
\textbf{~~~~Thermodynamic Volume~~~~}& \textbf{~~~~Thermodynamic Pressure~~~~}  \\
\hfill $\mathsf{V}_{\mathrm{th}} = \frac{4\pi}{3} r_+^3+\mathcal{O}(J^2)$~~~~~~~~~~~~~~  & \hfill $\mathsf{P}_{\mathrm{th}} = \frac{1}{64 \pi  \mathbf{G}_4 L^2 }\left(\frac{1-4\mathbf{s}\mathcal{X}_s}{\mathcal{X}_s^3 }\right)$~~~~~~~~~   \\ \hline
$\sim \text{Black Hole Volume (Lapse weighted)}$ & ~~~$\sim \left(\text{Extra Dimension Volume}\right)^{-3/2}$\\ \hline
\end{tabular}\label{table: Th Vol P}
\end{table}

The thermodynamic volume $\mathsf{V}_{\mathrm{th}}$, or equivalently the specific volume $v$, characterizes the length scale associated with the four-dimensional black hole. On the other hand, the thermodynamic pressure $\mathsf{P}_{\mathrm{th}}$ is governed by the effective cosmological constant, which in our model depends on the stabilized extra-dimension size (or equivalently on the instanton number). Thus, the black hole size and the compactification scale are related through the phase space of the extended thermodynamics or black hole chemistry. See the table \ref{table: Th Vol P} for the relation in a simplified case. Roughly speaking, the extended phase space includes a nontrivial conjugate connection between the black hole size and the extra dimension size. In the following part of this section, we will explain the physical implications of our result shown in Figure \ref{svx} and Figure \ref{PD01}.

We first consider the Schwarzschild-AdS case discussed in Section~\ref{Schwar-section}. Figures~\ref{sPD01} and~\ref{svx} show the thermodynamic competition between thermal AdS and black hole phases for a given extra-dimensional size $\mathcal{X}_s$. The threshold in Figure \ref{svx} marks the onset of global thermodynamic preference for the black hole phase over thermal AdS, rather than a lower bound on the existence of black hole solutions. As $\mathcal{X}_s$ increases, this threshold shifts toward larger black hole sizes. Black holes below the threshold are thermodynamically disfavored, although such solutions may still exist. Their possible transitions to thermal AdS or evolution toward larger black hole configurations depend on dynamical processes beyond the equilibrium analysis presented here.

\begin{figure}[t]
    \centering
    \begin{subfigure}{}
        \centering
        \includegraphics[width=60mm]{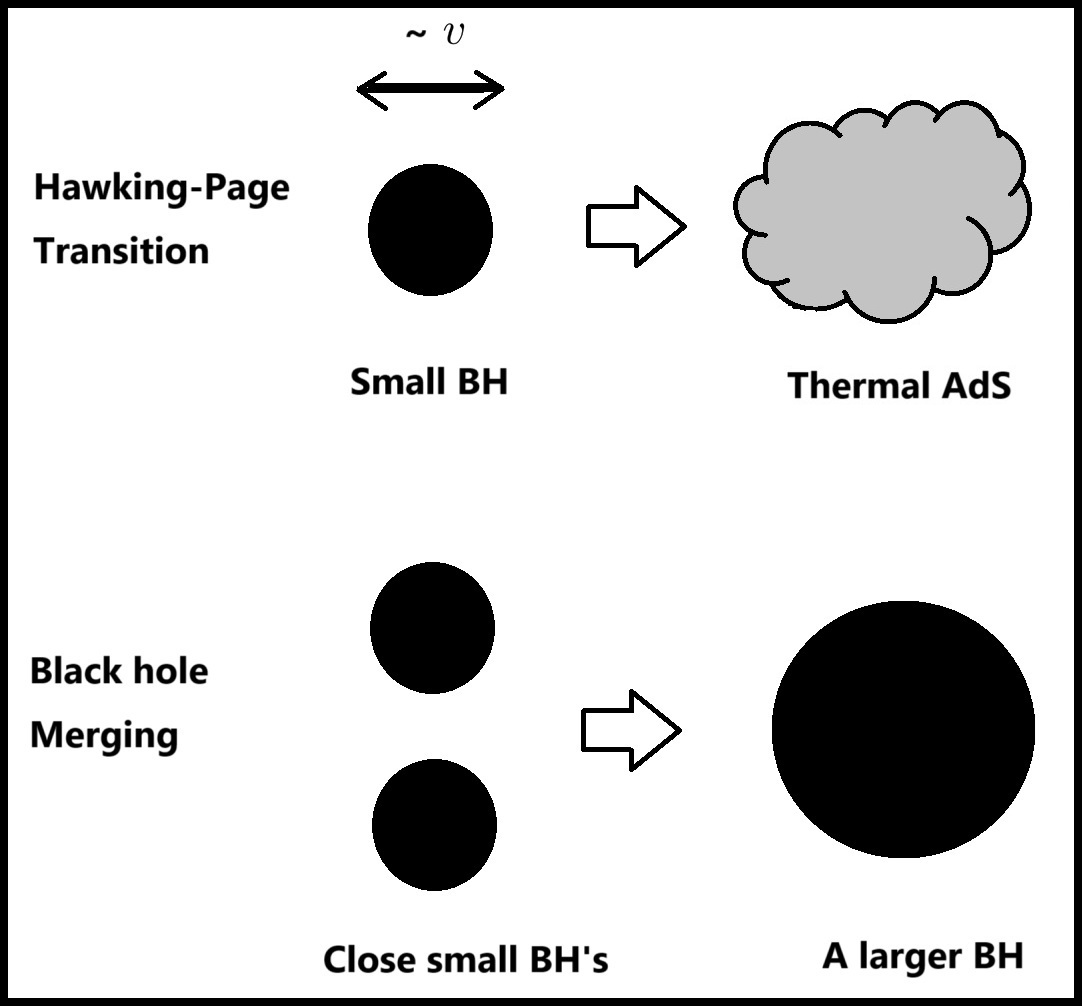}
    \end{subfigure}
    \begin{subfigure}{}
        \centering
        \includegraphics[width=60mm]{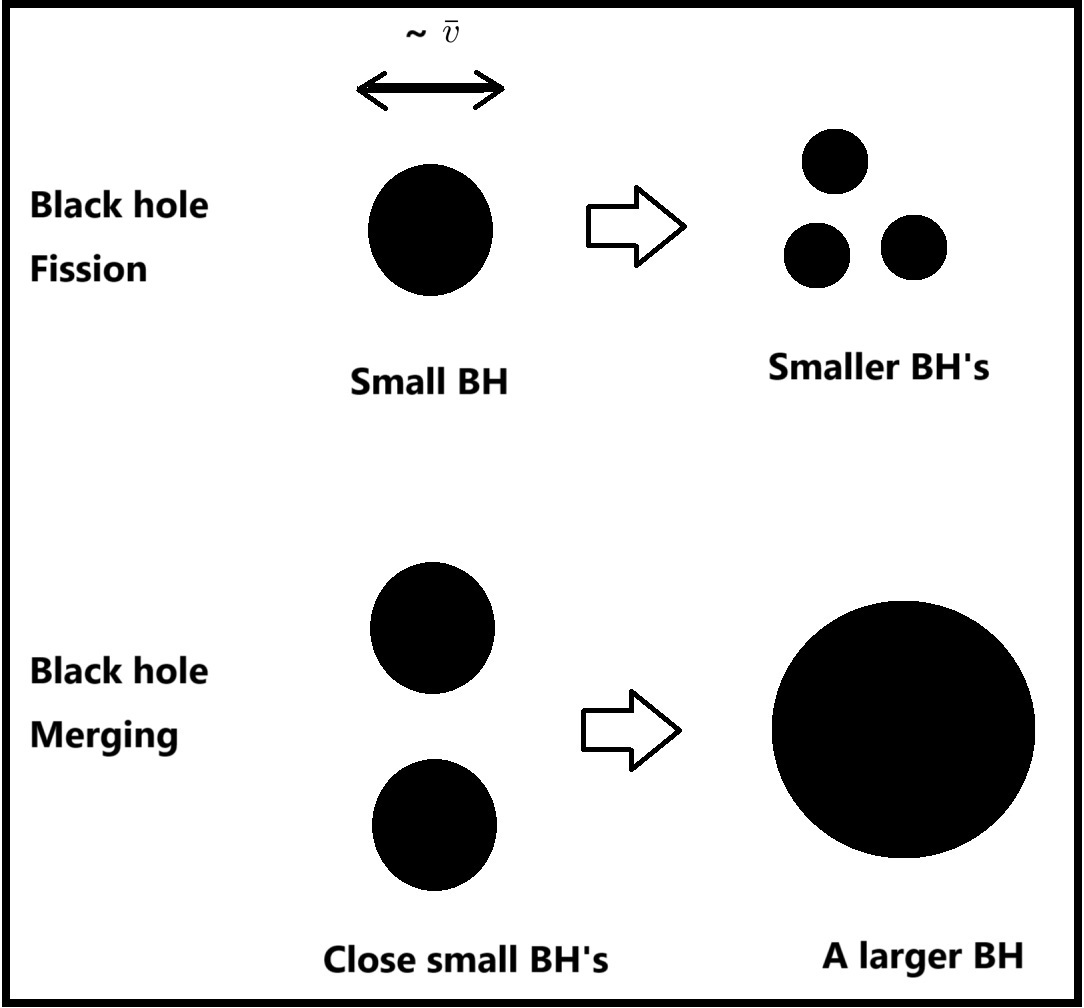}
    \end{subfigure}
    \caption{{\bf Schematic implications of the thermodynamic size bounds:} For Schwarzschild-AdS black holes (left), increasing the extra-dimensional size raises the minimum thermodynamically allowed black hole size; a sub-threshold configuration is therefore disfavored relative to thermal AdS or to a larger black hole configuration. For dyonic or Kerr-AdS black holes (right), an intermediate excluded region appears above the critical extra-dimensional size. Configurations in this region are not necessarily globally preferred thermodynamic phases. Possible transitions toward other configurations are shown schematically and require an independent dynamical analysis.}
    \label{fig: Cartoon}
\end{figure}

In contrast, dyonic and Kerr-AdS black holes exhibit distinct Small and Large black hole phases. When the extra-dimensional size $\mathcal{X}_s$ is below the critical value (marked by black dots in Figure \ref{PD01}), no intermediate coexistence region appears. Once $\mathcal{X}_s$ exceeds the critical value, however, a thermodynamically disfavored region develops, as indicated by the colored areas in the right panels of Figure \ref{PD01}. This region should not be interpreted as a range in which black hole solutions cannot exist. Rather, it characterizes the thermodynamic preference between the competing black hole phases. Depending on the thermodynamic conditions, a non-dominant branch may be metastable or locally unstable. Possible transitions toward other configurations depend on the available dynamical processes and conserved charges. Figure \ref{fig: Cartoon} illustrates such possibilities schematically, without implying a specific dynamical mechanism.

\begin{figure}[t]
    \centering
    \begin{subfigure}{}
        \centering
        \includegraphics[width=100mm]{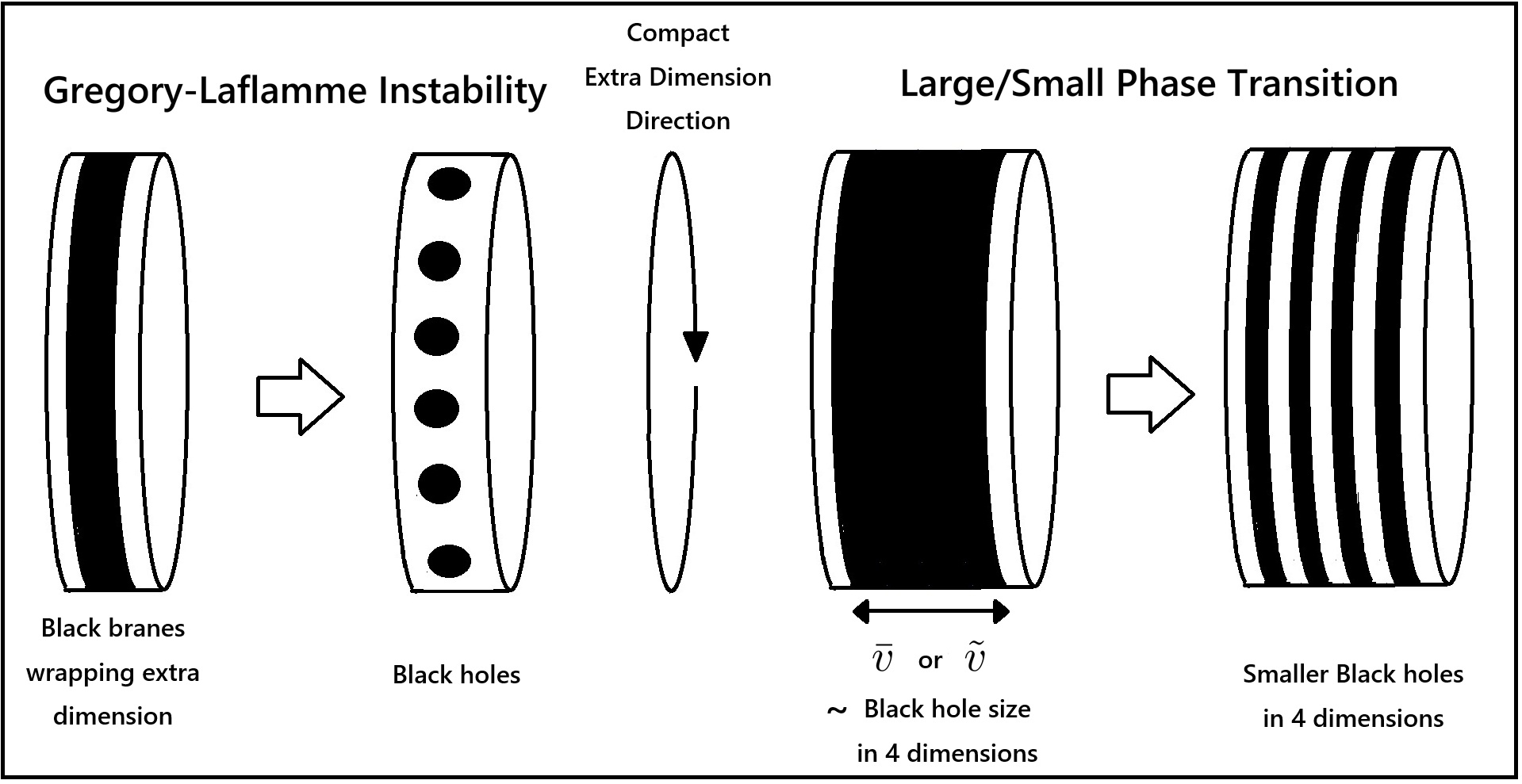}
    \end{subfigure}
  \caption{{\bf Schematic comparison between the Gregory–Laflamme instability and the thermodynamic size restriction from the Large/Small black hole transition.}}
    \label{fig: Cartoon GL-Ours}
\end{figure}

The size restrictions found above are reminiscent of the Gregory–Laflamme instability~\cite{Gregory:1993vy}. The mechanism, however, is different: the restriction arises not from a perturbative instability of a black string or black brane along the compact directions, but from the effective four-dimensional thermodynamic pressure generated by the compact internal space (see Figure~\ref{fig: Cartoon GL-Ours}). In this broader sense, our result provides another example in which compact extra dimensions impose nontrivial constraints on thermodynamically preferred configurations of black objects.

In our construction, the compactification scale determines the thermodynamic pressure, while the thermodynamic volume is its conjugate variable. This motivates a further geometric question: whether the thermodynamic volume can be represented by a distinguished surface in the four-dimensional black hole geometry. In the next section, we address this question and subsequently discuss its possible holographic interpretation.

\section{A Constrained Minimal-Area Representation of Thermodynamic Volume}\label{sec:thermodynamic-volume-surface}

The thermodynamic volume is defined as the quantity conjugate to the thermodynamic pressure, which appears in the first law and the Smarr relation. Its geometric interpretation, however, is not automatic.  In particular, several inequivalent notions of volume can be introduced in a curved spacetime.  The purpose of this section is to formulate a geometric prescription that reproduces the event horizon in the static charged (dyonic) black hole and admits a nontrivial rotating continuation in the Kerr-AdS black hole.

A geometric interpretation of thermodynamic volume is not unique, particularly in a rotating spacetime. However, one may consider an ADM decomposition for the timelike direction, with the shift vector implying the rotation. In the following analysis, we use the lapse-weighted spatial volume as follows:
\begin{align}\label{intro volume functional}
\mathsf{V}[\Sigma] = \int_{\Omega_\Sigma} d^3 x  \sqrt{-g}\,= \int_{\Omega_\Sigma}dr d\theta d\phi\, \mathcal{N} \sqrt{h} \,,
\end{align}
where $\Omega_\Sigma$ denotes the region bounded by a closed
codimension-two surface $\Sigma$ on a constant-time domain. In addition, $\mathcal{N}$ and $h$ are the decomposed lapse function and the determinant of the decomposed spatial metric under the ADM formalism. This prescription is closely related to invariant volume constructions for stationary black holes \cite{Parikh:2005qs, Ballik:2013uia}. It should be distinguished from the Riemannian proper spatial volume $\int_{\Omega_\Sigma}\sqrt{h}\,d^3x$ on a constant-time slice. We will refer to this invariant enclosed volume $\mathsf{V}[\Sigma]$ as the invariant lapse-weighted (LW) volume from now on.

Using this invariant LW volume, we propose a constrained minimal-area prescription for a geometric representative of the thermodynamic volume:
\begin{align}\label{intro thermodynamic volume surface}
\Sigma_{\rm th}=\underset{\Sigma}{\operatorname{argmin}}\left\{A[\Sigma]\;\middle|\; \mathsf{V}[\Sigma]=\mathsf{V}_{\mathrm{th}}\right\}.
\end{align}
In this section, we implement this prescription for the class of closed axisymmetric surfaces that enclose the black hole. For regular surfaces surrounding spherically symmetric black holes, the surfaces satisfying (\ref{intro thermodynamic volume surface}) are the event horizons of the black holes, such as Schwarzschild, charged, and dyonic black holes.

For the Kerr-AdS black hole, however, the same prescription (\ref{intro thermodynamic volume surface}) has a nontrivial rotating contribution that makes $\Sigma_{\mathrm{th}}$ differ from the horizon. A constant-radius profile is no longer, in general, a solution of the constrained Euler-Lagrange equation for the area functional. We compare the thermodynamic volume with the LW volumes enclosed by two natural surfaces in the rotating geometry: the outer event horizon and the ergosphere. For nonzero rotation, we recover the known result that the LW volume enclosed by the event horizon is strictly smaller than $\mathsf{V}_{\mathrm{th}}$~\cite{Cvetic:2010jb}. This motivates seeking a surface that encloses the horizon, with the ergosphere providing a further natural reference. Figure \ref{fig:vergo-vth-region} compares the LW volume enclosed by the ergosphere with $\mathsf{V}_{\mathrm{th}}$, revealing a nontrivial dependence on the Kerr-AdS black-hole parameters. Furthermore, the constrained minimal surface can be studied in terms of the angular momentum $J$ of the Kerr-AdS black hole. This construction gives a simple physical interpretation of the effect of rotation.  In the static geometry, the event horizon simultaneously serves as the entropy surface and the thermodynamic-volume surface. Rotation separates these two roles: the event horizon continues to carry the Bekenstein-Hawking entropy, whereas $\Sigma_{\rm th}$ provides a distinct geometric representative of the thermodynamic volume.

Possible holographic interpretations of the constrained surface, particularly in connection with the surface/state correspondence and MERA, will be discussed in Section \ref{Holo Interp}.

\subsection{Static charged (dyonic) black hole: recovery of the event horizon}
\label{subsec:static-thermodynamic-volume-surface}

We first apply the prescription~\eqref{intro thermodynamic volume surface} to the static charged (dyonic) black hole discussed in Section~\ref{subsec:dyonic}. The corresponding metric is given by (\ref{Dyonic Ansatz}) with $U(r)$ in \eqref{dyonicU}. For the electrically charged black hole, one may reduce $2q^2$ to $q^2$. The precise normalization of the electric and magnetic charges does not affect the geometric argument below.

Consider a closed axisymmetric surface described by $r=R(\theta)$.  Its area, its LW volume, and the constrained functional are given by
\begin{align}
&A[R]=2\pi\int_0^\pi d\theta\,\sin\theta\,R(\theta)\sqrt{R(\theta)^2+\frac{R'(\theta)^2}{U(R(\theta))}},\nonumber\\
&\mathsf{V}[R]=\frac{2\pi}{3}\int_0^\pi d\theta\,\sin\theta\,R(\theta)^3,\nonumber\\
&\mathcal I[R,\lambda]=A[R]-\lambda\left(\mathsf{V}[R]-\mathsf{V}_{\mathrm{th}}\right).
\end{align}
The Euler-Lagrange equation of the last functional admits a constant-radius profile $R(\theta)=R_0$, with Lagrange multiplier $\lambda=2/R_0$.  The lapse-weighted volume enclosed by the minimal surface is then $\mathsf{V}(R_0)=4\pi R_0^3/3$. Since the thermodynamic volume of the dyonic AdS black hole is $\mathsf{V}_{\mathrm{th}}=4\pi r_+^3/3$, the fixed-volume condition determines $R_0=r_+$. The detailed variation of the static functional is collected in Appendix~\ref{app:minsurf-functional}. Therefore, $\Sigma_{\mathrm{th}}$ is equivalent to the horizon for the charged (dyonic) black hole. The Schwarzschild-AdS result follows immediately as the neutral limit.

The thermodynamic-volume surface is not introduced as an additional geometric object in the charged (dyonic) black hole. Rather, the event horizon itself already satisfies the constrained area-minimization principle. This observation provides the starting point for the analysis of the Kerr-AdS black hole.

The minimal surfaces in the asymptotically AdS geometries have reasonable physical interpretations. A famous surface is the Ryu-Takayanagi surface anchored at the boundary of the bulk geometries, whose area gives the entanglement entropy \cite{Ryu:2006bv, Ryu:2006ef}. In this particular case, a minimal surface corresponds to a reduced density operator or a mixed state. An extension of this perspective has been proposed in \cite{Miyaji:2015yva}. We will return to this view later to get an intuition for our thermodynamic volume surface $\Sigma_{\mathrm{th}}$. Also, we will see how the thermodynamic volume can be deformed from the horizon under the rotation of the black hole in the following sections.

\subsection{Kerr-AdS: Volumes enclosed by the horizon and ergosurface}

In general, the thermodynamic volume corresponds to a minimal surface through the LW volume defined in (\ref{intro volume functional}). In the non-rotating charged black hole cases, the only surface that has a physically nontrivial meaning is the horizon. Furthermore, the thermodynamic volume surface $\Sigma_{\mathrm{th}}$ introduced in (\ref{intro thermodynamic volume surface}) coincides with the horizon. On the other hand, the thermodynamic volume (\ref{Vth-J}) of the Kerr-AdS black hole implies that the corresponding surface differs from the horizon. In addition, there is another physically nontrivial surface in this geometry. It is known as the ergosphere. Thus, it is desirable to compare the event horizon, the ergosphere, and the thermodynamic-volume surface $\Sigma_{\mathrm{th}}$, and their LW volumes in the Kerr-AdS black hole. We will discuss these in this section.

For a closed axisymmetric surface $r=R(\theta)\equiv r_+ \xi(\theta)$, the enclosed LW volume becomes
\begin{align}\label{Volume-functional}
\mathsf{V}[\xi]=\frac{2\pi r_+^3}{\Xi}\int_0^\pi d\theta\,\sin\theta\left[\frac{\xi(\theta)^3}{3}+
\frac{a^2}{r_+^2}\xi(\theta)\cos^2\theta\right].
\end{align}
The condition $\mathsf{V}[\xi]=\mathsf{V}_{\mathrm{th}}$ admits infinitely many $R(\theta)$-surface profiles. We will apply the minimal surface prescription proposed in the previous section.  Before imposing the prescription, we compare $V_{\mathrm{th}}$ with the LW volumes enclosed by the two most distinguished surfaces in the Kerr-AdS black hole. They are the horizon and ergosurface.

The LW volume enclosed by the event horizon is $\mathsf{V}_{\rm horizon}=4\pi r_+(r_+^2+a^2)/(3\Xi)$, which coincides with the geometric volume considered in~\cite{Cvetic:2010jb}. Comparing this expression with the thermodynamic volume, we recover the relation
\begin{align}\label{DeltaV}
\Delta \mathsf{V}
\equiv \mathsf{V}_{\mathrm{th}}-\mathsf{V}_{\rm horizon}
=\frac{2\pi a^2(r_+^2+a^2)\left(3-\tilde{\Lambda}_{\rm eff}r_+^2\right)}{9r_+\Xi^2}
=\frac{4\pi G_4}{3}\,\frac{J^2}{M}.
\end{align}
Since $\tilde{\Lambda}_{\rm eff}=-3/L_{\rm eff}^2<0$ and $\Xi>0$, this difference is strictly positive for nonzero rotation:
\begin{align}
\mathsf{V}_{\mathrm{th}}>\mathsf{V}_{\rm horizon}
\qquad(a\neq0).
\end{align}
The event horizon continues to carry the Bekenstein--Hawking entropy, but its enclosed LW volume no longer reproduces the thermodynamic volume once rotation is turned on.

\begin{figure}[t]
    \centering
    \includegraphics[width=90mm]{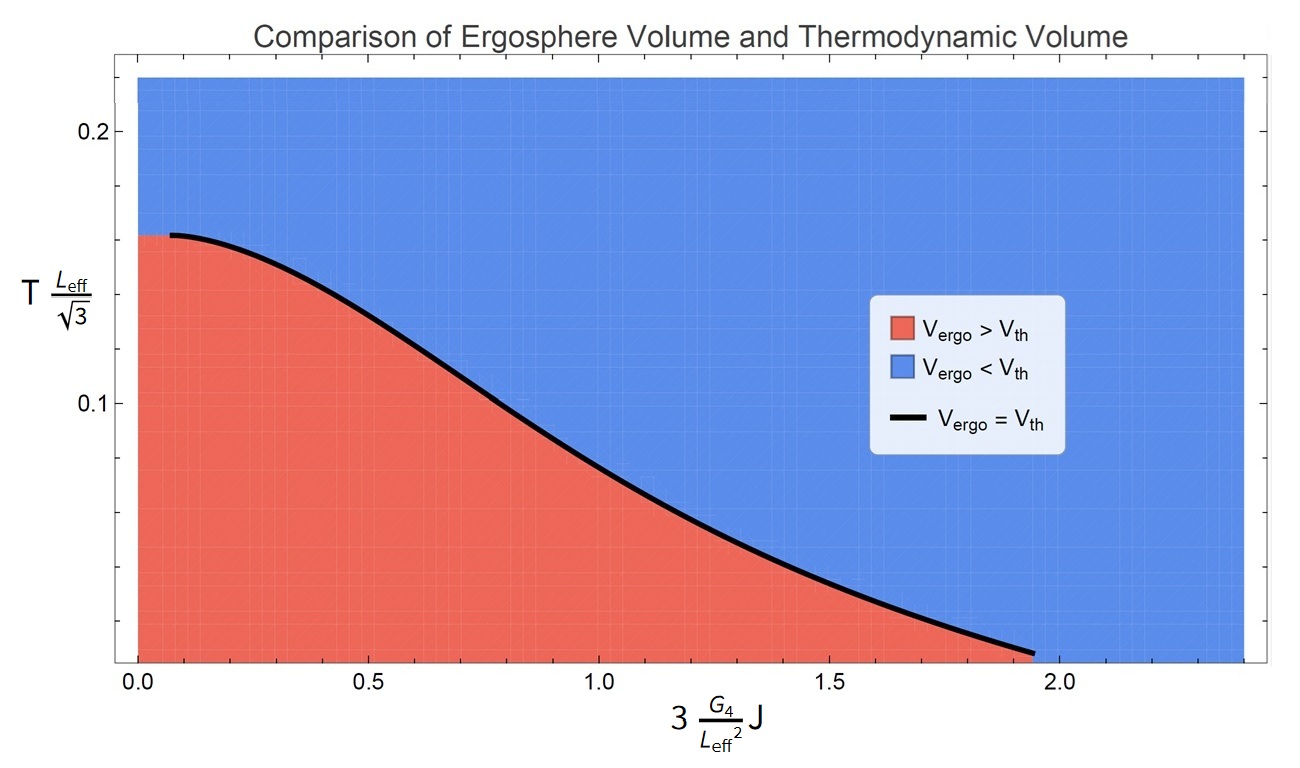}
    \caption{Numerical comparison of \(\mathsf{V}_{\rm ergo}\) and \(\mathsf{V}_{\mathrm{th}}\) over the physical Kerr-AdS solution space parametrized by \((J,T)\) with a normalization for massless expression. The red and blue regions correspond to \(\mathsf{V}_{\rm ergo}>\mathsf{V}_{\mathrm{th}}\) and \(\mathsf{V}_{\rm ergo}<\mathsf{V}_{\mathrm{th}}\), respectively. The black solid curve is the critical line \(\mathsf{V}_{\rm ergo}=\mathsf{V}_{\mathrm{th}}\). The ordering between the ergosphere volume and the thermodynamic volume is therefore parameter-dependent.
    }
    \label{fig:vergo-vth-region}
\end{figure}

We next consider the ergosphere. The outer stationary-limit surface $r=r_e(\theta)$ is defined by $g_{tt}=0$, or equivalently by $\Delta_r(r_e)=a^2\Delta_\theta(\theta)\sin^2\theta$. See (\ref{Kerr-AdS-metric}). Its enclosed LW volume is obtained by substituting $R(\theta)=r_e(\theta)$ into equation~\eqref{Volume-functional}.  Since $r_e(\theta)$ is determined only implicitly, the resulting volume must be evaluated numerically. At fixed $\tilde{\Lambda}_{\rm eff}$, we compare $\mathsf{V}_{\rm ergo}$ and $\mathsf{V}_{\mathrm{th}}$ over the physical Kerr--AdS solution space parametrized by $(J, T)$. Figure~\ref{fig:vergo-vth-region} shows that there is no universal ordering between the two volumes.  The equation $\mathsf{V}_{\rm ergo}=\mathsf{V}_{\mathrm{th}}$ defines a nontrivial critical line in the $(J, T)$ plane.  Across this line, the ordering changes from $\mathsf{V}_{\rm ergo}>\mathsf{V}_{\mathrm{th}}$ to $\mathsf{V}_{\rm ergo}<\mathsf{V}_{\mathrm{th}}$.

In summary, neither of the two natural surfaces in the Kerr--AdS geometry provides a universal representative of the thermodynamic volume. The LW volume enclosed by the horizon is strictly smaller than $\mathsf{V}_{\mathrm{th}}$ for a nonvanishing rotation, whereas the ordering between the ergosphere volume and $\mathsf{V}_{\mathrm{th}}$ depends on the black-hole parameters. The static horizon principle must therefore be continued to a more general off-horizon surface.

\subsection{Kerr-AdS: Minimal surface constrained by $\mathsf{V}_{\mathrm{th}}$}
\label{Amin}

We now apply the same prescription used in the static charged black holes to the Kerr-AdS black hole. Within the class of closed axisymmetric surfaces $r=r_+\xi(\theta)$, we minimize the area under the fixed-volume condition $\mathsf{V}[\xi]=\mathsf{V}_{\mathrm{th}}$. The constrained functional is
\begin{align}
\mathcal I[\xi,\lambda]=\mathsf{A}[\xi]-\lambda\left(\mathsf{V}[\xi]-\mathsf{V}_{\mathrm{th}}\right).
\label{eq:constrained-functional-main}
\end{align}
The explicit area functional (\ref{eq:appendix-kerr-area}) and the corresponding Euler-Lagrange equation are collected in Appendix~\ref{app:minsurf-functional}. In contrast to the static charged geometry, a constant-radius profile is not, in general, a solution of the full Kerr-AdS constrained equation. The thermodynamic-volume surface, therefore, becomes a nontrivial rotating continuation of the static event horizon.

The slow-rotation limit can nevertheless be treated analytically. Since the geometry is invariant under reversal of the rotation direction, $a\rightarrow -a$, the surface profile admits an expansion in even powers of $a$. Expanding the constrained Euler--Lagrange equation together with the fixed-volume condition through quartic order, we obtain (\ref{eq:appendix-xi-a4-final}). The detailed derivation is presented in Appendix~\ref{app:slow-rotation-expansion}. At quadratic order, the surface undergoes only a uniform radial displacement and therefore remains a constant-radius surface.  The first genuine shape deformation appears at quartic order and is purely quadrupolar. In particular, the difference between the profiles at the rotation axis and the equatorial plane is given by (\ref{eq:appendix-pole-equator}). Thus, although the leading rotational correction merely shifts the surface away from the horizon, the next-order correction already captures the angular deformation observed in the fully numerical solution.

\begin{figure}[t]
\centering
    \includegraphics[width=60mm]{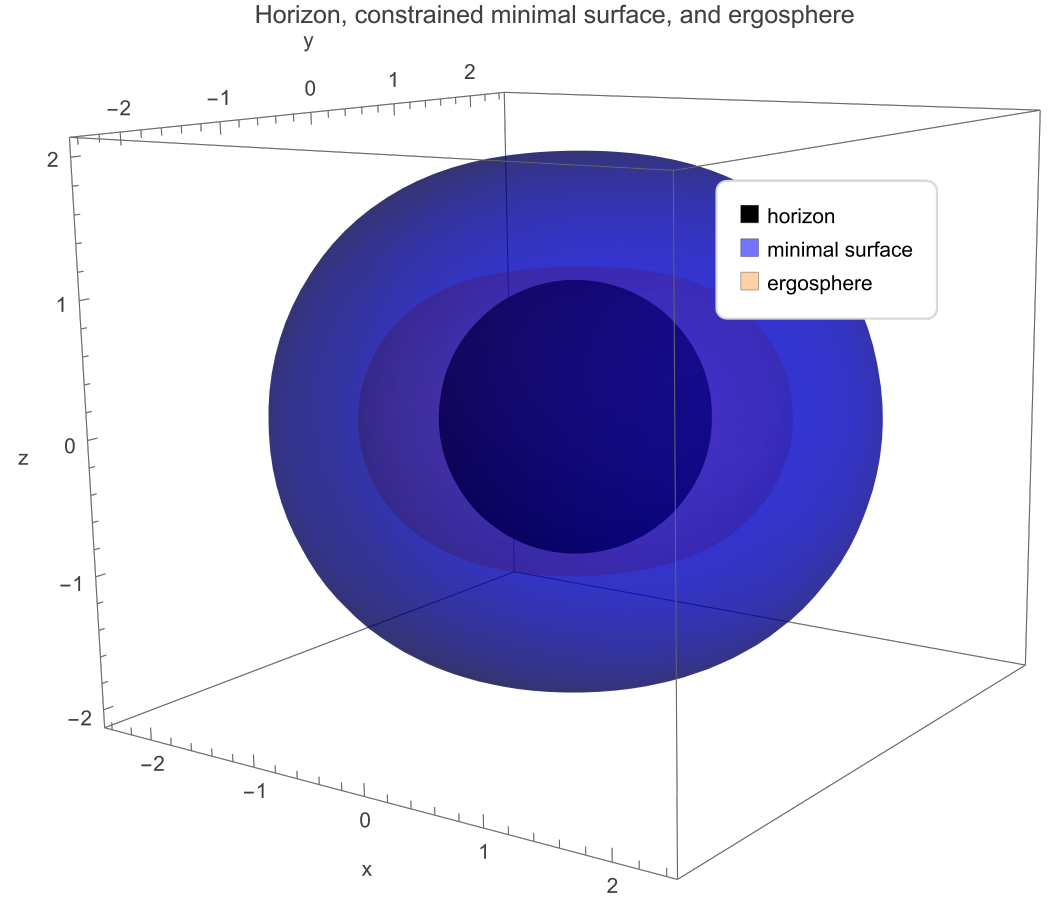}
    \includegraphics[width=90mm]{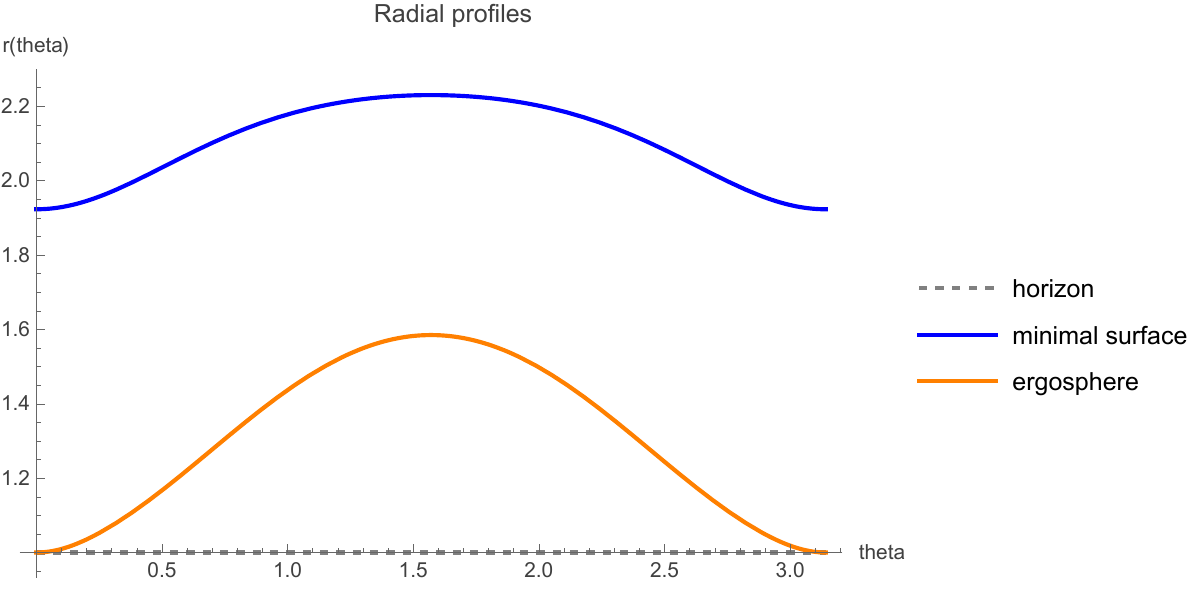}
\caption{
Comparison between the event horizon, the ergosphere, and the numerically obtained constrained minimal-area surface for the representative Kerr-AdS parameters \(\tilde{\Lambda}_{\mathrm{eff}}=-1\), \(a=1.4\), and \(r_+=1\). The thermodynamic-volume surface is obtained by minimizing the area functional under the fixed-volume condition \(\mathsf{V}[\xi]=\mathsf{V}_{\mathrm{th}}\). The left panel shows the full three-dimensional surfaces, while the right panel shows the meridional cross-section in a plane containing the rotation
axis.}
\label{fig:minsurf_horizon}
\end{figure}

For finite rotation, we solve the constrained variational problem numerically.  We discretize the profile on the northern hemisphere, use reflection symmetry across the equatorial plane, impose regularity at the axis and the equator, and minimize the area under the condition $\mathsf{V}[\xi]=\mathsf{V}_{\mathrm{th}}$.  Within the class of closed axisymmetric surfaces considered here, the numerical minimization selects an off-horizon representative enclosing the black hole. A representative solution is shown in Figure~\ref{fig:minsurf_horizon}. The thermodynamic-volume surface encloses the horizon and differs from both the horizon and the ergosphere. The figure therefore illustrates the nontrivial rotating continuation of the static case.

\subsection{Holographic Interpretation}\label{subsec:holographic-interpretation-sigma}
\label{Holo Interp}

So far, we have considered a minimal surface supported by the LW volume, which is identified with the thermodynamic volume. We will interpret this volume and area from the perspective of the surface/state correspondence studied in \cite{Miyaji:2015yva}. The authors of this paper propose a correspondence between convex surfaces in asymptotically AdS geometries and states defined by cutting links in tensor networks, particularly the Multi-scale Entanglement Renormalization Ansatz (MERA) and its continuous version, cMERA.

In this correspondence, the volume enclosed by a surface can be heuristically associated with the number of disentanglers in a MERA network. However, the geometric interpretation of MERA is not unique. In particular, Czech et al. \cite{Czech:2015kbp} argued that MERA is more naturally related to kinematic space than directly to a constant-time slice of the AdS bulk. Furthermore, our lapse-weighted (LW) volume includes the lapse function associated with a chosen stationary time slicing and therefore differs from the ordinary spatial volume. We thus propose a schematic interpretation of the LW volume as an effective, redshift-weighted measure of disentangling operations, rather than identifying it directly with the number of disentanglers in a specific tensor network. This lapse weighting is reminiscent of the gravitational redshift appearing in the relation between Hawking and Tolman temperatures.

\begin{figure}[t]
\centering
    \includegraphics[width=100mm]{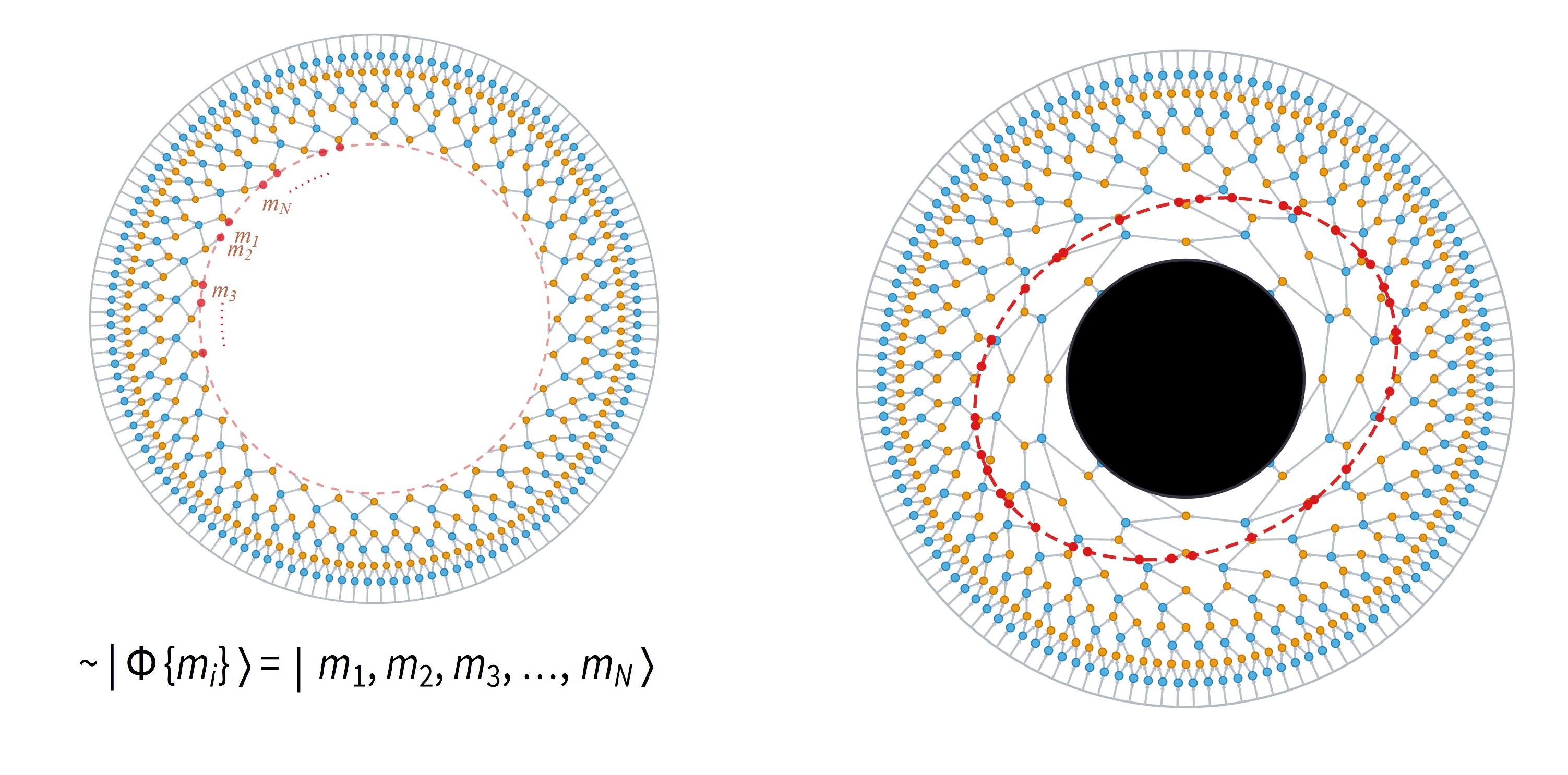}
    \caption{{\bf Left:} A schematic plot for the state/surface correspondence in \cite{Miyaji:2015yva}. ~{\bf Right:} The horizon interior (Black circle) and a thermodynamic volume surface (Dashed line).}
\label{fig:mera_pure}
\end{figure}

In more detail, a state, denoted by $\left.|\Phi(m_1,\ldots,m_N)\right>$, in this correspondence \cite{Miyaji:2015yva} is constructed by cutting links with a surface. See Figure \ref{fig:mera_pure}. In such a configuration, the surface area is proportional to the number of links cut by this surface. Furthermore, this surface area does not depend on an observer. On the other hand, a mixed state is constructed using this pure state with some weighting coefficients $\mathbb{C}_{\{m_i\},\{m'_j\}}$. This surface state is also used to construct a density operator for a mixed state as follows:
\begin{align}\label{mixed MERA}
\rho_{\rm mixed} = \sum_{\{m_i\},\{m'_j\}} \mathbb{C}_{\{m_i\},\{m'_j\}}\left.|\Phi\{m_i\}\right> \left< \Phi\{m'_j\} |\right.\,.
\end{align}
Also, \cite{Miyaji:2015yva} discussed methods to obtain these coefficients.

A representative mixed state is the thermal state of the CFT. This state is achieved by cutting along the horizon surface, with the Boltzmann factors serving as the weighting coefficients (\ref{mixed MERA}). The LW volume of this horizon surface $\mathsf{V}_{\rm horizon}$ may be an `effective' number of disentanglers, including the red shift effect inside the horizon cutting surface. The `effective' means that $\mathsf{V}_{\rm horizon} (=\frac{4}{3}\pi r_+^3)$ is the default value for the horizon. Now, consider a surface larger than and surrounding the horizon. See the right figure in Figure \ref{fig:mera_pure}. This surface cuts more links than the horizon, and we denote the effective number of additional disentanglers between this surface and the horizon by $\Delta\mathsf{V}$ in (\ref{DeltaV}). I.e.,
\begin{align}
\Delta\mathsf{V} \sim \left(
\begin{array}{c}
 \text{Effective number of disentanglers between}  \\
 \text{the horizon and a convex surface}  \\
\end{array}
\right)v_{cell}\,,
\end{align}
where $v_{cell}$ is an effective volume per disentangler, whose normalization depends on the tensor-network regularization and is not determined by our gravitational construction.

Once you pick a thermodynamic volume value corresponding to a particular $\Delta\mathsf{V}$, you can choose many cutting surfaces that give rise to $\Delta\mathsf{V}$. Among these surfaces, it is possible to choose one that yields the minimal cut links. This minimal cutting surface corresponds to the minimal surface constrained by the thermodynamic volume. This selected surface is conjectured to approach the minimal surface constrained by volume in the previous section \ref{Amin} in the cMERA limit.

To summarize what we found, let us denote the thermodynamic-volume surface and the horizon surface in the rotating geometry by $\Sigma_{\mathrm{th}, a}$ and $\Sigma_{\mathrm{H}, a}$, where $a$ is the rotation parameter. The geometric result of the previous subsections can be summarized as
\begin{align}
\Sigma_{\mathrm{th},0}=\Sigma_{\mathrm{H},0},\qquad\Sigma_{\mathrm{th},a}\neq\Sigma_{\mathrm{H}, a},\qquad\mathsf{V}[\Sigma_{\mathrm{th},a}]=\mathsf{V}_{\mathrm{th}}(a)
\quad(a\neq0).
\label{eq:sigma-summary}
\end{align}
The event horizon $\Sigma_{\mathrm{H}, a}$ continues to determine the Bekenstein--Hawking entropy. In contrast, the off-horizon surface $\Sigma_{\mathrm{th}, a}$ provides a lapse-weighted geometric representative of the thermodynamic volume. Rotation, therefore, separates two roles that coincide in the static geometry: the entropy carrier and the thermodynamic-volume carrier.

It is important to state clearly what is not being claimed.  The surface $\Sigma_{\mathrm{th}, a}$ is not identified with a Ryu--Takayanagi surface or a quantum extremal surface.  It is not anchored on a boundary subregion, and it is not selected by extremizing a generalized entropy.  Nor do we claim that it is a genuine holographic-screen leaf in the sense of Bousso and Engelhardt
\cite{Bousso:2015qqa}.  In that framework, a holographic screen is foliated by codimension-two leaves satisfying a marginal null-expansion condition. By contrast, $\Sigma_{\mathrm{th},a}$ is selected by constrained area minimization under the fixed-volume condition $\mathsf{ V}[\Sigma_{\mathrm{th},a}]=\mathsf{V}_{\mathrm{th}}(a)$. The deformation begins at quadratic order in the rotation parameter, in accordance with the invariance of the geometry under $a\rightarrow-a$.  This is also qualitatively consistent with the general expectation that distinguishability measures between nearby states begin quadratically in a smooth deformation parameter. Relative entropy provides a natural language for this statement in holography \cite{Jafferis:2015del}.  We do not, however, identify the relative entropy with an area difference or a volume difference involving $\Sigma_{\mathrm{th}, a}$.

The relation $\mathsf{V}[\Sigma_{\mathrm{th},a}]=\mathsf{V}_{\mathrm{th}}(a)$ is also suggestive in view of the connections between thermodynamic volume and holographic complexity explored in the literature \cite{Zhang:2024mxb, Susskind:2014rva, Stanford:2014jda, Brown:2015lvg, Couch:2016exn, AlBalushi:2020rqe, AlBalushi:2020heq}.

\section{Conclusion and Discussion}\label{sec:conclusion}

In this work, we studied a higher-dimensional realization of black hole chemistry in an eight-dimensional Einstein--Yang--Mills--Maxwell theory with compact internal spaces supporting Yang--Mills instantons. After dimensional reduction to the four-dimensional Einstein frame, the size modulus of the compact internal space is stabilized by the internal curvature, the higher-dimensional cosmological constant, and the instanton contribution. The stabilized value $\mathcal X_s$ determines the effective four-dimensional cosmological constant $\tilde{\Lambda}_{\rm eff}$. Thus, in the AdS branch, the thermodynamic pressure (\ref{Pth-Lambda}) is not introduced as an independent parameter, but is generated dynamically by the compactification data.

The vacuum structure of the reduced theory depends on the parameters of the higher-dimensional model. In particular, the effective four-dimensional vacuum can be either anti-de Sitter or de Sitter. In the present work, we focused on the anti-de Sitter branch for the black-hole-chemistry analysis. We considered Schwarzschild-AdS, dyonic AdS, and Kerr-AdS black holes, and re-expressed their phase structures in terms of the stabilized size of the extra dimensions. In all cases, increasing $\mathcal X_s$ lowers the transition temperature and enlarges the range of excluded black hole sizes. This shows that the thermodynamic phase structure of the effective four-dimensional black hole is sensitive to the size of the compact internal space.

For Schwarzschild-AdS black holes, the Hawking--Page transition gives a lower bound on the thermodynamically allowed black hole size for a given extra-dimensional size. As $\mathcal X_s$ increases, this lower bound also increases. Therefore, configurations below the threshold are thermodynamically disfavored relative to thermal AdS or to larger black hole configurations. For dyonic and Kerr-AdS black holes, the Small/Large black hole transition leads to an intermediate excluded region above the critical extra-dimension size. Configurations in this region should not be regarded as allowed equilibrium states in the corresponding ensemble. Rather, if an appropriate dynamical channel is available and the conserved charges are respected, such configurations are expected to evolve toward one of the stable branches.

These thermodynamic size restrictions are reminiscent of the Gregory--Laflamme instability in the broad sense that the relative scale between a black object and a compact direction plays an important role. The mechanism discussed here, however, is different. We have not analyzed perturbative instabilities of black strings or black branes. Instead, the size restriction arises from the effective four-dimensional pressure determined by the compact internal space. It would be interesting to investigate whether there is a more direct dynamical relation between our thermodynamic size bounds and higher-dimensional perturbative instabilities.

We also investigated the geometric meaning of the Kerr-AdS thermodynamic volume. For static charged black holes, the thermodynamic volume can be reproduced by integrating the LW volume element over the region inside the event horizon. In the rotating case, however, the situation is more subtle. We compared $\mathsf{V}_{\mathrm{th}}$ with two natural geometric volumes: the LW volume enclosed by the event horizon and the volume enclosed by the ergosphere. We found that the horizon volume is strictly smaller than the thermodynamic volume when rotation is nonzero. In contrast, the ordering between the ergosphere volume and the thermodynamic volume depends on the black hole parameters. Thus, neither the horizon nor the ergosphere provides a universal geometric representative of the Kerr-AdS thermodynamic volume.

Motivated by this observation, we proposed a constrained minimal-area surface $\Sigma_{\mathrm{th}, a}$ as a representative of the thermodynamic volume. The surface is defined within the class of closed axisymmetric surfaces enclosing the black hole and satisfying the fixed-volume condition
\begin{align}
    \mathsf{V}[\Sigma_{\mathrm{th},a}]=\mathsf{V}_{\mathrm{th}}(a).
\end{align}
Among such surfaces, $\Sigma_{\mathrm{th}, a}$ is selected by minimizing the area.   In the slow-rotation expansion, this surface is a constant-radius surface slightly outside the event horizon, while for finite rotation it appears as a numerically obtained spheroid. This construction shows that, for rotating Kerr-AdS black holes, the entropy carrier and the thermodynamic-volume carrier are geometrically distinct: the event horizon $\Sigma_{\mathrm{H}, a}$ determines the Bekenstein--Hawking entropy. At the same time, the off-horizon surface $\Sigma_{\mathrm{th}, a}$ represents the thermodynamic volume.

We further discuss a possible holographic interpretation of this surface using the so-called surface/state correspondence \cite{Miyaji:2015yva}. This correspondence proposal is based on a conjecture between spacetimes and tensor networks \cite{Swingle:2009bg, Nozaki:2012zj, Swingle:2012wq}. A convex surface cuts links, which determines a state. We quantify the LW volume with our thermodynamic-volume minimal surface. As a result, the thermodynamic volume based on the LW volume is related to the number of disentanglers between the horizon and the thermodynamic-volume surface.

The surface $\Sigma_{\mathrm{th}, a}$ is not a Ryu--Takayanagi surface. It is not anchored to a boundary subregion, nor does a marginal null-expansion condition define it. Nevertheless, as a closed codimension-two surface selected by thermodynamic data, it may be viewed as a possible screen-like representative of the rotating black hole state, in the spirit of the surface/state correspondence. From this viewpoint, the rotation deformation separates the entropy surface from the thermodynamic-volume surface. The condition $\mathsf{V}[\Sigma_{\mathrm{th}, a}]=\mathsf{V}_{\mathrm{th}}(a)$ also suggests a possible CV-type interpretation of the thermodynamic volume. However, we do not claim an exact holographic complexity formula.

Future work has several directions. First, in the present paper, we restricted our attention to constant values of the compactification modulus. It would be important to relax this assumption and study black holes with a dynamical warping factor or moduli profile. Such solutions would clarify how the compactification dynamics responds locally to black hole charge, rotation, and thermodynamic pressure. Second, the relation between our thermodynamic size bounds and Gregory--Laflamme-type instabilities deserves a more direct higher-dimensional analysis. In particular, it would be interesting to study whether the excluded regions in the four-dimensional phase diagrams have a counterpart in the perturbative spectrum of the corresponding higher-dimensional configurations.

Third, the de Sitter branch of the reduced theory remains to be explored. The present black-hole-chemistry analysis was carried out mainly in the AdS branch, where the pressure-volume interpretation is standard. Since the same compactification mechanism also allows de Sitter vacua for appropriate parameters, it would be useful to determine whether an analogous extended thermodynamic interpretation can be developed in that context. Finally, the geometric and holographic status of the thermodynamic-volume surface $\Sigma_{\mathrm{th}, a}$ should be studied further. In particular, computing its null expansions may clarify whether it is related to genuine holographic screens. In addition, a more precise treatment of relative entropy or holographic complexity may sharpen the proposed screen-like interpretation.

\section*{Acknowledgement}

We thank Yunseok Seo, Sang-Heon Yi, and Sang-A Park for their helpful discussions. K. K. Kim thanks Nobuyoshi Ohta for a helpful comment on the magnetic black hole. This work is supported by the Basic Science Research Program through the NRF grant No. NRF-2019R1A2C 1007396 (K. K. Kim) and S-2024-0539-000 (J. Ho). K. K. Kim and J. Ho are members of the Korea Research Network for Theoretical Physics(KNeT-TP). We thank members of KNeT-TP for helpful discussions. We thank the participants in the Topical Research Program [APCTP--026--T04] held at APCTP in Pohang, Korea, for fruitful discussions. We acknowledge the hospitality at APCTP, where part of this work was done.

\appendix

\section*{Appendix}
\section{$\mathbb{CP}^2$ and $\mathbb{S}^2 \times \mathbb{S}^2$ as extra dimensions}

\subsection{$\mathbb{CP}^2$ as extra dimensions}\label{section cp2}

Let us consider $\mathbb{CP}^2$ as the extra-dimensional space. In this case, $h_{ab}$ in the metric ansatz (\ref{metricansatz}) can be written as
\begin{align}
    h_{ab}{dy}^a{dy}^b=\frac{1}{u}\left(r^2\sigma_1^2+r^2\sigma_2^2\right)+\frac{1}{u^2}\left(r^2\sigma_3^2+{dr}^2\right)=e^a\otimes e^a
\end{align}
with
\begin{align}
    u(r)=1+\frac{\lambda\, r^2}{6}\,,
\end{align}
where $\lambda$ is a constant representing the size of $\mathbb{CP}^2$, and $r^2=\sum_{a=1}^4 y_a y_a$. The coordinates $y_a$ parameterize $\mathbb{R}^4$, and the $\sigma_i$ are the left-invariant one-forms on the group manifold of $SU(2) \cong\mathbb{S}^3$, for $i=1, 2, 3$. We set $\lambda=\frac{1}{4}$ so that $\mathbb{CP}^2$ has unit Ricci scalar, similar to the $\mathbb{S}^4$ case. The vierbeins can be expressed as
\begin{align}
&e^1=\frac{1}{r\sqrt{u(r)}}(-y_4 dy_1 -y_3 dy_2 + y_2 dy_3 +y_1 dy_4) \nonumber \\
&e^2=\frac{1}{r\sqrt{u(r)}}(y_3 dy_1 -y_4 dy_2 -y_1 dy_3 +y_2 dy_4) \nonumber \\
&e^3=\frac{1}{ru(r)}(-y_2 dy_1 + y_1 dy -y_4 dy_3 + y_3 dy_4) \nonumber \\
&e^4=\frac{1}{ru(r)}(y_1 dy_1 + y_2 dy_2 + y_3 dy_3 + y_4 dy_4 )\,.
\end{align}
We then introduce a self-dual $SU(2)$ gauge field configuration given by
\begin{align}
A=A^3\tau^{3},~A^3
=-\frac{2}{\sqrt{6}}\frac{H}{16} r e^3\,.
\end{align}
The corresponding field strength is
\begin{align}
F=F^3\tau^{3},~F^3
=\frac{2}{\sqrt{6}}\frac{H}{8}(e^1\wedge e^2+ e^3\wedge e^4)\,,
\end{align}
Consequently, one finds $\text{Tr} F_{ab} F^{ab}=-\frac{H^2}{24}$. One can now perform dimensional reduction under this self-dual configuration. The resulting effective action for the $\mathbb{CP}^2$-compactification takes the same form as (\ref{effactions4}), with $\ell_H^2=\frac{\pi \mathbf{G}_{(8)} H^2}{6 g_{\text{YM}}^{2}}$.

\subsection{\(\mathbb{S}^2 \times \mathbb{S}^2\) as extra dimensions}\label{section s2s2}

Here, two $\mathbb{S}^2$'s in $\mathbb{S}^2\times \mathbb{S}^2$ have the same radius. In addition, $\mathbb{S}^2\times \mathbb{S}^2$ can be generalized to non-Einstein manifolds with different radii.

Another nonsingular compact manifold consistent with our ansatz, (\ref{metricansatz}) and (\ref{gaugeansatz}), is \(\mathbb{S}^2 \times \mathbb{S}^2\). Let us choose four angle coordinates as $(\theta_1,\phi_1,\theta_2,\phi_2)$, and the radius of each sphere is given by $\varrho_1$ and $\varrho_2$, respectively. The metric is then given by:
\begin{align}
ds^2_{\mathbb{S}^2 \times \mathbb{S}^2} = \varrho_1^2 \left(d\theta_1^2 + \sin^2\theta_1\, d\phi_1^2\right) + \varrho_2^2 \left(d\theta_2^2 + \sin^2\theta_2\, d\phi_2^2\right)\,.
\end{align}
A natural gauge field configuration is
\begin{align}
A = A^3 \tau^3, \quad A^3 = -\frac{H_1}{2} \cos\theta_1\, d\phi_1 - \frac{H_2}{2} \cos\theta_2\, d\phi_2\,,
\end{align}
and the corresponding field strength becomes
\begin{align}
F = F^3 \tau^3, \quad F^3 = \frac{H_1}{2} \sin\theta_1\, d\theta_1 \wedge d\phi_1 + \frac{H_2}{2} \sin\theta_2\, d\theta_2 \wedge d\phi_2\,.
\end{align}
This gauge field satisfies the equation of motion, but it is not (anti-)self-dual unless \(\frac{H_1}{H_2} = \pm \left(\frac{\varrho_1}{\varrho_2}\right)^2\).

To investigate the structure of the equations of motion within a more general framework, we extend the Lorentzian 
$(3+1)$-dimensional spacetime to $(d+1)$ dimensions. The equations of motion derived from the original action (\ref{original action}) can be decomposed into the $(d+1)$-dimensional Lorentzian spacetime and two $\mathbb{S}^2$ components as follows:
\begin{align}
&\mathcal{R}_{\mu \nu}  - \frac{1}{2} g_{\mu \nu} \left( \mathcal{R} + R_{(4)} - 2 \Lambda  
+ \frac{4 \pi \mathbf{G}_{(D)}}{g_{\text{YM}}^2} \text{Tr} F_{a b} F^{a b}  \right) 
=8\pi \mathbf{G}_{(D)}\, T^{U(1)}_{\mu\nu}~\label{eom_s2s2},\\
&R_{ab}^{(1)}  - \frac{1}{2} h_{ab}^{(1)} \left( \mathcal{R} + R_{(4)} - 2 \Lambda  
- \frac{4 \pi \mathbf{G}_{(D)}}{e^2} \mathcal{F}^2 \right)=8\pi \mathbf{G}_{(D)}\, T^{SU(2)}_{ab}~,\label{1st S2}\\
&R_{cd}^{(2)}  - \frac{1}{2} h_{cd}^{(2)} \left( \mathcal{R} + R_{(4)} - 2 \Lambda  
- \frac{4 \pi \mathbf{G}_{(D)}}{e^2} \mathcal{F}^2 \right)=8\pi \mathbf{G}_{(D)}\, T^{SU(2)}_{cd}~,\label{2nd S2}
\end{align}
where $R_{ab}^{(\alpha)}$ ($\alpha=1,2$) denotes the Ricci tensor of the $\alpha$-th 2-sphere, and the $h_{ab}^{(\alpha)}$ is its corresponding metric. Each sphere constitutes an Einstein manifold; however, their product is not Einstein unless $\varrho_1=\varrho_2$. $T^{SU(2)}_{ab}$ is proportional to the metric of each $\mathbb{S}^2$ component. Using $R_{ab}^{(\alpha)}=\frac{1}{2}h_{ab}^{(\alpha)}R^{(\alpha)}$, equations (\ref{1st S2}) and (\ref{2nd S2}) can be reformulated into a non-self-dual condition for the fluxes and the $(d+1)$-dimensional curvature as follows:
\begin{align}
&\left(\frac{1}{\varrho_1^2}-\frac{1}{\varrho_2^2}\right)=\frac{2 \pi \mathbf{G}_{(D)}}{g_{\text{YM}}^2}\left(\frac{H_1^2}{\varrho_1^4}-\frac{H_2^2}{\varrho_2^4}\right)\label{conditions2s2_1},\\
&\mathcal{R}=\frac{4 \pi \mathbf{G}_{(D)} \mathcal{F}^2}{e^2}+2\Lambda-\left(\frac{1}{\varrho_1^2}+\frac{1}{\varrho_2^2}\right)\,. \label{conditions2s2_2}
\end{align}

We now examine the trace of equation (\ref{eom_s2s2}), expressed as:
\begin{align}
   \mathcal{R}=&-\left(\frac{d+1}{d-1}\right)\left[2\left(\frac{1}{\varrho_1^2}+\frac{1}{\varrho_2^2}\right)-2\Lambda-\frac{2\pi \mathbf{G}_{(D)}}{g_{\text{YM}}^2}\left(\frac{H_1^2}{\varrho_1^4}+\frac{H_2^2}{\varrho_2^4}\right)\right]\nonumber\\
  &+\frac{4\pi \mathbf{G}_{(D)}}{e^2}\left(\frac{3-d}{1-d}\right)\mathcal{F}^2.
\end{align}
To ensure consistency with equation (\ref{conditions2s2_2}), we derive the $(d+1)$-dimensional curvature and $\mathcal{F}^2$ as follows:
\begin{align}
&\mathcal{R}=4 \Lambda -\frac{1}{2} (d+5) \left(\frac{1}{\varrho _2^2}+\frac{1}{\varrho _1^2}\right)+ \frac{(1+d)\pi\mathbf{G}_{(D)}}{g_{\text{YM}}^2}\left(\frac{{H_1}^2}{\varrho _1^4}+\frac{{H_2}^2}{\varrho _2^4}\right)\,, \nonumber\\
&\mathcal{F}^2=\frac{e^2 \Lambda }{2 \pi  \mathbf{G}_{(D)}} -\frac{(d+3) e^2}{8 \pi  \mathbf{G}_{(D)}}\left(\frac{1}{\varrho _2^2}+\frac{1}{\varrho _1^2}\right) +\frac{(d+1) e^2}{4 g_{\text{YM}}^2}\left(\frac{H_1^2}{\varrho _1^4}+\frac{H_2^2}{\varrho _2^4}\right)\,.
\end{align}
In the generic case, $\mathcal{F}^2$ is a nontrivial function of coordinates. Consequently, we restrict our analysis to the case where $\mathcal{F}^2=0$, yielding:
\begin{align}
\mathcal{R}=\frac{2 (d+1)}{d+3}\left[\Lambda -\frac{\pi  \mathbf{G}_{(D)}}{g_{\text{YM}}^2}\left(\frac{H_1^2}{\varrho _1^4}+\frac{H_2^2}{\varrho _2^4}\right) \right]\,.
\end{align}
This result demonstrates that a straightforward extension of the non-Einstein $\mathbb{S}^2\times\mathbb{S}^2$ is feasible.

We now derive the 4-dimensional reduced action to assess the stability of vacua with the constant warping factor $e^{2f}$. To dynamically account for the size of the extra dimensions, we propose the following ansatz:  
\begin{align}
&{ds}^2=g_{\mu \nu }{dx}^{\mu }{dx}^{\nu } + L^2 e^{2f_1(x)}  h_{ab}^{(1)}(y) {dy}_1^a {dy}_1^b
+ L^2 e^{2f_2(x)}  h_{cd}^{(2)}(y) {dy}_2^c {dy}_2^d\,,\nonumber\\
& A=A_{a}(y){dy}^{a}=\left(-\frac{H_1}{2} \cos{\theta_1} d\phi_1 
    -\frac{H_2}{2}\cos{\theta_2}  d\phi_2 \right)\tau^3 ,\;\mathcal{A}=\mathcal{A}_{\mu}(x){dx}^{\mu}\,,
\end{align} 
where $h_{ab}^{(i)}$ is the metric of the $i$-th sphere with the unit curvature radius. Using this ansatz, we partially integrate out the extra-dimensional components of the original action, yielding:
\begin{align}
S_{tot}
&=S_{gravity}+S_{gauge} \nonumber\\
&=\frac{1}{2\kappa_4^2} \int d^4 x \sqrt{-g} \left[e^{2(f_1+f_2)} \left( \mathcal{R} -\frac{2\mathbf{s}}{L^2} -\ell_e^2 \mathcal{F}_{\mu \nu} \mathcal{F}^{\mu \nu} \right) \right. \nonumber\\
&~~~~~~~~~~~~~~~ + 2 e^{2(f_1+f_2)} \left(\nabla_{\mu}f_1\nabla^{\mu}f_1+\nabla_{\mu}f_2\nabla^{\mu}f_2+4\nabla_{\mu}f_1\nabla^{\mu}f_2\right)\nonumber \\
&~~~~~~~~~~~~~~~ \left.+\frac{1}{2 L^2}e^{2f_1}+\frac{1}{2 L^2}e^{2f_2} - \frac{\ell_{H_1}^2}{2L^4} e^{2(f_2-f_1)}-\frac{\ell_{H_2}^2}{2L^4} e^{2(f_1-f_2)}  \right],
\end{align} 
where $2\kappa_4^2=\frac{16\pi \mathbf{G}_{(8)}}{\mathcal{V}_1\mathcal{V}_2}$ with $\mathcal{V}_i = \int d^2 y_{i} L^2 \sqrt{h^{(i)}}$ and we set $R[h^{(i)}_{ab}]=1/2$. The length parameter is defined as $\ell_{H_i}^2\equiv\frac{\pi G_{(8)} H_i^2}{4 g_{\text{YM}}^{2}}$. When $H_1=H_2$ and $f_1=f_2=f$, this action reduces to the $\mathbb{S}^4$ case (\ref{reduce action}). Applying a Weyl transformation, $g_{\mu\nu}=e^{-2(f_1+f_2)}\tilde{g}_{\mu\nu}$, the action becomes:
\begin{align}
\mathcal{I}
&=\frac{1}{2\kappa^2}\int d^{4} x \sqrt{-\tilde{g}}\left[\tilde{\mathcal{R}} -\ell_e^2 e^{2(f_1+f_2)}\tilde{\mathcal{F}}_{\mu \nu} \tilde{\mathcal{F}}^{\mu \nu} \right. \nonumber \\
&~~~~~~~~~~\left.-4\left(\tilde{\nabla}_{\mu}f_1\tilde{\nabla}^{\mu}f_1+\tilde{\nabla}_{\mu}f_1\tilde{\nabla}^{\mu}f_2+\tilde{\nabla}_{\mu}f_2\tilde{\nabla}^{\mu}f_2\right)-V(\mathcal{X}_1,\mathcal{X}_2)\right]\,, \label{effaction_s2s2}
\end{align}
with the potential given by
\begin{align}
V(\mathcal{X}_1,\mathcal{X}_2)=\frac{2\mathbf{s}}{L^2}\frac{1}{\mathcal{X}_1\mathcal{X}_2}-\frac{1}{2L^2}\frac{1}{\mathcal{X}_1^2\mathcal{X}_2}-\frac{1}{2L^2}\frac{1}{\mathcal{X}_1\mathcal{X}_2^2}
 +\frac{\ell_{H_1}^2}{2L^4}\frac{1}{\mathcal{X}_1^3\mathcal{X}_2}+\frac{\ell_{H_2}^2}{2L^4}\frac{1}{\mathcal{X}_1\mathcal{X}_2^3}\,,
\end{align}
where $\mathcal{X}_1 = e^{2f_1}$ and $\mathcal{X}_2 = e^{2f_2}$. The potential $V(\mathcal{X}_1,\mathcal{X}_2)$ characterizes the vacuum structure on this compactification. Certain extrema of the potential correspond to stable vacua.  When $f_1=f_2$ and $H_1=H_2$, this potential reduces to that of the $\mathbb{S}^4$ case. We focus on this scenario; however, an extension to extra dimensions with differing $\mathbb{S}^2$ sizes remains feasible.

\section{D=4 Euclidean Kerr-AdS Black Hole}

We briefly summarize the thermodynamics of the Euclidean Kerr-AdS black hole in four dimensions with a curvature radius $L$. The analytic continuation of the time coordinate and the rotation parameter is given by $t = i \tau$ and $a = i \hat{a}$ between the Euclidean and Lorentzian systems. The Euclidean metric is 
\begin{align}
ds^2 = \frac{\hat{\rho}^2 }{\hat{\Delta}_r} dr^2 +\frac{\hat{\Delta}_r \hat{\rho}^2}{(r^2-\hat{a}^2)^2}\omega^2  +\frac{\hat{\rho}^2}{\hat{\Delta}_\theta}\left( d\theta^2 + \sin^2 \theta \,\tilde{\omega}^2 \right)\,
\end{align}
where
\begin{align}
&\hat{\rho}^2 = r^2 - \hat{a}^2 \cos^2 \theta~,~\hat{\Delta}_r=(r^2 -\hat{a}^2)\left( 1+\frac{r^2}{L^2} \right)- 2 \mathbf{G}_4 m r\\
&\hat{\Delta}_\theta = 1 +\frac{\hat{a}^2}{L^2} \cos^2 \theta~,~\hat{\Xi}=1+\frac{\hat{a}^2}{L^2}\,, \\
&\omega = \frac{r^2-\hat{a}^2}{\hat{\rho}^2}\left(d\tau - \frac{\hat{a} \sin^2\theta}{\hat{\Xi}}d\phi  \right)~,~\tilde{\omega}= \hat{\Delta}_\theta \frac{r^2 -\hat{a}^2}{\hat{\Xi}\hat{\rho}^2}\left(d\phi +\frac{\hat{a}\hat{\Xi}}{r^2-\hat{a}^2}d\tau \right)\,.
\end{align}

Now, we consider the near-horizon geometry by introducing the following coordinate:
\begin{align}
\hat{\Delta}_r  \sim  (r-r_+)  \hat{\Delta}_r'(r_+)=\frac{\alpha}{4} \eta^2 \,,
\end{align} 
where $\alpha=\hat{\Delta}_r'(r_+)$. Then, the metric near the horizon is given by
\begin{align}
ds^2 &\sim \frac{\hat{\rho}_+^2 }{\alpha} d\eta^2 +\frac{\alpha\eta^2 \hat{\rho}_+^2}{4(r_+^2-\hat{a}^2)^2}\omega_+^2  +\frac{\hat{\rho}_+^2}{\hat{\Delta}_\theta}\left( d\theta^2 + \sin^2 \theta \,\tilde{\omega}_+^2 \right)\nonumber\\
&=ds_{\text{1st}}^2 + ds_{\text{2nd}}^2\,,
\end{align}
where
\begin{align}
\omega_+ = \frac{r_+^2-\hat{a}^2}{\hat{\rho}_+^2}\left(d\tau - \frac{\hat{a} \sin^2\theta}{\hat{\Xi}}d\phi  \right)~\text{and}~\tilde{\omega}_+= \hat{\Delta}_\theta \frac{r_+^2 -\hat{a}^2}{\hat{\Xi}\hat{\rho}_+^2}\left(d\phi +\frac{\hat{a}\hat{\Xi}}{r_+^2-\hat{a}^2}d\tau \right)
\end{align}
with
\begin{align}
\hat{\rho}_+^2 = r_+^2 -\hat{a}^2 \cos^2\theta\,. 
\end{align}
The second part of the metric is
\begin{align}
ds_{\text{2nd}}^2 = \frac{\hat{\rho}_+^2}{\hat{\Delta}_\theta}d\theta^2 +   \frac{\hat{\rho}_+^2}{\hat{\Delta}_\theta} \left(\hat{\Delta}_\theta \frac{r_+^2 -\hat{a}^2}{\hat{\Xi}\hat{\rho}_+^2}\right)^2 \sin^2\theta d\psi^2\,,
\end{align}
where we introduce an angle coordinate $\psi = \phi + \frac{\hat{a}\hat{\Xi}}{r_+^2 - \hat{a}^2}\tau$. The regularity can be justified by the regularity at $\theta=0$ and $\theta=\pi$. The metric near these points is
\begin{align}
ds_{\text{2nd}}^2 \sim \frac{r_+^2 -\hat{a}^2}{1+\frac{\hat{a^2}}{L^2}}\left(d\theta^2 + \sin^2\theta d\psi^2 \right)\,.
\end{align}
Therefore, the regularity requires the identification of $\psi=0$ and $\psi = 2\pi$.

Now, let us focus on the first part at a point $\theta$ and $\psi$. At this point, $d\phi = - \frac{\hat{a}\hat{\Xi}}{r_+^2 - \hat{a}^2}d\tau$. The $\omega_+$ becomes
\begin{align}
\omega_+ = \frac{r_+^2-\hat{a}^2}{\hat{\rho}_+^2}\left(1 + \frac{\hat{a}^2 \sin^2\theta}{r_+^2-\hat{a}^2}  \right)d\tau=d\tau\,.
\end{align}
Using this, the first part of the metric

\begin{align}
ds_{\text{1st}}^2 = \frac{\hat{\rho}_+^2 }{\alpha} d\eta^2 +\frac{\alpha\eta^2 \hat{\rho}_+^2}{4(r_+^2-\hat{a}^2)^2}\omega_+^2 = \frac{\hat{\rho}_+^2 }{\alpha}\left(d\eta^2 + \frac{\eta^2 \alpha^2}{4(r_+^2-\hat{a}^2)^2}d\tau^2 \right)\,.
\end{align}
To avoid a conical singularity, the period of $\tau$, equivalent to the inverse temperature, is identified by
\begin{align}
T = \frac{\hat{\Delta}'_r(r_+)}{4\pi(r_+^2-\hat{a}^2)}\,.
\end{align}
Also, $\phi$ has a period $\Omega' / T$ for the regular behavior, where $\Omega'=\frac{\hat{a}\,\hat{\Xi}}{r_+^2-\hat{a}^2}$. This is identified with the angular velocity, measured relative to a frame rotating at infinity. Therefore, the angular velocity is given by
\begin{align}
\Omega = \Omega' + \frac{a}{L^2}\,.
\end{align}
The explicit expression of the temperature is
\begin{align}
T = \frac{r_+}{4 \pi  \left(r_+^2-\hat{a}^2\right)}\left(1+\frac{\hat{a}^2}{r_+^2}+\frac{3}{L^2} \left(r_+^2-\frac{\hat{a}^2}{3}\right)\right)\,.
\end{align}
The corresponding expression in the Lorentzian signature is obtained by $\hat{a}\to -i a$. In addition, the conjugate variable entropy is
\begin{align}
S = \frac{1}{4 \mathbf{G}_4}\int_0^{2\pi} d\phi \int_0^{\pi}d\theta\sqrt{g_{\theta\theta}g_{\phi\phi}}=\frac{\pi }{\mathbf{G}_4}\left(\frac{r_+^2+a^2}{1-\frac{a^2}{L^2}}\right)\,.
\end{align}

\section{Constrained Area Functional for the Thermodynamic-Volume Surface}

\label{app:minsurf-functional}

In this section, we present the constrained variational problem used in section~\ref{sec:thermodynamic-volume-surface}. We first explicitly show that the prescription reproduces the event horizon of a static black hole. We then derive the corresponding functional for the Kerr-AdS geometry, discuss its slow-rotation expansion, and summarize the boundary conditions used in the numerical construction. Throughout this appendix, the enclosed volume is defined by the LW volume (\ref{intro volume functional}) rather than by the Riemannian proper spatial volume on a constant-time slice.


\subsection{Static black hole metric}

\label{app:static-constrained-functional}

We begin with the static AdS black hole metric discussed in Section~\ref{subsec:dyonic}. Its metric form is
\begin{align}
ds^2=-U(r)\,dt^2+\frac{dr^2}{U(r)}+r^2\left(d\theta^2+\sin^2\theta\,d\phi^2\right),
\label{eq:appendix-static-metric}
\end{align}
where the explicit form of \(U(r)\) is given in equation~\eqref{dyonicU}. The argument below does not depend on the precise normalization of the electric and magnetic charges.

Consider a closed axisymmetric surface described by $r=R(\theta)$. The induced metric on this surface is
\begin{align}\label{eq:appendix-static-induced-metric}
ds^2_{\Sigma}=\left[R(\theta)^2+\frac{R'(\theta)^2}{U(R(\theta))}\right]d\theta^2+R(\theta)^2\sin^2\theta\,d\phi^2 .
\end{align}
Its area and enclosed LW volume are therefore
\begin{align}\label{eq:appendix-static-area-volume}
\mathsf{A}[R]&=2\pi\int_0^\pi d\theta\,\sin\theta\,R(\theta)\sqrt{R(\theta)^2+\frac{R'(\theta)^2}{U(R(\theta))}},\nonumber\\
\mathsf{V}[R]&=\frac{2\pi}{3}\int_0^\pi d\theta\,\sin\theta\,R(\theta)^3.
\end{align}
To impose the thermodynamic-volume constraint, we introduce
\begin{align}\label{eq:appendix-static-constrained-action}
\mathcal I[R,\lambda]=\mathsf{A}[R]-\lambda\left(\mathsf{V}[R]-\mathsf{V}_{\mathrm{th}}\right),
\end{align}
where \(\lambda\) is a Lagrange multiplier.

Using $x=-\cos\theta$, the area functional can be written as
\begin{align}
\mathsf{A}[R]=2\pi \int_{-1}^1 dx  R(x)\sqrt{R(x)^2 + (1-x^2)\frac{R'(x)^2}{U(R(x))}}\,.
\end{align}
The corresponding Euler-Lagrange equation leads to
\begin{align}
&R''-\frac{R'}{2}\left(R'\frac{ \partial_R U(R)}{U(R)}+\frac{4 x}{1-x^2}\right)  -\frac{3}{R}\left(R'\right)^2-\frac{x}{R^2}\frac{\left(R'\right)^3}{ U(R)}-\frac{2 R }{1-x^2}U(R)\nonumber\\
&+\lambda\frac{  U(R)}{R \left(1-x^2\right)}  \left(R^2+\frac{\left(1-x^2\right) R'(x)^2}{U(R)}\right)^{3/2} =0\,.
\end{align}
For a constant-radius profile $R(x)=R_0$ outside the horizon, the constrained Euler-Lagrange equation gives $\lambda=2/R_0$. The fixed-volume condition $V(R_0)=V_{\rm th}=4\pi r_h^3/3$ then determines $R_0=r_h$, understood as the limiting surface approached from outside the horizon. Therefore, the horizon surface $R_0=r_h$ is a stationary surface. The volume enclosed by the horizon also satisfies $\mathsf{V}[r_h]= \mathsf{V}_{\mathrm{th}}$. Hence, the thermodynamic-volume surface coincides with the event horizon in the static geometry. The Schwarzschild--AdS result follows immediately.

\subsection{Kerr--AdS constrained functional}
\label{app:kerr-constrained-functional}

We now consider the Kerr--AdS geometry.  On a constant-time slice, the spatial line element is
\begin{align}
dl^2=\frac{\rho^2}{\Delta_r}\,dr^2+\frac{\rho^2}{\Delta_\theta}\,d\theta^2&+\frac{\sin^2\theta}{\rho^2\Xi^2}
\left[\Delta_\theta(r^2+a^2)^2-a^2\Delta_r\sin^2\theta\right]d\phi^2 ,
\label{eq:appendix-kerr-spatial-metric}
\end{align}
For a closed axisymmetric surface $r=r_+\xi(\theta)$, Its area functional is
\begin{align}
\mathsf{A}[\xi]=\frac{2\pi}{\Xi}\int_0^\pi d\theta\,\sin\theta\,\left[\sqrt{\left[\frac{1}{\Delta_\theta}+\frac{\dot{r}^2}{\Delta_r}\right]\left[\Delta_\theta(r^2+a^2)^2-a^2\Delta_r\sin^2\theta\right]}\,\right]_{r=r_+\xi(\theta)}.
\label{eq:appendix-kerr-area}
\end{align}
where 
\begin{align}
&\Delta_r =r_+^2 \frac{r_+^2}{L_{\rm eff}^2}(\xi-1) \left[\xi  \left(\frac{L_{\rm eff}^2}{r_+^2}+\xi ^2+\xi +1\right)-\frac{a^2}{r_+^2}  \left(\frac{L_{\rm eff}^2}{r_+^2}-\xi \right) \right]\nonumber\\
&\Delta_\theta=1-\frac{a^2}{L_{\rm eff}^2}\cos^2\theta,~~\rho^2=r_+^2\xi(\theta)^2+a^2\cos^2\theta,~~\Xi=1- \frac{a^2}{L_{\rm eff}^2}\,,
\label{eq:appendix-kerr-definitions}
\end{align}
and the dot denotes the derivative with respect to $\theta$. One may use dimensionless expressions given as 
\begin{align}\label{dimensionless a L}
\mathsf{a}=a/r_+~,~~\mathsf{L}=L_{\rm eff}/r_+~,~~{\Delta}_\xi(\xi;\mathsf{a},\mathsf{L})=\Delta_r/r_+^2~.
\end{align}
Then, the dimensionless area functional is given by
\begin{align}\label{Area dless}
\tilde{\mathsf{A}}[\xi]=\frac{\mathsf{A}[\xi]}{r_+^2}=2\pi\int_0^\pi d\theta\,\frac{\sin\theta}{1-\frac{\mathsf{a}^2}{\mathsf{L}^2}}\mathbb{S}\,,
\end{align}
where
\begin{align}
&\mathbb{S}=\sqrt{\left[\frac{1}{\Delta_\theta}+\frac{\dot{\xi}^2}{{\Delta}_\xi}\right]\left[\Delta_\theta(\xi^2+\mathsf{a}^2)^2-\mathsf{a}^2\Delta_\xi\sin^2\theta\right]}, \nonumber\\
&\Delta_\theta=1-\frac{\mathsf{a}^2}{\mathsf{L}^2}\cos^2\theta,~\Delta_\xi=\frac{(\xi-1)}{\mathsf{L}^2} \left[\left(1+\xi+\xi^2+\mathsf{L}^2+\mathsf{a}^2\right)\xi  -\mathsf{a}^2\mathsf{L}^2 \right].
\end{align}

In addition, we define the dimensionless volume from (\ref{Volume-functional}) as follows:
\begin{align}
\tilde{\mathsf{V}}[\xi]\equiv\frac{\mathsf{V}[\xi]}{r_+^3}=2\pi\int_0^\pi d\theta\,\frac{\sin\theta}{1-\frac{\mathsf{a}^2}{\mathsf{L}^2}}
\left[
\frac{\xi(\theta)^3}{3}+\mathsf{a}^2\,\xi(\theta)\cos^2\theta\right].
\label{eq:appendix-kerr-volume}
\end{align}
The relevant dimensionless functional $\tilde{\mathcal{I}}$ defined by $\tilde{\mathcal{I}}[\xi,\tilde{\lambda}]\equiv\mathcal{I}[\xi,\lambda]/r_+^2$ can be written as follows:
\begin{align}
\tilde{\mathcal{I}}[\xi,\tilde{\lambda}]=\tilde{\mathsf{A}}[\xi] - \tilde{\lambda}\left(\tilde{\mathsf{V}}[\xi]-\tilde{\mathsf{V}}_{\rm th}\right)\,,
\end{align}
where $\tilde{\lambda}=\lambda\, r_+$, and $\tilde{\mathsf{V}}_{\rm th}$ is the dimensionless version of (\ref{Vth-J}) given as
\begin{align}
\tilde{\mathsf{V}}_{\rm th}= \frac{4 \pi}{3}   \frac{1+ {\mathsf{a}^2} }{\left(1-\frac{\mathsf{a}^2}{\mathsf{L}^2}\right)^2}  \left(1+\frac{\mathsf{a}^2}{2}- \frac{\mathsf{a}^2}{2\mathsf{L}^2} \right)\,.
\end{align}
The equation of motion for the constrained functional $\tilde{\mathcal{I}}[\xi,\tilde{\lambda}]$ is given as follows:
\begin{align}\label{eom xi}
&\frac{d}{d(\cos\theta)}\left[  \frac{\sin\theta\,\dot{\xi}}{\Delta _{\xi}} \left(\left(\mathsf{a}^2+\xi ^2\right)^2 \Delta _{\theta }-\mathsf{a}^2 \Delta _{\xi } \sin^2\theta\right)  \right]=\tilde{\lambda } \left(\mathsf{a}^2 \cos^2\theta+\xi ^2\right)\nonumber\\
&+\frac{\mathbb{S}}{2}\left(\frac{ \Delta_{\theta }\dot{\xi}^2   }{\Delta _{\xi}+ \Delta_{\theta }\dot{\xi}^2  }\right) \frac{\partial_\xi\Delta_\xi}{\Delta_{\xi}} - \frac{4\, \xi  \left(\mathsf{a}^2+\xi ^2\right) \Delta _{\theta }-\mathsf{a}^2 \sin^2\theta\,\partial_\xi \Delta_{\xi} }{2\,\mathbb{S}} \left(\frac{1}{\Delta _{\theta }}+\frac{\dot{\xi }^2}{\Delta _{\xi }}\right)\,.
\end{align}
In addition to this, we consider variation with respect to $\tilde{\lambda}$(or $\lambda$) to impose the fixed-volume condition
\begin{align}
\mathsf{V}[\xi]=\mathsf{V}_{\mathrm{th}} ~(\tilde{\mathsf{V}}[\xi]=\tilde{\mathsf{V}}_{\rm th}).
\label{eq:appendix-kerr-volume-constraint}
\end{align}
Unlike the nonrotating geometries, a constant-radius profile does not solve the full finite-rotation equation (\ref{eom xi}) in general. Nevertheless, the leading slow-rotation correction can be obtained analytically.

\subsection{Slow-rotation expansion}

\label{app:slow-rotation-expansion}

We now solve the constrained problem perturbatively for small rotation.  Since the Kerr--AdS geometry is invariant under reversal of the rotation direction, \(a\rightarrow-a\), the surface profile admits an expansion in even powers of $\mathsf{a}$:
\begin{align}
\xi(\theta)
=
1+\mathsf{a}^2u(\theta)+\mathcal{O}(\mathsf{a}^4).
\label{eq:appendix-small-a-profile}
\end{align}
The horizon condition \(\Delta_r(r_+)=0\) implies 
\begin{align}
\Delta_\xi(\xi)=\mathsf{a}^2\left(1+\frac{3}{\mathsf{L}^2}\right)u(\theta)+\mathcal{O}(\mathsf{a}^4).
\label{eq:appendix-delta-r-expansion}
\end{align}
This relation is important because the radial component of the spatial metric diverges at the horizon.  Although \(\xi'(\theta)=\mathcal{O}(\mathsf{a}^2)\), the combination \(\xi'^2/\Delta_\xi(\xi)\) contributes already at order \(\mathsf{a}^2\).

Expanding the LW volume \eqref{eq:appendix-kerr-volume} gives
\begin{align}
\tilde{\mathsf{V}}[\xi]=\frac{4\pi }{3}+2\pi \mathsf{a}^2\left[\int_0^\pi d\theta\,
\sin\theta\,u(\theta)+\frac{2}{3}+\frac{2}{3\mathsf{L}^2}\right]+\mathcal{O}(\mathsf{a}^4).
\label{eq:appendix-volume-expansion}
\end{align}
The Kerr--AdS thermodynamic volume in equation~\eqref{Vth-J} expands as
\begin{align}
\tilde{\mathsf{V}}_{\mathrm{th}}=\frac{4\pi}{3}+2\pi \mathsf{a}^2
\left[1+\frac{1}{\mathsf{L}^2}\right]+\mathcal{O}(\mathsf{a}^4).
\label{eq:appendix-vth-expansion}
\end{align}
The fixed-volume condition, therefore, becomes 
\begin{align}
\int_0^\pi d\theta\,\sin\theta\,u(\theta)=\frac{1}{3}\left(1 +\frac{1}{\mathsf{L}^2}\right).
\label{eq:appendix-small-a-volume-constraint}
\end{align}

We next expand the area functional.  Using equation~\eqref{eq:appendix-delta-r-expansion}, one obtains
\begin{align}
\tilde{\mathsf{A}}[\xi]={}&4\pi +2\pi \mathsf{a}^2\Bigg[2\int_0^\pi d\theta\,\sin\theta\,u(\theta)+2+
\frac{2}{\mathsf{L}^2}\nonumber\\
&\hspace{27mm}+\frac{1}{2\left(1+ 3/\mathsf{L}^2\right)}\int_0^\pi d\theta\,\sin\theta\,
\frac{\dot{u}(\theta)^2}{u(\theta)}\Bigg]+\mathcal{O}(\mathsf{a}^4).
\label{eq:appendix-area-expansion}
\end{align}
The first integral in equation~\eqref{eq:appendix-area-expansion} is fixed by the volume constraint~\eqref{eq:appendix-small-a-volume-constraint}.
Consequently, the profile dependence of the area at this order is entirely contained in
\begin{align}
\delta \tilde{\mathsf{A}}_{\rm shape}=\frac{\pi \mathsf{a}^2}{1+ 3/\mathsf{L}^2}\int_0^\pi d\theta\,\sin\theta\,\frac{\dot{u}(\theta)^2}{u(\theta)}.
\label{eq:appendix-shape-dependent-area}
\end{align}
For the enclosing surfaces considered here, the leading outward displacement satisfies \(u(\theta)>0\).  Writing \(u(\theta)=w(\theta)^2\), the shape-dependent contribution becomes
\begin{align}
\delta \tilde{\mathsf{A}}_{\rm shape}
=\frac{4\pi \mathsf{a}^2}{1+ 3/\mathsf{L}^2}\int_0^\pi d\theta\,\sin\theta\,\dot{w}(\theta)^2.
\label{eq:appendix-shape-dependent-area-w}
\end{align}
This contribution is manifestly non-negative. The area is therefore minimized when \(w(\theta)\), and hence
\(u(\theta)\), is constant.  The volume constraint then fixes
\begin{align}
u(\theta)=\frac{1}{6}\left(1 + 1/\mathsf{L}^2 \right).
\label{eq:appendix-small-a-u}
\end{align}
Substituting this result into equation~\eqref{eq:appendix-small-a-profile}
gives
\begin{align}
\xi(\theta)=1+\mathsf{a}^2\frac{\left(1 + 1/\mathsf{L}^2 \right)}{6}+\mathcal{O}(\mathsf{a}^4).
\label{eq:appendix-small-a-final-profile}
\end{align}

The result above shows that the leading rotational correction is a
uniform displacement.  To determine the first nontrivial angular
deformation, we extend the perturbative ansatz to quartic order:
\begin{equation}
    \xi(\theta)=1+\mathsf{a}^2\frac{\left(1 + 1/\mathsf{L}^2 \right)}{6}+\mathsf{a}^{4}v(\theta)+\mathcal{O}(\mathsf{a}^{6}).
    \label{eq:appendix-xi-a4-ansatz}
\end{equation}
We also expand the Lagrange multiplier as
\begin{equation}
    \tilde{\lambda}= 2+ \mathsf{a}^{2}\tilde{\lambda}_{2} +\mathcal{O}(\mathsf{a}^{4}).
    \label{eq:appendix-lambda-expansion}
\end{equation}
At this order, it is convenient to substitute \eqref{eq:appendix-xi-a4-ansatz} directly into the exact constrained Euler-Lagrange equation~\eqref{eom xi}. Although $v(\theta)$ enters the profile at order $\mathsf{a}^{4}$, its derivatives already contribute to the first nontrivial expansion of the Euler-Lagrange equation because $\Delta_{\xi}(\xi(\theta))=\mathcal{O}(\mathsf{a}^{2})$ near the horizon.

After using \eqref{eq:appendix-xi-a4-ansatz}, the resulting equation can be written as
\begin{align}
\frac{6\left( \ddot{v}(\theta)+\cot\theta\,\dot{v}(\theta)\right)}{\left(1 + 1/\mathsf{L}^2\right) \left(1+ 3/\mathsf{L}^2 \right)}+\left(1-1/\mathsf{L}^2\right)P_{2}(\cos\theta)+ \tilde{\lambda}_{2}+\frac{4}{3}\left(1 +1/\mathsf{L}^2\right)=0.
    \label{eq:appendix-v-equation}
\end{align}
The angular source contains only the $\ell=0$ and $\ell=2$ axisymmetric harmonics.  Regularity at the poles and reflection symmetry across the equatorial plane therefore allow us to write
\begin{equation}
v(\theta)=v_{0}+v_{2}P_{2}(\cos\theta).
    \label{eq:appendix-v-decomposition}
\end{equation}
Using a Legendre polynomial identity\footnote{$\left(\frac{d^{2}}{d\theta^{2}}+\cot\theta\frac{d}{d\theta}\right) P_{2}(\cos\theta) = -6P_{2}(\cos\theta)$, where $P_{2}(\cos\theta)=\frac{1}{2}\left(3\cos^{2}\theta-1\right)$.}, the $\ell=0$ component of \eqref{eq:appendix-v-equation} fixes the correction to the Lagrange multiplier:
\begin{equation}
    \tilde{\lambda}_{2}= -\frac{4}{3}\left(1+1/\mathsf{L}^2\right).
    \label{eq:appendix-lambda2}
\end{equation}
Furthermore, the $\ell=2$ component determines the quadrupolar coefficient:
\begin{equation}
v_{2}=\frac{1}{36}\left(1-1/\mathsf{L}^4\right)\left(1+3/\mathsf{L}^2\right).
    \label{eq:appendix-v2}
\end{equation}
The constant mode $v_{0}$ does not enter the differential part of the Euler-Lagrange equation at this order.  It is instead determined by expanding the fixed-volume condition $\tilde{\mathsf{V}}[\xi]=\tilde{\mathsf{V}}_{\mathrm{th}}$ through order $\mathsf{a}^{4}$. This gives 
\begin{equation}
v_{0}=\frac{1}{36}\left(1+1/\mathsf{L}^2\right)\left(3 +5/\mathsf{L}^2 \right).
    \label{eq:appendix-v0}
\end{equation}
Therefore, we obtain the solution upto the quartic order as follows:
\begin{align}
\xi(\theta)={}&1+\frac{\mathsf{a}^{2}}{6}\left(1+1/\mathsf{L}^2\right)\nonumber\\
    &+\frac{\mathsf{a}^{4}}{36}\Bigg[\left(1+1/\mathsf{L}^2\right)\left(3+ 5/\mathsf{L}^2\right)+\left(1-1/\mathsf{L}^4\right)\left(1+3/\mathsf{L}^2\right)P_{2}(\cos\theta)\Bigg]+\mathcal{O}(\mathsf{a}^{6}).
    \label{eq:appendix-xi-a4-final}
\end{align}
One can see that the constant-radius solution found at quadratic order is therefore not preserved at quartic order. The first angular deformation is an axisymmetric quadrupole. The pole-equator difference is
\begin{equation}
\xi(0)-\xi\left({\pi}/{2}\right)=\frac{\mathsf{a}^{4}}{24}\left(1-1/\mathsf{L}^4\right)\left(
        1+ 3/\mathsf{L}^2\right)+\mathcal{O}(\mathsf{a}^{6}).
    \label{eq:appendix-pole-equator}
\end{equation}
This result provides an analytic explanation for the nonconstant profile obtained in the finite-rotation numerical construction. As shown in Figure \ref{fig:minsurf_horizon} and the approximated solution (\ref{eq:appendix-xi-a4-final}), the minimal surface shows shapes protruding more on the polar sides and the equator side for small rotation and finite rotation, respectively.

\subsection{Boundary conditions for the numerical construction}

\label{app:numerical-implementation}

For a finite rotation, the constrained equation \eqref{eom xi} is solved numerically together with the fixed-volume condition~\eqref{eq:appendix-kerr-volume-constraint}.  We impose reflection symmetry across the equatorial plane and solve the problem on the
interval \(0\leq\theta\leq\pi/2\).  Regularity at the rotation axis and reflection symmetry at the equator require
\begin{align}\label{eq:appendix-numerical-boundary-conditions}
\xi'(0)=0,\qquad\xi'\!\left({\pi}/{2}\right)=0.
\end{align}
The value of the Lagrange multiplier is determined simultaneously with the surface profile by enforcing the fixed-volume condition. The resulting solution is the constrained minimal-area representative within the class of closed axisymmetric surfaces considered in this work.

\providecommand{\href}[2]{#2}\begingroup\raggedright

\endgroup


\begin{thebibliography}{10}




\bibitem{Kastor:2009wy}
D.~Kastor, S.~Ray and J.~Traschen,
``Enthalpy and the Mechanics of AdS Black Holes,''
Class. Quant. Grav. \textbf{26}, 195011 (2009)
[arXiv:0904.2765 [hep-th]].

\bibitem{Dolan:2010ha}
B.~P.~Dolan,
``The cosmological constant and the black hole equation of state,''
Class. Quant. Grav. \textbf{28}, 125020 (2011)
[arXiv:1008.5023 [gr-qc]].

\bibitem{Dolan:2011xt}
B.~P.~Dolan,
``Pressure and volume in the first law of black hole thermodynamics,''
Class. Quant. Grav. \textbf{28}, 235017 (2011)
[arXiv:1106.6260 [gr-qc]].

\bibitem{Cvetic:2010jb}
M.~Cvetic, G.~W.~Gibbons, D.~Kubiznak and C.~N.~Pope,
``Black Hole Enthalpy and an Entropy Inequality for the Thermodynamic Volume,''
Phys. Rev. D \textbf{84}, 024037 (2011)
[arXiv:1012.2888 [hep-th]].

\bibitem{Kubiznak:2012wp}
D.~Kubiznak and R.~B.~Mann,
``P-V criticality of charged AdS black holes,''
JHEP \textbf{07}, 033 (2012)
[arXiv:1205.0559 [hep-th]].

\bibitem{Kubiznak:2016qmn}
D.~Kubiznak, R.~B.~Mann and M.~Teo,
``Black hole chemistry: thermodynamics with Lambda,''
Class. Quant. Grav. \textbf{34}, no.6, 063001 (2017)
[arXiv:1608.06147 [hep-th]].

\bibitem{Hawking:1982dh}
S.~W.~Hawking and D.~N.~Page,
``Thermodynamics of Black Holes in anti-De Sitter Space,''
Commun. Math. Phys. \textbf{87}, 577 (1983)

\bibitem{Kubiznak:2014zwa}
D.~Kubiznak and R.~B.~Mann,
``Black hole chemistry,''
Can. J. Phys. \textbf{93}, no.9, 999-1002 (2015)
[arXiv:1404.2126 [gr-qc]].

\bibitem{Witten:1998zw}
E.~Witten,
``Anti-de Sitter space, thermal phase transition, and confinement in gauge theories,''
Adv. Theor. Math. Phys. \textbf{2}, 505-532 (1998)
[arXiv:hep-th/9803131 [hep-th]].

\bibitem{Henneaux:1984ji}
M.~Henneaux and C.~Teitelboim,
``THE COSMOLOGICAL CONSTANT AS A CANONICAL VARIABLE,''
Phys. Lett. B \textbf{143}, 415-420 (1984)

\bibitem{Teitelboim:1985dp}
C.~Teitelboim,
``THE COSMOLOGICAL CONSTANT AS A THERMODYNAMIC BLACK HOLE PARAMETER,''
Phys. Lett. B \textbf{158}, 293-297 (1985)

\bibitem{Creighton:1995au}
J.~D.~E.~Creighton and R.~B.~Mann,
``Quasilocal thermodynamics of dilaton gravity coupled to gauge fields,''
Phys. Rev. D \textbf{52}, 4569-4587 (1995)
[arXiv:gr-qc/9505007 [gr-qc]].

\bibitem{Altamirano:2013ane}
N.~Altamirano, D.~Kubiznak and R.~B.~Mann,
``Reentrant phase transitions in rotating anti\textendash{}de Sitter black holes,''
Phys. Rev. D \textbf{88}, no.10, 101502 (2013)
[arXiv:1306.5756 [hep-th]].

\bibitem{Frassino:2014pha}
A.~M.~Frassino, D.~Kubiznak, R.~B.~Mann and F.~Simovic,
``Multiple Reentrant Phase Transitions and Triple Points in Lovelock Thermodynamics,''
JHEP \textbf{09}, 080 (2014)
[arXiv:1406.7015 [hep-th]].

\bibitem{Altamirano:2013uqa}
N.~Altamirano, D.~Kubiz\v{n}\'ak, R.~B.~Mann and Z.~Sherkatghanad,
``Kerr-AdS analogue of triple point and solid/liquid/gas phase transition,''
Class. Quant. Grav. \textbf{31}, 042001 (2014)
[arXiv:1308.2672 [hep-th]].

\bibitem{Wei:2014hba}
S.~W.~Wei and Y.~X.~Liu,
``Triple points and phase diagrams in the extended phase space of charged Gauss-Bonnet black holes in AdS space,''
Phys. Rev. D \textbf{90}, no.4, 044057 (2014)
[arXiv:1402.2837 [hep-th]].

\bibitem{Dehghani:2020blz}
A.~Dehghani, S.~H.~Hendi and R.~B.~Mann,
``Range of novel black hole phase transitions via massive gravity: Triple points and $N$-fold reentrant phase transitions,''
Phys. Rev. D \textbf{101}, no.8, 084026 (2020)
[arXiv:2009.07980 [hep-th]].

\bibitem{Altamirano:2014tva}
N.~Altamirano, D.~Kubiznak, R.~B.~Mann and Z.~Sherkatghanad,
``Thermodynamics of rotating black holes and black rings: phase transitions and thermodynamic volume,''
Galaxies \textbf{2}, 89-159 (2014)
[arXiv:1401.2586 [hep-th]].

\bibitem{Johnson:2014yja}
C.~V.~Johnson,
``Holographic Heat Engines,''
Class. Quant. Grav. \textbf{31}, 205002 (2014)
[arXiv:1404.5982 [hep-th]].

\bibitem{Dolan:2014vba}
B.~P.~Dolan, A.~Kostouki, D.~Kubiznak and R.~B.~Mann,
``Isolated critical point from Lovelock gravity,''
Class. Quant. Grav. \textbf{31}, no.24, 242001 (2014)
[arXiv:1407.4783 [hep-th]].

\bibitem{Hennigar:2016xwd}
R.~A.~Hennigar, R.~B.~Mann and E.~Tjoa,
``Superfluid Black Holes,''
Phys. Rev. Lett. \textbf{118}, no.2, 021301 (2017)
[arXiv:1609.02564 [hep-th]].

\bibitem{Wei:2015iwa}
S.~W.~Wei and Y.~X.~Liu,
``Insight into the Microscopic Structure of an AdS Black Hole from a Thermodynamical Phase Transition,''
Phys. Rev. Lett. \textbf{115}, no.11, 111302 (2015)
[erratum: Phys. Rev. Lett. \textbf{116}, no.16, 169903 (2016)]
[arXiv:1502.00386 [gr-qc]].


\bibitem{Mancilla:2024spp}
R.~Mancilla,
``Generalized Euler Equation from Effective Action: Implications for the Smarr Formula in AdS Black Holes,''
[arXiv:2410.06605 [hep-th]].



\bibitem{Astorino:2016ybm}
M.~Astorino,
``Thermodynamics of Regular Accelerating Black Holes,''
Phys. Rev. D \textbf{95}, no.6, 064007 (2017)
[arXiv:1612.04387 [gr-qc]].

\bibitem{Anabalon:2018ydc}
A.~Anabal\'on, M.~Appels, R.~Gregory, D.~Kubiz\v{n}\'ak, R.~B.~Mann and A.~Ovg\"un,
``Holographic Thermodynamics of Accelerating Black Holes,''
Phys. Rev. D \textbf{98}, no.10, 104038 (2018)
[arXiv:1805.02687 [hep-th]].

\bibitem{Anabalon:2018qfv}
A.~Anabal\'on, F.~Gray, R.~Gregory, D.~Kubiz\v{n}\'ak and R.~B.~Mann,
``Thermodynamics of Charged, Rotating, and Accelerating Black Holes,''
JHEP \textbf{04}, 096 (2019)
[arXiv:1811.04936 [hep-th]].

\bibitem{Wu:2022xmp}
J.~Wu and R.~B.~Mann,
``Thermodynamically stable phases of asymptotically flat Lovelock black holes,''
Class. Quant. Grav. \textbf{40}, no.14, 145009 (2023)
[arXiv:2212.08673 [hep-th]].

\bibitem{Dolan:2013ft}
B.~P.~Dolan, D.~Kastor, D.~Kubiznak, R.~B.~Mann and J.~Traschen,
``Thermodynamic Volumes and Isoperimetric Inequalities for de Sitter Black Holes,''
Phys. Rev. D \textbf{87}, no.10, 104017 (2013)
[arXiv:1301.5926 [hep-th]].

\bibitem{Mbarek:2018bau}
S.~Mbarek and R.~B.~Mann,
``Reverse Hawking-Page Phase Transition in de Sitter Black Holes,''
JHEP \textbf{02}, 103 (2019)
[arXiv:1808.03349 [hep-th]].

\bibitem{Simovic:2018tdy}
F.~Simovic and R.~B.~Mann,
``Critical Phenomena of Charged de Sitter Black Holes in Cavities,''
Class. Quant. Grav. \textbf{36}, no.1, 014002 (2019)
[arXiv:1807.11875 [gr-qc]].

\bibitem{Mbarek:2016mep}
S.~Mbarek and R.~B.~Mann,
``Thermodynamic Volume of Cosmological Solitons,''
Phys. Lett. B \textbf{765}, 352-358 (2017)
[arXiv:1611.01131 [hep-th]].

\bibitem{Quijada:2023fkc}
C.~Quijada, A.~Anabal\'on, R.~B.~Mann and J.~Oliva,
``Triple points of gravitational AdS solitons and black holes,''
Phys. Rev. D \textbf{110}, no.2, L021902 (2024)
[arXiv:2308.16341 [hep-th]].

\bibitem{Maldacena:1997re}
J.~M.~Maldacena,
``The Large N limit of superconformal field theories and supergravity,''
Adv. Theor. Math. Phys. \textbf{2}, 231-252 (1998)
[arXiv:hep-th/9711200 [hep-th]].

\bibitem{Karch:2015rpa}
A.~Karch and B.~Robinson,
``Holographic Black Hole Chemistry,''
JHEP \textbf{12}, 073 (2015)
[arXiv:1510.02472 [hep-th]].

\bibitem{Sinamuli:2017rhp}
M.~Sinamuli and R.~B.~Mann,
``Higher Order Corrections to Holographic Black Hole Chemistry,''
Phys. Rev. D \textbf{96}, no.8, 086008 (2017)
[arXiv:1706.04259 [hep-th]].

\bibitem{Cong:2021fnf}
W.~Cong, D.~Kubiznak and R.~B.~Mann,
``Thermodynamics of AdS Black Holes: Critical Behavior of the Central Charge,''
Phys. Rev. Lett. \textbf{127}, no.9, 091301 (2021)
[arXiv:2105.02223 [hep-th]].

\bibitem{Kim:2018nzy}
K.~K.~Kim and B.~Ahn,
``Thermodynamic Volume in AdS/CFT,''
EPJ Web Conf. \textbf{168}, 07003 (2018)

\bibitem{Ahmed:2023snm}
M.~B.~Ahmed, W.~Cong, D.~Kubiz\v{n}\'ak, R.~B.~Mann and M.~R.~Visser,
``Holographic Dual of Extended Black Hole Thermodynamics,''
Phys. Rev. Lett. \textbf{130}, no.18, 181401 (2023)
[arXiv:2302.08163 [hep-th]].

\bibitem{Mann:2024sru}
R.~B.~Mann,
``Recent Developments in Holographic Black Hole Chemistry,''
JHAP \textbf{4}, no.1, 1-26 (2024)
[arXiv:2403.02864 [hep-th]].

\bibitem{Randjbar-Daemi:1983xth}
S.~Randjbar-Daemi, A.~Salam and J.~A.~Strathdee,
``Stability of Instanton Induced Compactification in Eight-dimensions,''
Nucl. Phys. B \textbf{242}, 447-472 (1984)

\bibitem{Kim:2018mfv}
K.~K.~Kim, S.~Koh and H.~S.~Yang,
``Expanding Universe and Dynamical Compactification Using Yang-Mills Instantons,''
JHEP \textbf{12}, 085 (2018)
[arXiv:1810.12291 [hep-th]].

\bibitem{Kim:2023gbm}
K.~K.~Kim, S.~Koh and G.~Tumurtushaa,
``Dynamical Compactification with Matter,''
JHEP \textbf{06}, 181 (2023)
[arXiv:2303.13758 [hep-th]].

\bibitem{Ho:2023jqz}
J.~Ho, K.~K.~Kim, S.~Koh and H.~S.~Yang,
``Generalization of instanton-induced inflation and dynamical compactification,''
JHEP \textbf{11}, 050 (2023)
[arXiv:2309.02056 [hep-th]].

\bibitem{Ho:2023wcg}
J.~Ho, K.~K.~Kim and H.~S.~Yang,
``Einstein Structure of Squashed Four-Spheres,''
[arXiv:2309.05335 [math.DG]].










\bibitem{Dolan:1980jb}
B.~P.~Dolan,
``General SU(2) Classical Instanton Configurations in Curved Space Times,''
J. Phys. A \textbf{14}, 1205 (1981)



\bibitem{Hyun:2017nkb}
S.~Hyun, J.~Jeong, S.~A.~Park and S.~H.~Yi,
``Thermodynamic Volume and the Extended Smarr Relation,''
JHEP \textbf{04}, 048 (2017)
[arXiv:1702.06629 [hep-th]].



\bibitem{Arkani-Hamed:2006emk}
N.~Arkani-Hamed, L.~Motl, A.~Nicolis and C.~Vafa,
``The String landscape, black holes and gravity as the weakest force,''
JHEP \textbf{06}, 060 (2007)
[arXiv:hep-th/0601001 [hep-th]].





\bibitem{Kachru:2003aw}
S.~Kachru, R.~Kallosh, A.~D.~Linde and S.~P.~Trivedi,
``De Sitter vacua in string theory,''
Phys. Rev. D \textbf{68}, 046005 (2003)
[arXiv:hep-th/0301240 [hep-th]].
 


\bibitem{Gregory:1993vy}
R.~Gregory and R.~Laflamme,
``Black strings and p-branes are unstable,''
Phys. Rev. Lett. \textbf{70}, 2837-2840 (1993)
[arXiv:hep-th/9301052 [hep-th]].

\bibitem{Miyaji:2015yva}
M.~Miyaji and T.~Takayanagi,
``Surface/State Correspondence as a Generalized Holography,''
PTEP \textbf{2015} (2015) no.7, 073B03
doi:10.1093/ptep/ptv089
[arXiv:1503.03542 [hep-th]].

\bibitem{Bousso:2015qqa}
R.~Bousso and N.~Engelhardt,
``Proof of a New Area Law in General Relativity,''
Phys. Rev. D \textbf{92} (2015) no.4, 044031 doi:10.1103/PhysRevD.92.044031
[arXiv:1504.07660 [gr-qc]].

\bibitem{Jafferis:2015del}
D.~L.~Jafferis, A.~Lewkowycz, J.~Maldacena and S.~J.~Suh,
``Relative entropy equals bulk relative entropy,''
JHEP \textbf{06} (2016), 004
doi:10.1007/JHEP06(2016)004
[arXiv:1512.06431 [hep-th]].

\bibitem{AlBalushi:2020rqe}
A.~Al Balushi, R.~A.~Hennigar, H.~K.~Kunduri and R.~B.~Mann,
``Holographic Complexity and Thermodynamic Volume,''
Phys. Rev. Lett. \textbf{126} (2021) no.10, 101601
doi:10.1103/PhysRevLett.126.101601
[arXiv:2008.09138 [hep-th]].

\bibitem{AlBalushi:2020heq}
A.~Al Balushi, R.~A.~Hennigar, H.~K.~Kunduri and R.~B.~Mann,
``Holographic complexity of rotating black holes,''
JHEP \textbf{05} (2021), 226
doi:10.1007/JHEP05(2021)226
[arXiv:2010.11203 [hep-th]].

\bibitem{Couch:2016exn}
J.~Couch, W.~Fischler and P.~H.~Nguyen,
``Noether charge, black hole volume, and complexity,''
JHEP \textbf{03} (2017), 119
doi:10.1007/JHEP03(2017)119
[arXiv:1610.02038 [hep-th]].

\bibitem{Susskind:2014rva}
L.~Susskind,
``Computational Complexity and Black Hole Horizons,''
Fortsch. Phys. \textbf{64} (2016), 24-43
doi:10.1002/prop.201500092
[arXiv:1403.5695 [hep-th]].

\bibitem{Stanford:2014jda}
D.~Stanford and L.~Susskind,
``Complexity and Shock Wave Geometries,''
Phys. Rev. D \textbf{90} (2014) no.12, 126007
doi:10.1103/PhysRevD.90.126007
[arXiv:1406.2678 [hep-th]].

\bibitem{Brown:2015lvg}
A.~R.~Brown, D.~A.~Roberts, L.~Susskind, B.~Swingle and Y.~Zhao,
``Complexity, action, and black holes,''
Phys. Rev. D \textbf{93} (2016) no.8, 086006
doi:10.1103/PhysRevD.93.086006
[arXiv:1512.04993 [hep-th]].




\bibitem{Zhang:2024mxb}
M.~Zhang, J.~Sun and R.~B.~Mann,
``Generalized holographic complexity of rotating black holes,''
JHEP \textbf{09} (2024), 050
doi:10.1007/JHEP09(2024)050
[arXiv:2401.08571 [hep-th]].

\bibitem{Parikh:2005qs}
M.~K.~Parikh,
``The Volume of black holes,''
Phys. Rev. D \textbf{73} (2006), 124021
doi:10.1103/PhysRevD.73.124021
[arXiv:hep-th/0508108 [hep-th]].

\bibitem{Ballik:2013uia}
W.~Ballik and K.~Lake,
``Vector volume and black holes,''
Phys. Rev. D \textbf{88} (2013) no.10, 104038
doi:10.1103/PhysRevD.88.104038
[arXiv:1310.1935 [gr-qc]].



\bibitem{Ryu:2006bv}
S.~Ryu and T.~Takayanagi,
``Holographic derivation of entanglement entropy from AdS/CFT,''
Phys. Rev. Lett. \textbf{96}, 181602 (2006)
doi:10.1103/PhysRevLett.96.181602
[arXiv:hep-th/0603001 [hep-th]].


\bibitem{Ryu:2006ef}
S.~Ryu and T.~Takayanagi,
``Aspects of Holographic Entanglement Entropy,''
JHEP \textbf{08}, 045 (2006)
doi:10.1088/1126-6708/2006/08/045
[arXiv:hep-th/0605073 [hep-th]].


\bibitem{Swingle:2009bg}
B.~Swingle,
``Entanglement Renormalization and Holography,''
Phys. Rev. D \textbf{86}, 065007 (2012)
[arXiv:0905.1317 [cond-mat.str-el]].



\bibitem{Nozaki:2012zj}
M.~Nozaki, S.~Ryu and T.~Takayanagi,
``Holographic Geometry of Entanglement Renormalization in Quantum Field Theories,''
JHEP \textbf{10}, 193 (2012)
[arXiv:1208.3469 [hep-th]].


\bibitem{Swingle:2012wq}
B.~Swingle,
``Constructing holographic spacetimes using entanglement renormalization,''
[arXiv:1209.3304 [hep-th]].



\bibitem{Czech:2015kbp}
B.~Czech, L.~Lamprou, S.~McCandlish and J.~Sully,
``Tensor Networks from Kinematic Space,''
JHEP \textbf{07}, 100 (2016)
[arXiv:1512.01548 [hep-th]].






\end{thebibliography}
\end{document}